\documentclass[a4paper,aps,pra,10pt,onecolumn,notitlepage,superscriptaddress,dvipsnames,allowfontchangeintitle,
tightenlines,floatfix,showpacs,accepted=2020-10-06]{quantumarticle}
\pdfoutput=1
\usepackage{authblk}
\usepackage[top=1.2in, bottom=1.7in, left=1.2in, right=1.2in]{geometry}

\usepackage[british]{babel}
\usepackage[utf8]{inputenc}
\usepackage[pdftex]{graphicx,color}
\usepackage{amsmath}
\usepackage{amssymb}
\usepackage{epsfig}
\usepackage{amsthm}
\usepackage{tikz}
\usepackage{wrapfig}
\usepackage{comment}
\usepackage{latexsym,revsymb}

\usepackage[numbers, square, comma, sort&compress]{natbib}
\definecolor{vdarkgreen}{RGB}{0,51,0}
\definecolor{vdarkblue}{RGB}{0,0,51}
\definecolor{vdarkred}{RGB}{0,100,0}
\usepackage{hyperref}

\definecolor{special1}{HTML}{CA3542}
\newcommand{\one}[1]{{\color{black} #1}}

\newcommand{\two}[1]{{\color{black} #1}}

\definecolor{special3}{HTML}{F5793A}
\newcommand{\three}[1]{{\color{black} #1}}
\newcommand{\hlt}[1]{\two{\textbf{\emph{#1}}}}

\usepackage{mathtools}

\newcommand{\vertiii}[1]{{\left\vert\kern-0.25ex\left\vert\kern-0.25ex\left\vert #1 
    \right\vert\kern-0.25ex\right\vert\kern-0.25ex\right\vert}}
 
\newcommand{\ket}[1]{| #1 \rangle} 
\newcommand{\bra}[1]{\langle #1 |} 
\newcommand{\f}[2]{\textstyle{\frac{#1}{#2}}}
\newcommand{\g}[1]{\mathbf{#1}}

\newcommand{\proj}[1]{| #1 \rangle\langle #1 |}

\newcommand{\m}[1]{\mathcal{#1}}
\newcommand{\tr}[1]{\mathrm{tr}\! \left[#1\right]}

\makeatletter
\newtheorem*{rep@theorem}{\rep@title}
\newcommand{\newreptheorem}[2]{%
\newenvironment{rep#1}[1]{%
 \def\rep@title{#2 \ref{##1}}%
 \begin{rep@theorem}}%
 {\end{rep@theorem}}}
\makeatother

\newtheorem{theorem}{Theorem}
\numberwithin{theorem}{subsection}
\newreptheorem{theorem}{Theorem}

\newtheorem{corollary}[theorem]{Corollary}
\newreptheorem{corollary}{Corollary}
\newtheorem{lemma}[theorem]{Lemma}
\numberwithin{theorem}{subsection} 
\newreptheorem{lemma}{Lemma}
\newtheorem{Definition}[theorem]{Definition}
\numberwithin{theorem}{subsection}

\numberwithin{theorem}{subsection} 
\newreptheorem{example}{Example}

\reversemarginpar

\newcommand\blfootnote[1]{%
  \begingroup
  \renewcommand\thefootnote{}\footnote{#1}%
  \addtocounter{footnote}{-1}%
  \endgroup
}

\title{\textbf{A review of Quantum Cellular Automata}}

\date{}

\author{Terry Farrelly$^{*}$}

\affiliation{Institut f\"{u}r Theoretische Physik, Leibniz Universit\"{a}t Hannover, 
30167 Hannover, Germany}
\affiliation{ARC Centre for Engineered Quantum Systems, School of Mathematics and Physics, University of Queensland, Brisbane, QLD 4072, Australia}

\begin{document}

\maketitle
\blfootnote{$^{*}$ farreltc@tcd.ie}

\vspace{-1.2cm}

\begin{abstract}
Discretizing spacetime is often a natural step towards modelling physical systems.  For quantum systems, if we also demand a strict bound on the speed of information propagation, we get quantum cellular automata (QCAs).  These originally arose as an alternative paradigm for quantum computation, though more recently they have 
\one{found application in understanding topological phases of matter and have}
been proposed as models of periodically driven (Floquet) quantum systems, where QCA methods were used to classify their phases.
QCAs have also been used as a natural discretization of quantum field theory, and some interesting examples of QCAs have been introduced that become interacting quantum field theories in the continuum limit.  This review discusses all of these applications, as well as some other interesting results on the structure of quantum cellular automata, including the tensor-network unitary approach, the index theory and higher dimensional classifications of QCAs.
\end{abstract}
\newpage
\tableofcontents
\newpage

\section{Introduction}
\label{sec:introduction}
\hlt{Cellular automata} (CAs) are fascinating systems:\ despite having extremely simple dynamics, we often see the emergence of highly complex behaviour.  CAs were first introduced by von Neumann \cite{vN66,Burks70}, who wanted to find a model that was universal for computation and could in some sense replicate itself \cite{Delorme99}.  Building on a suggestion by Ulam, von Neumann considered systems with discrete variables on a two-dimensional lattice evolving over discrete timesteps via a local update rule.  Von Neumann's program would eventually be successful, as CAs exist that can efficiently simulate Turing machines and display self replication.  An example of a CA where this can be seen is Conway's game of life \cite{BCG82}, which also involves a two-dimensional lattice of bits.  There one sees remarkable patterns and dynamics, and it indeed includes configurations that can self replicate.

In general a cellular automaton is a $d$-dimensional lattice of bits (or a more general finite set of  variables) that updates over discrete timesteps.  The update rule is the same everywhere and updates a bit based only on its state and those of its neighbours.  An important example of a cellular automaton is known as the rule $110$ CA.  In this case, we have a one-dimensional array of bits, and the updated state of each bit after one timestep depends on its previous state and that of its two nearest neighbours.  The update rule is best summarized by the table below.
\begin{center}
    \begin{tabular}{ | c | c | c | c | c | c | c | c | c |}
    \hline
    Initial state of a bit and its neighbours & 111 & 110 & 101 & 100 & 011 & 010 & 001 & 000 \\ \hline
    New state of middle bit & 0 & 1 & 1 & 0 & 1 & 1 & 1 & 0\\
    \hline
    \end{tabular}
\end{center}
The name comes from the result of the update rule \cite{Wolfram83}.  Treating $01101110$, which is the bottom row of the table above specifying the update rule, as binary and converting to decimal gives $110$.  This CA can simulate a Turing machine \cite{Cook04}, and it was shown later that this can even be done efficiently (with only polynomial overhead) \cite{NW06}.
\begin{figure}[!ht]
{\centering
\resizebox{11cm}{!}{\includegraphics{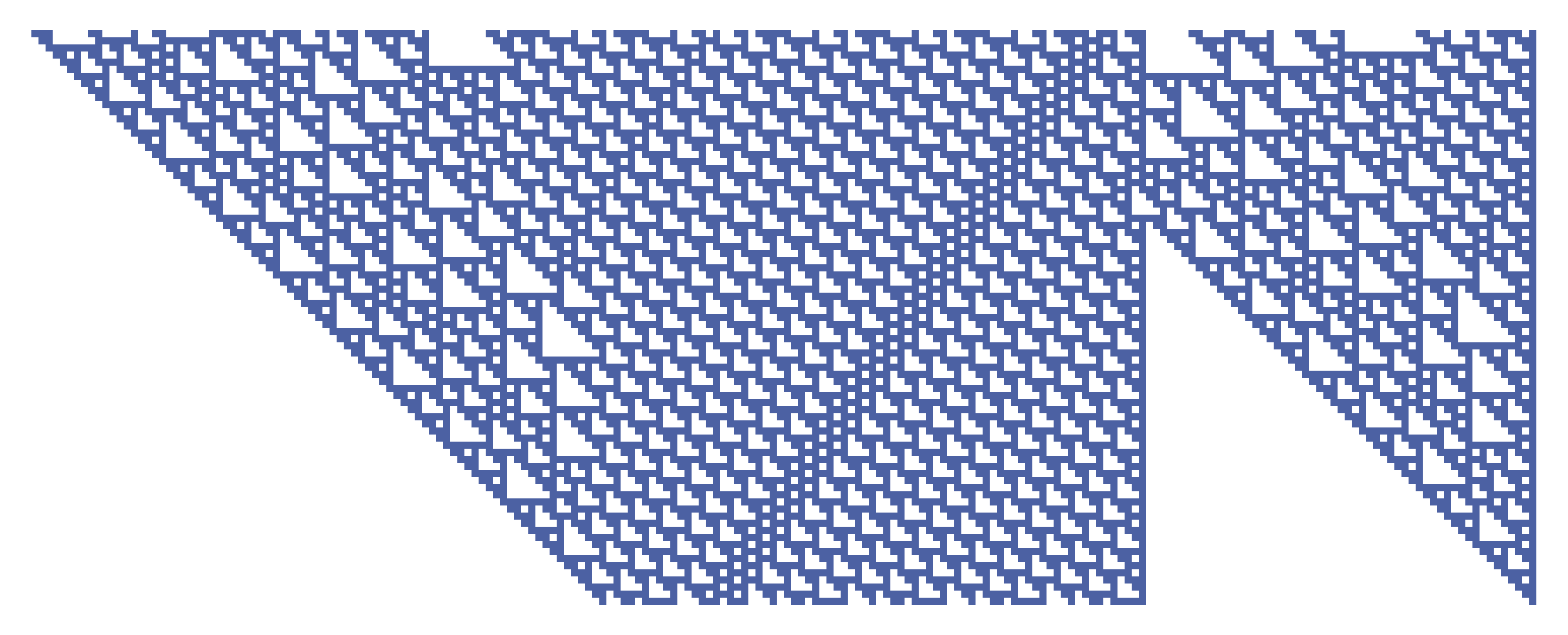}} \caption[Example of a CA]{An example of a CA evolution for the rule $110$ CA, with time going up.  Bits with value $1$ are represented by the blue squares, while $0$ is represented by white squares. \label{fig:CA}}
}
\end{figure}

It would turn out that CAs have many practical applications, which include traffic models, fluid flows, biological pattern formation and reaction-diffusion systems \cite{Chopard12}.  A special case of CAs are lattice gas cellular automata, which were used to model fluid dynamics, and indeed, with the right choice of model, one recovers the Navier-Stokes equation \cite{Succi01,Wolf-Gladrow00}.  These have been replaced by lattice Boltzmann models \cite{Succi01,Wolf-Gladrow00}, which use continuous functions instead of discrete variables at the lattice sites.  More ambitiously, CAs have been put forward as discrete models of physics \cite{AG12game}, having many desirable properties, such as locality and homogeneity of the dynamics.  But, while this is a tempting idea, physics is fundamentally quantum, and CAs cannot correctly describe, e.g., Bell inequality violation, which results from quantum entanglement.\footnote{Of course, it could be possible if, e.g., one allowed superluminal signalling, but any such workaround would take us away from CAs and probably give rise to a rather obscure model.}

\hlt{Quantum cellular automata} (QCAs) are the quantum version of CAs.  The initial rough idea can be traced back at least as far as \cite{Feynman82}, where it was proposed that to simulate quantum physics it makes sense to consider \emph{quantum} computers as opposed to classical.  Indeed, it is a widely held belief that quantum simulations of physics will probably be the first application of a quantum computer \cite{Trabesinger12}.  Initial specific models for QCAs were then given in, e.g., \cite{Margolus86,Lloyd93,Watrous95}.  In \cite{Watrous95} QCAs were introduced as an alternative paradigm for quantum computation and were shown to be universal, meaning they could efficiently simulate a quantum Turing machine.
In the earlier paper \cite{Lloyd93}, a model of quantum computation was given that could be described as a classically controlled QCA, meaning that over each discrete timestep we have a choice of global translationally invariant unitaries to apply.  The proposal suggested applying homogeneous electric fields of different frequencies to a one-dimensional \two{line} of systems (e.g., a row of molecules), and it was argued that this would be a realizable model of quantum computation.  A potential benefit of this type of quantum computation is that it involves homogeneous global operations on all qudits, in contrast to the circuit model, where single or few qubit addressability is essential, and this is sometimes difficult in, e.g., trapped ions.

Perhaps surprisingly, the route from CAs to QCAs was not so straightforward.  Various roadblocks arose, and naive approaches, such as simply extending the classical evolution by linearity to get something quantum, do not always work.  Another problem is that in CAs each cell can be updated simultaneously, but we may need to copy the original state of a cell, so that we can update its neighbours after it has been updated.  This is impossible in the quantum setting because of the no-cloning theorem \cite{Nielsen00,BW03}.  To get around many of these problems, there were various constructive approaches, typically defining QCAs as layers of finite-depth circuits, possibly in combination with shifting some qudits left or right.  It was only later that an axiomatic definition of QCAs was given, which captured the spirit of CAs, while ensuring that the dynamics were quantum \cite{SW04}.  It was not initially clear how useful this definition was in dimensions greater than one, at least until it was shown in \cite{ANW11} that any QCA defined in this axiomatic way can be written as a local finite-depth circuit by appending local ancillas.  With this it seems that a sensible definition of QCAs has been found, and any of the constructive approaches satisfy it.  Essentially, the axiomatic definition (which we will use) of a QCA is that it comprises a spatial lattice with quantum systems at each site, and it evolves over discrete timesteps via a unitary\footnote{Or automorphism of the observable algebra for truly infinite systems, but we will get back to this later.} that is locality preserving.  Locality preservation is the discrete analogue of relativistic causality, which means that in the Heisenberg picture local operators get mapped to local operators.  We will go into more detail about all this in section \ref{sec:Definition}.

One ambitious application of QCAs is as discrete models of physics.  In fact, once spacetime is taken to be discrete, and a maximum speed of propagation of information is assumed, then also assuming unitarity leaves us with a QCA.  In contrast, it is clear that continuous-time dynamics via a local Hamiltonian on a lattice will not suffice, as there is no strict upper bound on information propagation (see section \ref{sec:Ham}).
Whether or not physics can truly be discretized is a nontrivial and interesting question in its own right \cite{Tong12}.  And while roadblocks such as fermion doubling may arise and would still have to be dealt with, QCAs have nevertheless been proposed as an interesting class of discrete models of physics \cite{AG12game}.  

In a more specific setting, QCAs have been considered as discretized quantum field theories, i.e., as an alternative regularization of quantum field theory.  They have a couple of nice properties in this regard:\ they have a strict upper bound on the speed of propagation of information and the lattice provides a cutoff, which allows us to regulate the infinities that need to be handled carefully in quantum field theory.  At the moment however, it seems unlikely that QCAs can offer any advantages for performing calculations in comparison with, e.g., dimensional regularization.  Still, we may be interested in QCA discretizations of quantum field theory for simulating physics on a quantum computer.  This is more promising, and it may offer a new perspective on quantum simulations of quantum field theory.  (In fact, the discrete-time path integral for bosonic lattice field theories (such as non-abelian gauge theory) is a QCA \cite{FS20}.)  Indeed, much progress was made in understanding quantum field theory in the 70s by Wilson, while he was trying to understand how to simulate QFT on a classical computer \cite{Wilson93}.

Recently, in condensed matter physics, QCAs have been proposed as models for quantum lattice systems evolving via time-dependent periodic Hamiltonians (i.e., Floquet systems).  Typically, such systems do not quite fulfil the locality preserving component of the definition of a QCA, but because of Lieb-Robinson bounds \cite{LR72}, they do approximately.  In this context, QCA techniques have been useful for classifying chiral phases of Floquet systems \cite{PFM16,FPP17}.  The key to this is the index theory for one-dimensional QCAs, which allows one to assign a number to one-dimensional QCAs quantifying the net information flow along the \two{line}.  And crucially this number is robust against perturbations.  By applying the index theory to the boundary dynamics of two-dimensional Floquet systems with many-body localization, we get a classification of different dynamical phases given by the index of the boundary theory.  \one{QCAs have other intriguing applications to understanding topological phases of matter, as in \cite{HFH18}, where it was shown that a QCA can be used to disentangle the ground state of a three dimensional toy model with interesting topological phases on two dimensional boundaries.  Whether this boundary topological order can be realized by a local commuting projector Hamiltonian is still unknown, but it is now known to be equivalent to whether this disentangling QCA is a trivial QCA (defined in section \ref{sec:partitioning}).}


There are a handful of older reviews of QCAs available \cite{04AT,Horowitz08,Wiesner09}, and a key reference that provides a thorough discussion of QCAs up to $2004$ is \cite{SW04}.  There are also a couple of theses on QCAs, which offer good reviews of specific aspects of the subject as well as some novel insights; see, e.g., \cite{vanDam96,Perez07,Vogts09,Arrighi09,Guetschow12}.  Since these were written, many new results have appeared, so the time is ripe for an up-to-date treatment of the topic.  Coincidentally, another review was written at the same time as this one \cite{Arrighi19}, which is quite complementary to this one, as it has a detailed discussion of intrinsic universality, a topic that is mentioned only briefly here in section \ref{sec:other}.

\subsection{Two examples of QCAs}
\label{sec:ex}
Let us look at two simple examples of QCAs that not only provide us with some intuition but also arise as components of most constructions of QCAs.  Before we start, note that, for continuous-time systems the dynamics is determined by the Schr\"{o}dinger equation, but we are going to work with discrete-time systems, so there is no Schr\"{o}dinger equation.  Instead, we work directly with the unitary operator that evolves the system every timestep.
\begin{figure}[!ht]
{\centering
\resizebox{7cm}{!}{\includegraphics{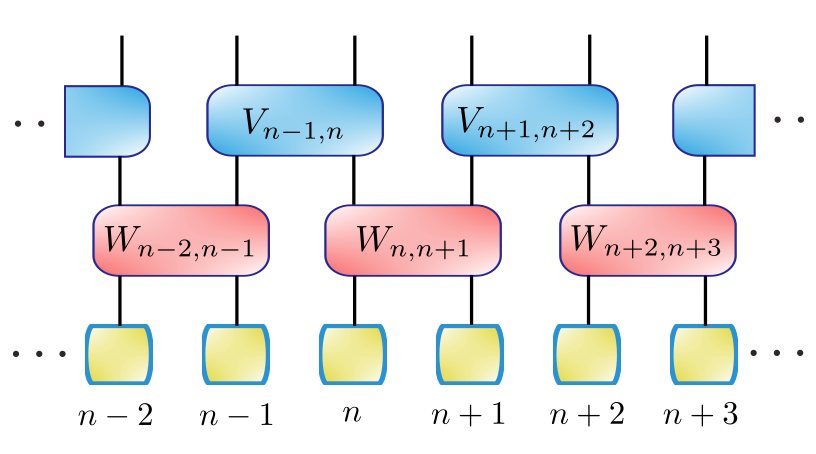}} \caption[Example of a QCA]{A simple example of a QCA unitary.  Qubits are represented by the yellow boxes, while the two-qubit local unitaries that implement the dynamics over one timestep are represented by the red and blue rectangles. \label{fig:QCA}}
}
\end{figure}

Take a discrete ring of qubits with position label $n\in\{0,...,N-1\}$, where $N$ is even and we have periodic boundary conditions, so we identify sites $N$ and $0$.  Consider applying a depth-two circuit of unitaries as in figure \ref{fig:QCA}.  Let $V_{2n,2n+1}$ be a unitary acting non trivially on qubits $2n$ and $2n+1$ only.  This means that it acts like the identity on all other qubits.  Similarly, let $W_{2n-1,2n}$ be a unitary acting on qubits $2n-1$ and $2n$.  Consider the unitary circuit 
\begin{equation}
 U=\prod_{n}V_{2n,2n+1}\prod_{m}W_{2m-1,2m}.
\end{equation}
Here, the ordering inside the products is unimportant because, e.g., $V_{2n,2n+1}$ and $V_{2k,2k+1}$ commute for all $n,k$.
As an example, let $V_{n,n+1}=W_{n,n+1}=C_{n}[X_{n+1}]$, which is the controlled-NOT unitary, familiar from quantum computing \cite{Nielsen00}.  Here $n$ labels the control qubit and $n+1$ labels the target qubit.  Explicitly,
\begin{equation}
 C_{n}[X_{n+1}]=\ket{0}_n\bra{0}\otimes\openone_{n+1}+\ket{1}_n\bra{1}\otimes X_{n+1},
\end{equation}
where $X_{n+1}$ denotes the Pauli $X$ operator on site $n+1$.  For completeness, the Pauli operators in the computational basis are
\begin{equation}
\begin{split}
 X & =\ket{0}\bra{1} + \ket{1}\bra{0}\\
 Y & =i(\ket{1}\bra{0} - \ket{0}\bra{1})\\
 Z & =\proj{0}- \proj{1}.
 \end{split}
\end{equation}
To see that the unitary $U$ takes local operators to local operators, it suffices to check that, for any Pauli operator (e.g., $X_n$) localized on site $n$, $u(X_n)=U^{\dagger}X_nU$ is localized on nearby sites.

\begin{figure}[!ht]
{\centering
\resizebox{9cm}{!}{\includegraphics{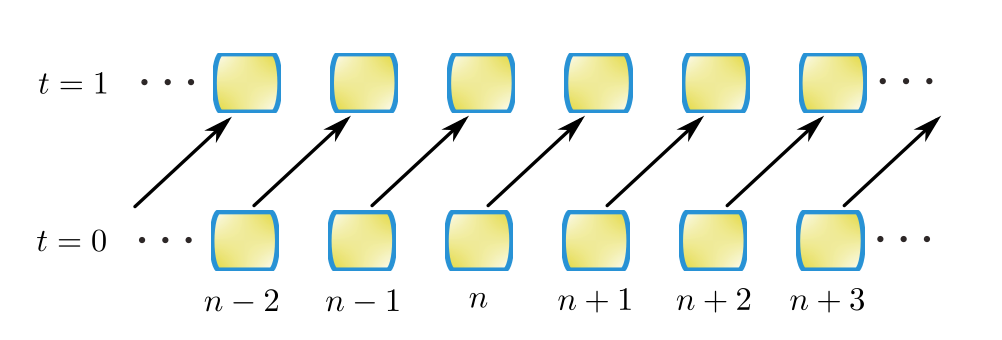}} \caption[A shift QCA]{A simple but important example of a QCA is a shift along the line. \label{fig:B}}
}
\end{figure}
While such constant-depth circuits form an important class of QCAs, we will see in section \ref{sec:ind} that there are QCAs that cannot be expressed in this form.  The quintessential example of this is a shift, which simply shifts each subsystem one step to the right as in figure \ref{fig:B}.  For qudits, the dynamics is given by
\begin{equation}
 s(A_n)= S^{\dagger}A_nS=A_{n-1},
\end{equation}
where $A_n$ is any operator on qudit $n$, and $S$ is the unitary implementing the shift.  This describes a shift of the qudits by one step to the right, and the fact that $A_n$ gets mapped to $A_{n-1}$ is because we are working in the Heisenberg picture.  Using the Heisenberg picture makes it easier to define what it means for the evolution to be locality preserving:\ one simply requires that the operators only spread a finite distance from where they started.  We will discuss this further in section \ref{sec:Definition}.  This second example of a QCA is simple yet surprisingly important.  It will play a major role in many of the discussions to come.

\subsection{Things that are not QCAs}
It is a little confusing that the name quantum cellular automata or other similar sounding names have been used for several unrelated concepts, so it will help to go through what we do \textit{not} mean by QCAs in this review.

One example of this occurs in \cite{Bial94}, where the systems studied are called unitary cellular automata.  These are actually single-particle systems, and these would be called discrete-time quantum walks according to current terminology (see, e.g., \cite{Strauch06,CRW13}).  Similarly, other works from this period refer to discrete-time quantum walks as quantum cellular automata or unitary cellular automata, e.g., \cite{GZ88,Meyer96a,Meyer96}.  In fact, we can often view these discrete-time quantum walks as the dynamics of the one-particle sector of a QCA \cite{SW04,Vogts09}, which is something we will return to in section \ref{sec:QCAs_and_particles}.

Another potential source of confusion are quantum dot cellular automata, which were first introduced in \cite{LTP93,TL94}.  These were originally referred to as quantum cellular automata, and, while the name quantum dot cellular automata seems to have become popular, the abbreviation QCA is often used for these.  Quantum dot cellular automata are a new paradigm for implementing \textit{classical} computing using quantum dots instead of conventional semiconductor-based integrated circuits.  However, a proposal has also been made to use such an array of quantum dots to implement a \textit{quantum} computer \cite{TL01}, though the idea here is to use the circuit model of quantum computation, which typically uses many different few-qubit gates as opposed to uniform global operations.  This particular quantum dot architecture is referred to in \cite{TL01} as a coherent quantum dot cellular automata.

Another similar-sounding but different model are continuous-time quantum cellular automata \cite{PC05}.  These also go by the name Hamiltonian QCAs \cite{NW08} or continuous cellular automata \cite{VC08}.  These do not fit our definition of quantum cellular automata because they evolve via Hamiltonians in continuous time, though they are an interesting model of quantum computation.

QCAs are also not to be confused with attempts to replace quantum mechanics by classical cellular automata \cite{Hooft16}, which is referred to as the cellular automaton interpretation of quantum mechanics.  It is worth mentioning that this proposal considers \textit{classical} cellular automata, which are obviously not quantum, and therefore cannot reproduce the correlations due to quantum entanglement without, e.g., correlated measurement choices.  Describing physics (and specifically quantum physics) by classical cellular automata was also proposed in \cite{Wolfram02}, though this also has problems in describing entanglement \cite{Aaronson04}.

\section{Definition}
\label{sec:Definition}
There had been many different proposals for a concrete definition of QCAs, but the theory was put on a sound mathematical footing in  \cite{RW96,SW04}.  It is worth highlighting that any of the previous proposals for a definition of a QCA (when they make sense, i.e., they involve locality-preserving unitaries or automorphisms), such as local unitary QCAs \cite{PC07}, satisfy the definition we give here.

QCAs include reversible CAs as a subset \cite{Margolus86,Margolus91}, which may be a useful fact, in analogy to how models in classical statistical physics are special cases of quantum systems and are often in the same universality classes as physical materials.  But it is also worth bearing in mind that naive quantization of CAs \cite{AN08} can oddly lead to faster-than-classical signalling \cite{ANW17}.

\subsection{Systems}
\label{sec:systems}
QCAs are defined on quantum lattice systems, meaning we have a discrete spatial lattice $\Gamma$ with quantum systems at each lattice site (or cell).  Some authors consider more general graphs, e.g., \cite{ANW11}, but we will mostly stick with lattices here.  The lattices we consider may be infinite, i.e., $\mathbb{Z}^d$, or finite, possibly with periodic boundary conditions.  (Another approach we will consider are control spaces and families of QCAs, defined in the following section.)

We will take the quantum systems at each lattice site to either be finite-dimensional quantum systems, i.e., qudits, or we take them to be a finite number of fermion modes.\footnote{If we allow for appending a constant number of ancillas per site, fermionic QCAs and  qudit QCAs can both simulate each other efficiently \cite{Farrelly15}.}  After all, since QCAs are the quantum generalization of CAs, it makes sense to not just admit qudits but also other non-classical systems arising in quantum physics, such as fermions.   We will not deal with truly bosonic systems here, i.e., those having infinite-dimensional Hilbert spaces at each lattice site with the usual bosonic creation and annihilation operators.  However, QCAs of continuous variable systems were considered in \cite{KW07}.  These were a specific class of QCAs, called Gaussian QCAs, which evolve via Gaussian operations, i.e., those taking Gaussian states to Gaussian states.  This restriction makes the dynamics tractable.  Similar restrictions for finite dimensional systems at each site give rise to Clifford QCAs, which we discuss in more detail in section \ref{sec:Cliff}.

For finite lattice systems, the total Hilbert space is just the tensor product of the Hilbert spaces corresponding to each lattice site.  Each lattice site $\vec{n}$ has a finite-dimensional quantum system with Hilbert space $\mathcal{H}_{\vec{n}}$, so the total Hilbert space is $\mathcal{H}=\otimes_{\vec{n}\in \Gamma}\mathcal{H}_{\vec{n}}$.  Instead of Hilbert spaces, it will be easier for us to work with the algebra of observables.  We denote the algebra of observables acting on the system at site $\vec{n}$ by $\m{A}_{\vec{n}}$, which is isomorphic to $\m{M}_{d_{\vec{n}}}$, the algebra of $d_{\vec{n}}\times d_{\vec{n}}$ complex matrices.  The total observable algebra is then $\m{A}=\otimes_{\vec{n}\in \Gamma}\m{A}_{\vec{n}}$.  Given a subset of the lattice $\m{R}\subset \Gamma$, we define $\m{A}_{\m{R}}$ to be the operators localized on region $\m{R}$, meaning they act like the identity on all sites outside of $\m{R}$.

For infinite lattices of qudits, it does not make sense to simply take the infinite tensor product of the Hilbert spaces or algebras.  This is because we cannot make sense of, e.g., the inner product of $\ket{0}^{\otimes N}$ and $(i\ket{0})^{\otimes N}$ as $N\rightarrow \infty$.  There are a couple of ways around this:\ most notably, either we restrict the observables we allow (the quasi-local algebra approach), or we restrict the states we allow (the finite-unbounded configurations approach).  From a physical perspective, the former is more satisfying:\ we can only do local operations in a lab, so it makes sense to rule out something like $\sigma_z^{\otimes \infty}$ as an observable.  For this reason, the choice we make is the quasi-local algebra approach.  We will explain why this is the more general approach anyway (a QCA on finite-unbounded configurations is always a representation of a QCA on a quasi-local algebra) in section \ref{sec:Finite unbounded configurations}.

In fact, very little of the formal theory of quasi-local algebras is actually needed, just some of the basics.  It may also be reassuring to note that for QCAs it is usually possible to just restrict our attention to a finite but sufficiently large lattice anyway.  This is because the dynamics is strictly locality-preserving, as we discuss in the following section.  For this reason, \emph{everything that follows can be understood by considering finite lattices}.  Nevertheless, it is also interesting to consider the infinite case, as the quasi-local algebra construction is rather elegant, and allows one to employ ideas from the statistical mechanics of infinite systems \cite{RW96}.

For infinite lattices, we also denote the algebra describing the system at site $\vec{n}$ on the lattice by $\mathcal{A}_{\vec{n}}$, which is isomorphic to $\m{M}_{d_{\vec{n}}}$.  And more generally, for every finite lattice region $\m{R}$, there is an associated algebra $\mathcal{A}_{\m{R}}$ isomorphic to $\otimes_{\vec{n}\in \m{R}}\m{A}_{\vec{n}}$.  Such algebras contain the local elements, which act nontrivially on systems only in region $\m{R}$ and like the identity on the rest of the lattice.  The algebra corresponding to a subregion $\m{R}^{\prime}\subset \m{R}$ is identified with the subalgebra of $\m{A}_{\m{R}}$ acting nontrivially only on sites in $\m{R}^{\prime}$.  The algebra for the whole system is defined to be
\begin{equation}
 \m{A}=\overline{\bigcup_{\m{R}\subset\Gamma}\m{A}_{\m{R}}},
\end{equation}
where the line above denotes closure in the norm.\footnote{The norm is equivalent to the usual operator norm on every finite region.}  In other words, the algebra includes elements defined by Cauchy sequences of local elements, hence the name quasi-local algebra.

For finite lattices, states are just density matrices, whereas in the infinite case, states are assignments of density matrices to each finite region, with a consistency requirement:\ given density matrices $\rho_{\m{R}}$ and $\rho_{\m{R}^{\prime}}$ on $\m{R}$ and $\m{R}^{\prime}$ respectively, with $\m{R}^{\prime}\subset \m{R}$, we require that $\mathrm{tr}_{\m{R}\backslash \m{R}^{\prime}}[\rho_{\m{R}}]=\rho_{\m{R}^{\prime}}$.  (More formally, we could instead start by defining states as linear functionals of the algebra $\omega$, with the constraints $\omega(\openone)=1$ and $\omega(A)\geq 0$ for all positive $A\in\m{A}$, meaning $A=B^{\dagger}B$ for some $B\in\m{A}$.)

We introduce some more of the formalism of C$^*$-algebras in appendix \ref{app}.  A nice introduction to quasi-local algebras can also be found in \cite{Naaijkens13}, and for more detail there is also the two-volume classic standard reference \cite{BR97}.

\subsection{Dynamics}
\label{sec:dyna}
As mentioned earlier, to describe the dynamics of QCAs, it is usually preferable to work in the Heisenberg picture, with observables evolving in time.  This makes it much easier to define and understand a key feature of the dynamics:\ locality preservation.\footnote{This can be done in the Schr{\"o}dinger picture too \cite{ANW11}, but it is more awkward.}  Of course, it is sometimes useful to switch to the Schr{\"o}dinger picture, as we will in some cases, since it is simpler for, e.g., seeing the connection to quantum computation.

So over each timestep the QCA evolves via a unitary operator in the finite case or an automorphism of the algebra in the infinite case.  We denote the dynamics of a QCA by $u: A\rightarrow u(A)$, for any $A\in\m{A}$.  Automorphisms are the natural generalization of unitaries to infinite systems, since they satisfy all the desired properties of unitary maps:\ they preserve, e.g., commutation relations.  However, there need not be any element of the algebra $U\in\m{A}$ that implements this transformation via $u(A)=U^{\dagger}AU$.  In contrast, for finite lattices, there is always such a unitary $U$.  We will often write the evolution as $u(A)$ and speak of automorphisms, though these can be replaced by $U^{\dagger}AU$ and unitaries respectively for finite systems.  Note that $u(AB)=u(A)u(B)$ always, so it is enough to understand how single-site algebras evolve to understand the whole evolution.
\begin{figure}[!ht]
{\centering
\resizebox{9cm}{!}{\includegraphics{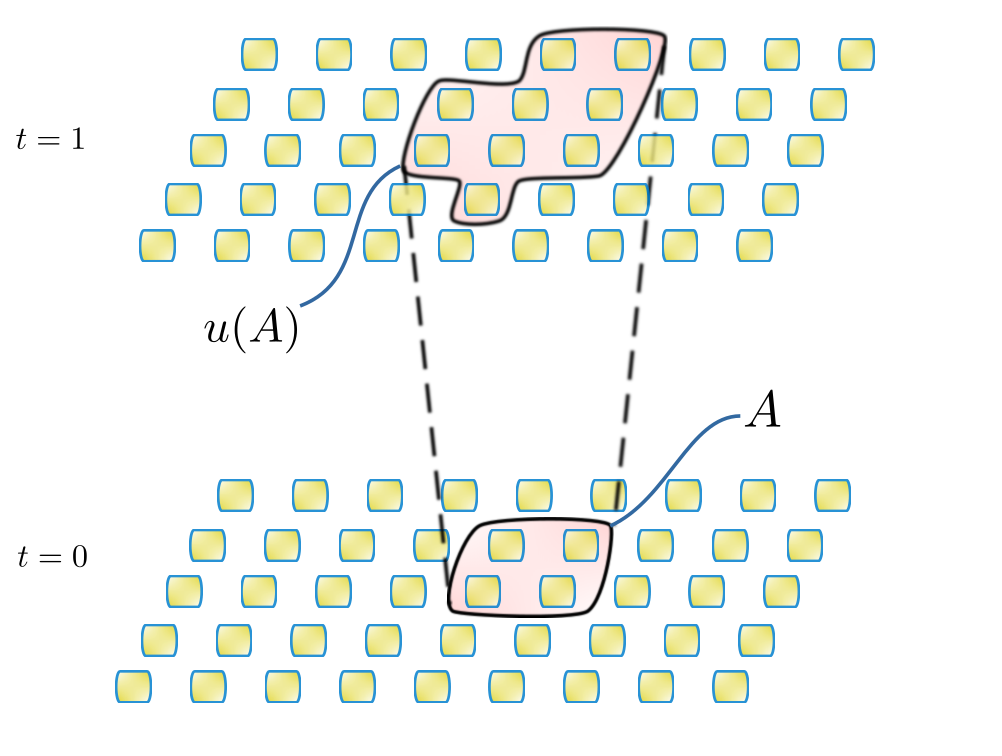}} \caption[Locality preservation]{A unitary or automorphism is locality preserving if it maps local operators to local operators (in the Heisenberg picture).  At $t=0$ an operator $A$ is localized on four sites, but after one timestep the updated operator $u(A)$ is localized on nine sites nearby.  \label{fig:C}}
}
\end{figure}

We also demand that the evolution is \hlt{locality preserving} (see figure \ref{fig:C}).  This means that local operators get mapped to operators localized on a region nearby.  Sometimes the locality preserving property is called \textit{causality} (e.g., in \cite{ANW11,FS13}) in analogy with relativistic causality.  More precisely, we have the following definition.
\begin{Definition}
 The dynamics of a QCA $u$ is locality preserving if there is some \hlt{range} $l\geq 0$ such that, for any $\vec{n}$ and any operator $A$ localized on $\vec{n}$, then $u(A)$ is localized on a region consisting only of sites $\vec{m}$ with $|\vec{n}-\vec{m}|\leq l$.  Here we use the Taxicab metric on the lattice.
\end{Definition}
We may also define the neighbourhood of a point $\vec{n}$, which we denote $\mathcal{N}(\vec{n})$.  Then $\mathcal{N}(\vec{n})$ is the smallest region on which the algebra $u(\mathcal{A}_{\vec{n}})$ is localized.  We can generalize this to describe the neighbourhood of a region $R$, denoted by $\mathcal{N}(R)$.

We can put all this together to give a summary of the definition of a QCA.
\begin{Definition}
 A QCA consists of a discrete hypercubic lattice, which may be finite or be $\mathbb{Z}^d$, with a finite quantum system at each site (qudits and/or fermion modes).  Evolution takes place over discrete timesteps via a locality-preserving automorphism (or unitary for finite systems).
\end{Definition}

We will not always assume that QCAs are translationally invariant \two{(which means that the dynamics commutes with shifts along any lattice direction)}, though this was traditionally assumed in many works.  Sometimes we will restrict to this case, as further interesting results then follow.  It would be more in the spirit of classical CAs, to always consider translationally invariant dynamics, but some of the most interesting QCA structure theorems do not require any form of translational invariance (see section \ref{sec:structure}).  As an aside, it would also be interesting to consider irreversible \two{(i.e., nonunitary)} QCAs, as in \cite{RW96,BW03,PC20}, though not much work has been done in this direction.  In this case, e.g., unitaries would be replaced by completely positive trace-preserving maps\two{, which are the most general possible dynamics consistent with quantum theory \cite{Nielsen00})}.

\one{
Finally, a useful alternative possibility to just fixing a lattice, described in \cite{FH19}, involves \hlt{control spaces}.
In that case, we start with a control space, which is typically a smooth manifold $\mathcal{M}$ with a metric $d(x,y)$.  The manifold can be unbounded or bounded and with or without boundary.  To define a QCA we embed the points of a discrete set $I$, which label lattice sites, into the control space via $\mathcal{I}:I\rightarrow \mathcal{M}$.  The lattice sites have quantum systems associated to them, and they inherit a notion of distance from the metric of the control space.

A \hlt{family of QCAs} is a sequence of QCAs labelled by $i\in\mathbb{N}$ with control space $\mathcal{M}$ with embeddings $\mathcal{I}_i$.  Each QCA has an automorphism $u_i$ with range $l_i$.  And we require that $l_i\rightarrow 0$ as $i\rightarrow 0$, which is because we are now measuring distance with the metric of the control space.  So intuitively, if we take some point $x$, then for larger $i$ the embedding will include more and more points within a fixed distance of $x$.  So we then require that the QCAs' range, as measured with this distance goes to zero as $i\rightarrow \infty$.  As an example, the control space could be a circle with circumference one, and the family of QCAs could correspond to finer and finer equally spaced points on the circle, e.g., $u_i$ could live on all sites separated by a distance $1/2^i$ on the circle.  More interesting examples of control spaces would be higher dimensional tori or the two-dimensional infinite plane with a hole at the origin.
}

\subsection{Global and local transition rules}
As we saw in section \ref{sec:introduction}, in the classical setting, CAs are usually defined by a \textit{local} transition rule.  This tells us how, e.g., the bit at site $n$ updates its state depending only on the states of its immediate neighbours.  In the previous subsection, however, we defined QCAs by a global transition rule---an automorphism or unitary of the observable algebra that preserves locality.  However, we may also specify a QCA by \textit{local} transition rules \cite{SW04}.  To go from a global rule to local rules, we define the local transition rules to be the maps $\alpha_{\vec{n}}: \mathcal{A}_{\vec{n}}\rightarrow \mathcal{A}_{\m{N}(\vec{n})}$ such that
\begin{equation}\label{eq:localrule}
\alpha_{\vec{n}}(A)=u(A)\ \mathrm{for\ any}\ A\in\mathcal{A}_{\vec{n}}.
\end{equation}
This $\alpha_{\vec{n}}$ tells us how observables localized on $\vec{n}$ spread out in $\m{N}(\vec{n})$.  (In the translationally invariant case, it is enough to specify $\alpha_{\vec{0}}$.)  To get something closer to the local rules of classical CAs, we could then switch to the Schr{\"o}dinger picture, though this description will be quite cumbersome.

In equation (\ref{eq:localrule}), we simply defined the local rule by using the global rule, so it is obvious that we get a sensible QCA and that the global dynamics uniquely specifies the local transition rules.  On the other hand, going the opposite direction is more interesting:\ we may start with some candidate for a local rule, some $\beta_{\vec{n}}$ taking $\mathcal{A}_{\vec{n}}$ to $\mathcal{A}_{\m{N}(\vec{n})}$ for every $\vec{n}$.  Then we define a candidate for a global rule $v$ to be linear, and to act on products of operators at different sites \two{in the following way.  If $A$ and $B$ are elements from $\mathcal{A}_{\vec{n}}$ and $\mathcal{A}_{\vec{m}}$ respectively,}
\begin{equation}
 v(AB)=\beta_{\vec{n}}(A)\beta_{\vec{m}}(B).
\end{equation}
Starting from such a local rule, the question is whether this candidate is a bona fide QCA?  For example, our rule may not be a valid automorphism; it may not even be invertible.  The first condition we need to ensure that we have a QCA is that $\beta_{\vec{n}}$ must be an isomorphism, so it must preserve the algebraic structure locally, mapping $\m{A}_{\vec{n}}$ to an isomorphic subalgebra of $\mathcal{A}_{\m{N}(\vec{n})}$ for every $\vec{n}$ (something like $\beta_{\vec{n}}(A)=0$ for all $A$ will obviously not do).  In fact, the only other condition we need is that, for any $\vec{n}\neq \vec{m}$, and any $A\in\mathcal{A}_{\vec{n}}$ and $B\in\mathcal{A}_{\vec{m}}$,
\begin{equation}\label{eq:localrulecomm}
 [\beta_{\vec{n}}(A),\beta_{\vec{m}}(B)]=0.
\end{equation}
These two conditions are enough to guarantee that the map defines an automorphism.  This was proved in \cite{SW04} for one-dimensional translationally invariant QCAs of qudits, but the proofs carry over straightforwardly to non-translationally invariant and higher dimensional systems.

For non-translationally invariant QCAs, checking that equation (\ref{eq:localrulecomm}) holds will usually be impossible, as we cannot check an infinite number of conditions in practice.  On the other hand, for translationally invariant systems, because the neighbourhoods are finite, we need only check that $[\beta_{\vec{0}}(A),\beta_{\vec{\two{n}}}(B)]=0$, for all $A\in\mathcal{A}_{\vec{0}}$ and $B\in\mathcal{A}_{\vec{n}}$, when the neighbourhoods of $\vec{0}$ and $\vec{n}$ overlap, i.e., $\mathcal{N}(\vec{0})\  \cap \mathcal{N}(\vec{n})\neq \emptyset$.  And we need only check this for a basis of elements of $\mathcal{A}_{\vec{0}}$ and $\mathcal{A}_{\vec{n}}$ for each $\vec{n}$ with $\mathcal{N}(\vec{0})\  \cap \mathcal{N}(\vec{n})\neq \emptyset$.  This means that only a finite number of commutators need to be computed, making it feasible to check if a local rule indeed gives a legitimate QCA.

In the translationally invariant case, there is also a natural correspondence between QCAs on infinite lattices and QCAs on finite lattices, which works because the dynamics is locality preserving.  This is called the \hlt{wrapping lemma} \cite{SW04}.  This means that any translationally invariant QCA on an infinite lattice is in one-to-one correspondence with a QCA on any finite lattice with periodic boundary conditions (and the same dimension), provided the finite lattice is big enough that the overlaps of the neighbourhoods of two sites are the same as for the infinite case.  This was recently generalized to fermions in \cite{PP18}.

\subsection{Finite, unbounded configurations}
\label{sec:Finite unbounded configurations}
An alternative way to deal with infinite systems and infinite tensor products (as opposed to the quasi-local algebra approach we use here) constructs a Hilbert space called the \hlt{Hilbert space of finite, unbounded configurations}.  This is used, for example, in \cite{ANW08}.  As we will see, this is covered by our definition, as these are always representations of a subset of QCAs defined in our sense.

The idea of the finite, unbounded configurations approach is to restrict the possible states the system can be in.  We start by assuming that each lattice site can be in one of a finite number of states,  and one of these states is a privileged state, called a quiescent state $q$.  We allow all configurations of the system where only a finite number of sites are not in the quiescent state, e.g., infinite strings like $...q\,q\,q\,a\,b\,c\,q\,q\,q...$ for a one-dimensional system.  Then we use these strings to label an orthonormal basis of a complex vector space.  To get a Hilbert space, we complete the vector space in the inner-product norm.  This Hilbert space is separable (i.e., it has a countable basis) and is sometimes called an incomplete infinite tensor product space.
This is the Hilbert space of finite, unbounded configurations.

Notice that this Hilbert space admits all operators corresponding to the quasi-local algebra of the system:\ we can perform any local operation we like, as this will only change the state in some finite region and will keep us in the Hilbert space of finite, unbounded configurations.  Next we define the dynamics to be determined by a locality-preserving unitary $U$ on this Hilbert space.  (We use the exact same notion of locality preserving as before in this setting, working in the Heisenberg picture.)  Requiring that $U$ is unitary, translationally invariant,\footnote{\two{This is usually but not always an assumption in the finite, unbounded configurations setting.}} and locality preserving means that $U$ must preserve the state $\ket{...qqqqq...}$.  The advantage of this formalism is that we can work with Hilbert spaces and unitaries, instead of algebras and automorphisms.  The disadvantage is that we exclude many interesting states and dynamics on the system (a workaround should be possible by increasing the local dimension in the finite, unbounded configurations setting).  For example, the unitary must always have the \emph{product} state $\ket{...qqqqq...}$ as a $+1$ eigenvector.

Instead of constructing the Hilbert space of finite, unbounded configurations and QCAs on them in this way, we can construct them as a representation of a QCA defined in the quasi-local algebra setting.  (See \cite{Shakeel18} for a similar construction in the translationally invariant case.)  Suppose we have a QCA with dynamics $u$, and let $\omega$ be some invariant state, so $\omega(u(A))=\omega(A)$ for all $A\in\m{A}$.  Here we are writing the expectation value of $A$ in state $\omega$ as $\omega(A)$ (for finite systems, this would be $\tr{\rho A}$ for some density operator $\rho$).  We can always construct a representation of the algebra where the state $\omega$ is represented in the resulting Hilbert space $\mathcal{H}$ by a normalized vector $\ket{\omega}$, with the representation of any operator $A$ being denoted $\pi(A)$.  (This uses the GNS representation of the algebra, which is described in appendix \ref{app}.)  Furthermore, because $u$ leaves the state $\omega$ invariant, there is a unique unitary $U$ that implements the dynamics in the representation, i.e., $U^{\dagger}\pi(A)U=\pi(u(A))$ and $U\ket{\omega}=\ket{\omega}$ (see appendix \ref{app} and \cite{BR97}).  To connect with the finite, unbounded configurations approach, let us suppose that $\omega$ is a pure product state.  Then the representation $\pi(A)$ is irreducible.  We can construct an orthonormal basis for $\m{H}$ by applying local operators $\pi(A)$ to $\ket{\omega}$, and because the Hilbert space must be complete, we get exactly the Hilbert space of finite, unbounded configurations.  As an example with qubits on a line, we could choose $\omega$ to be the $0$ state of all the qubits.  In the representation, we would have $\ket{\omega}=\ket{0...000...0}$.  Then, for example, the action of the Pauli $X$ operator at site $n$ is given by $\pi(X_n)\ket{0...000...0}=\ket{0...010...0}$.  An example of a QCA dynamics leaving $\omega$ invariant is the CNOT QCA in section \ref{sec:ex}.

We should emphasize that any QCA defined in the finite, unbounded configuration picture can be expressed in our formalism, as an automorphism on the quasi-local algebra.  This follows because the unitary implementing the dynamics for a finite, unbounded configuration also \textit{defines} an automorphism of the algebra of local operators that is locality-preserving and hence uniquely defines dynamics for a QCA in the quasi-local algebra picture.
So the finite, unbounded configurations QCAs can be viewed as special cases of our notion of QCAs.  But for some applications, it is useful and intuitive to use the finite, unbounded configurations picture.

To see that there are QCAs that cannot be expressed in terms of a Hilbert space of finite, unbounded configurations (at least without adding local ancillas), consider that a translationally invariant QCA need not have any invariant pure product state, something which we would actually expect to be the case in general.  If this is the case, then it is not directly representable in the finite, unbounded configuration picture.  As quantum field theories, for example, have highly entangled vacua \cite{SW85}, it seems clear that any ideas for discretizing such models with QCAs would not be naively representable by finite, unbounded configurations.

The finite, unbounded configurations approach to QCAs was used as a definition of a \hlt{Schr{\"o}dinger template} of a QCA in \cite{Shakeel18}.  We can easily generalize this by constructing a representation of the quasi-local algebra and dynamics using any invariant state (not necessarily a product state) of the automorphism as described above.  Then one has a picture of the QCA with a Hilbert space of states and dynamics represented by a unitary.  Indeed, it seems reasonable to view the invariant state in that case as a vacuum state and to think of the resulting Hilbert space as the space of localized excitations above the vacuum state.  From that point of view, it would make sense to call the resulting Hilbert space the Hilbert space of \textit{finite, unbounded excitations}.  And so we could interpret the main result of \cite{Shakeel18} as characterizing when a QCA admits a representation in terms of finite, unbounded configurations with an invariant \emph{product} state.

\section{More examples of QCAs}
There are several different specialized classes of QCAs that fit into our framework.  We will look at two below.  
The first class of QCAs involve applying shifts and local unitary circuits.  These typically involve some kind of partitioning scheme for the lattice.  Such QCAs have been particularly important, as they are manifestly unitary and have been used to show universality of QCAs for quantum computation \cite{Watrous95}.  Some discussions of the different models can also be found in \cite{SW04,PC05,SL13}.  The second class of QCAs that we will discuss are Clifford QCAs.  In this case, we restrict to dynamics that map strings of generalized Pauli operators to strings of generalized Pauli operators.  In contrast to the first class of QCAs, which are constructive by their definition, Clifford QCAs, are defined by a constraint on their dynamics, but it is this restriction makes this class of QCAs tractable.

\subsection{Partitioning schemes}
\label{sec:partitioning}
In section \ref{sec:ex} we saw two examples of QCAs:\ constant-depth circuits and shifts.  These basic building blocks can be combined to get various different classes of QCAs that appear in the literature.
\begin{figure}[!ht]
{\centering
\resizebox{5cm}{!}{\includegraphics{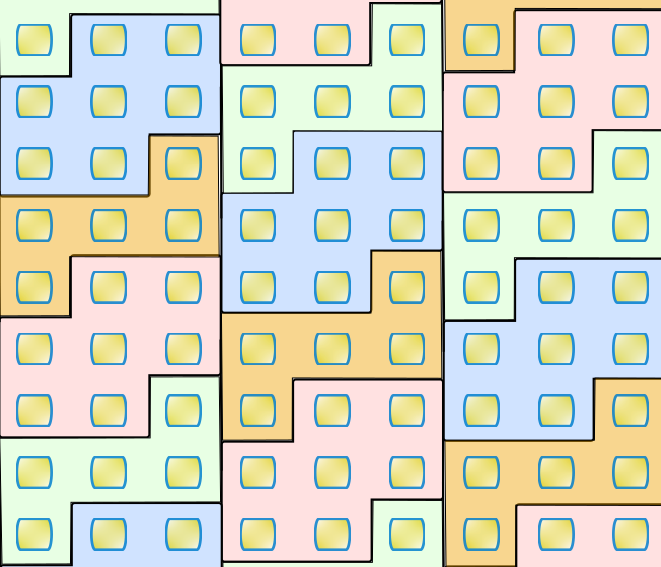}} \caption[Partitioning the lattice]{To define QCA dynamics it is often useful to partition the lattice into supercells by grouping sites together.  (The supercells can have any shape as long as they are finite.)  Then we apply unitaries to each system in the supercells.  More generally we consider several \emph{different} partitions and apply different unitaries to supercells in each partition in sequence to implement the QCA dynamics. \label{fig:D}}
}
\end{figure}

As a first step, we can generalize the \hlt{finite-depth circuit QCA} construction from section \ref{sec:ex} in the following way.
\two{
\begin{Definition}
Finite-depth circuit QCA:\ Consider $P$ different partitions of the lattice into supercells $\mathcal{C}^p_{\vec{n}}$, with $\vec{n}\in\mathbb{Z}^d$ and $p\in\{1,...,P\}$.  This is done in such a way that, for each partition, each cell on the lattice is contained in exactly one supercell (see figure \ref{fig:D}).  Then the dynamics is given by
\begin{equation}
 u_1\circ...\circ u_P,
\end{equation}
where each $u_p$ is a product of unitaries $U^p_{\vec{n}}$ localized on $\mathcal{C}^p_{\vec{n}}$.  In other words,
\begin{equation}\label{eq:supercell}
 u_p(\cdot)=\left(\prod_{\vec{n}}U^{p\dagger}_{\vec{n}}\right) \cdot \left(\prod_{\vec{n}}U^p_{\vec{n}}\right),
\end{equation}
where the order in the products does not matter because the unitaries act on non-overlapping regions and hence commute.
\end{Definition}
}

As an example of QCAs using such a partitioning scheme, let us look at the \hlt{block-partitioned QCAs} defined in \cite{BW03}, which use layers of conditional unitaries.  In the nearest-neighbour case with qubits on a one-dimensional system, we define the operator (acting nontrivially on sites $n-1$, $n$ and $n+1$ only)
\begin{equation}
 D^{ab}_n=\ket{a}\bra{a}_{n-1}\otimes v^{ab}_n\otimes \ket{b}\bra{b}_{n+1},
\end{equation}
where $a,b\in\{0,1\}$ and $v^{ab}_n$ is a unitary on site $n$.  Then we can define the conditional unitary operator
\begin{equation}
 V_n= \sum_{a,b\in\{0,1\}} D^{ab}_n.
\end{equation}
If the qubits $n+1$ and $n-1$ are in, e.g., the $0$ state, this implements the unitary $v^{00}$ on qubit $n$.  The QCA dynamics is then given by a depth-three circuit:\ first we apply $V_n$ to all cells with $n \mathrm{\ mod\ } 3 = 0$, then to all cells with $n \mathrm{\ mod\ } 3 = 1$, and finally to all cells with $n \mathrm{\ mod\ } 3 = 2$.  This way unitaries are never applied simultaneously on overlapping supercells.

The properties of these QCAs were studied in \cite{BW03}, where information transport along the \two{line} and entanglement generation were studied.  This partitioning scheme also makes the construction of non-unitary QCAs straightforward \cite{BW03}, by replacing the local unitaries by local completely positive trace-preserving maps.  This allows one to construct an irreversible QCA that gives a quantum version of the classical rule $110$ CA from section \ref{sec:introduction}.  To see that we need a non-unitary QCA to do this, notice that rule $110$ itself is irreversible.  This can be seen by considering the state with $0$ everywhere and the state with $1$ everywhere.  After one timestep, these both get mapped to the same state.

\begin{figure}[!ht]
{\centering
\resizebox{9cm}{!}{\includegraphics{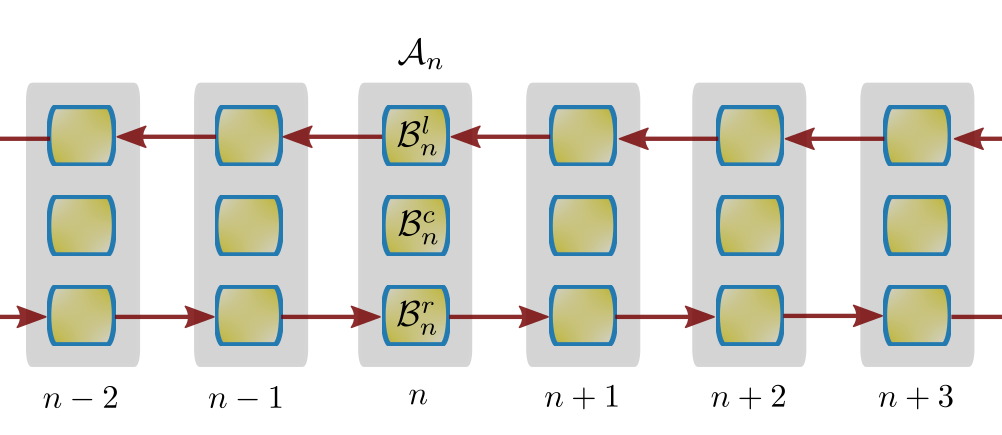}} \caption[Watrous QCAs]{A Watrous QCA shifts some subsystems left and some right while some remain stationary each timestep.  On-site unitaries also get applied before or after the shift step.  Each site has three qudits, represented by yellow boxes:\ $\mathcal{B}^l_{n}$, $\mathcal{B}^c_{n}$ and $\mathcal{B}^r_{n}$. \label{fig:E}}
}
\end{figure}

We can also generalize the shift along a line to shifts along lattice directions for higher dimensional systems.  Suppose the lattice has lattice basis vectors $\vec{e}_i$ where $i\in{1,...,d}$, then we define the shift along lattice direction $\vec{e}_i$ by
\begin{equation}
 s_i(\mathcal{A}_{\vec{n}})=\mathcal{A}_{\vec{n}-\vec{e}_i},
\end{equation}
which has no other effect than to shift the algebra by one step in the direction $\vec{e}_i$.  We may also consider partial shifts.  These would just shift a subalgebra instead of the whole algebra.  One useful example involves a one-dimensional lattice with qudits at each site.  Then we divide the cells into three subcells, which we label by $l, c$ and $r$.  So we may write $\mathcal{A}_{n}=\mathcal{B}^l_{n}\otimes\mathcal{B}^c_{n}\otimes \mathcal{B}^r_{n}$.  Then we define the conditional shift $\sigma$ to be
\begin{equation}
\begin{split}
 \sigma(\mathcal{B}^l_{n}) & =\mathcal{B}^l_{n+1}\\
 \sigma(\mathcal{B}^c_{n}) & =\mathcal{B}^c_{n}\\
 \sigma(\mathcal{B}^r_{n}) & =\mathcal{B}^r_{n-1}.
 \end{split}
\end{equation}
This particular conditional shift (we could define more general versions, particularly in higher lattice dimensions) is useful for constructing the partitioned QCAs in \cite{Watrous95,vanDam96}, which are usually referred to as \hlt{Watrous partitioned QCAs}.  After applying $\sigma$, we may also apply an automorphism $\lambda$ that is just a product of on-site unitaries.  The overall dynamics is as illustrated in figure \ref{fig:E}.  It is easy to see then that the dynamics is an automorphism.  Interestingly, this is a special case of quantum lattice gas automata, which we will return to in section \ref{sec:lattice}.  These comprise partial shifts (which are interpreted as propagating different particle types on the lattice) and on-site unitaries (which model interactions between particles).

Other examples of QCAs defined via partitioning schemes arise in the literature, e.g., local unitary QCAs \cite{PC07}.  As an aside, it is worth pointing out that many of these partitioning schemes were initially introduced in the framework of finite, unbounded configurations, but they also fit perfectly well into our framework.

\one{All examples in the section fall into a class of QCAs called \hlt{trivial} \cite{HFH18}.
\begin{Definition}
 A QCA $u$ is trivial if, possibly by adding ancillary degrees of freedom at each lattice site, we can decompose $u\otimes \openone$ into a finite-depth circuit QCA followed by some shifts, where $\openone$ acts on the ancillary degrees of freedom. 
\end{Definition}
Of course, the word trivial underplays the importance of such QCAs, as most of the interesting examples of QCAs that we will see are trivial.  On the other hand, this makes nontrivial QCAs, which we will return to in section \ref{sec:top_phases_2}, all the more interesting.}

\subsection{Clifford QCAs}
\label{sec:Cliff}
An important class of QCAs are \hlt{Clifford QCAs} \cite{SW04}.  Such models are efficiently classically simulatable, so Clifford QCAs cannot be universal for quantum computation (except in the unlikely event that quantum computers can be efficiently classically simulated).  Nevertheless, they have interesting properties, practical uses, and provide a nice playground to get some intuition about QCA dynamics.
\two{
\begin{Definition}
	Clifford QCAs are defined on lattices of qubits or qudits, where the dynamics is a Clifford operation, which means it takes any product of (generalized) Pauli operators to a multiple of a product of (generalized) Pauli operators.  (Generalized Pauli operators are defined below.)
\end{Definition}

Here we will mostly consider one-dimensional translationally invariant Clifford QCAs on qubits.}  Let us look at an example of such a Clifford QCA from \cite{SW04}.  The dynamics are determined by the automorphism $u_l$ acting as
\begin{equation}
\begin{split}\label{eq:Cliff_ex}
 u_l(X_n) & =Z_n\\
 u_l(Z_n) & =Z_{n-l}\otimes X_n\otimes Z_{n+l},
 \end{split}
\end{equation}
for a fixed $l\in\mathbb{Z}$.  Notice that this already determines the dynamics completely because $u_l(AB)=u_l(A)u_l(B)$ and $u_l$ is linear.  For example, with $l=1$ and writing $u=u_1$, we can easily calculate $u(Y_0\otimes Z_3)$ to get
\begin{equation}
\begin{split}
 u(Y_0\otimes Z_3) & = u(Y_0)u(Z_3)\\
 & = iu(Z_0)u(X_0)u(Z_3)\\
 & =-Z_{-1}\otimes Y_0\otimes Z_1 \otimes Z_2 \otimes X_3 \otimes Z_4,
 \end{split}
\end{equation}
where the last line follows from equation (\ref{eq:Cliff_ex}).

\begin{figure}[!ht]
{\centering
\resizebox{12cm}{!}{
\includegraphics{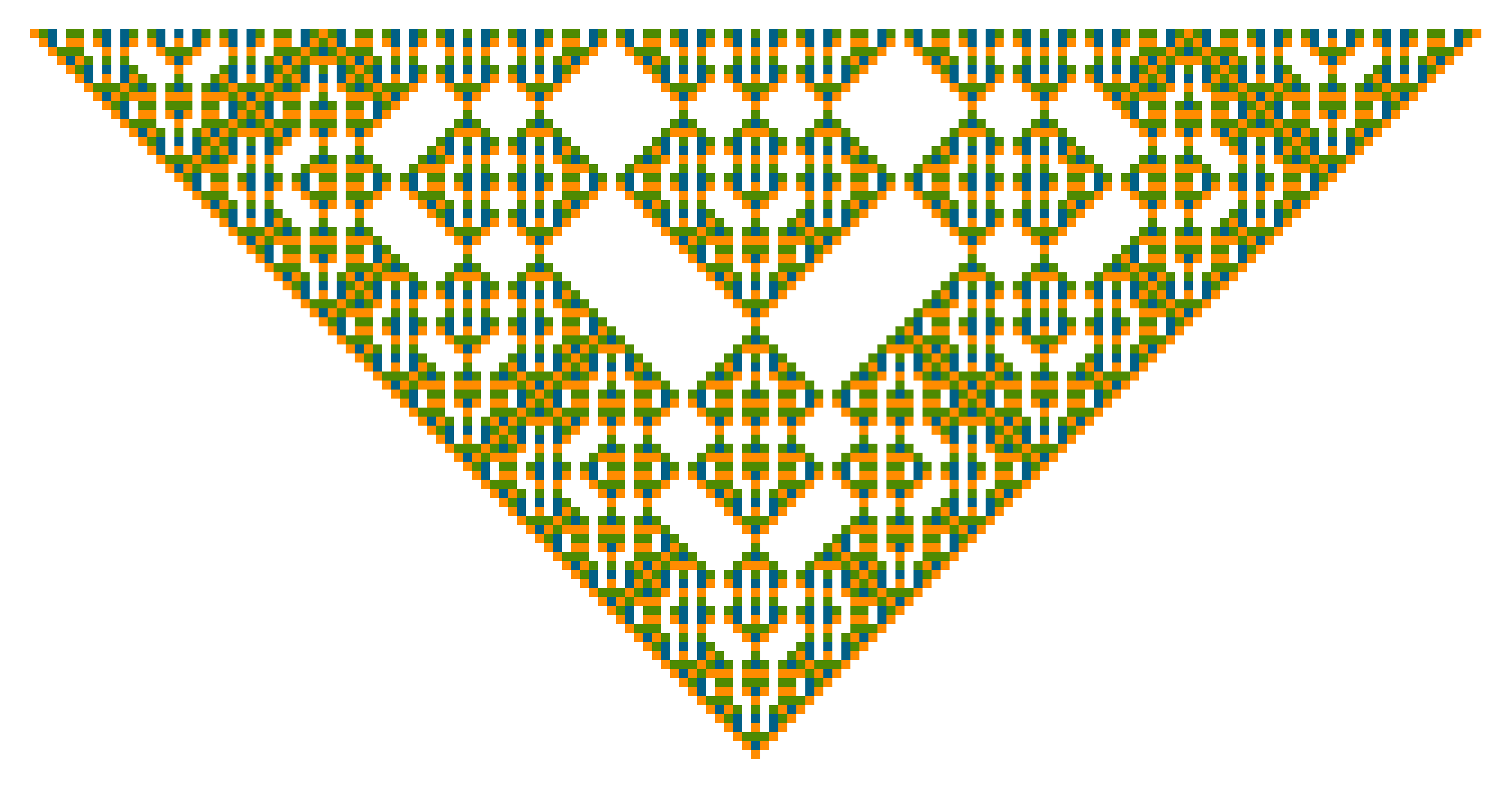}} \caption[CQCA plot]{An example of Clifford QCA dynamics with update rule fully specified by $X_n\rightarrow X_{n-1}\otimes Y_n\otimes X_{n+1}$ and $Z_n\rightarrow X_n$.  Each square represents an element in a Pauli string:\ white represents $\openone$, $X$ is yellow, $Y$ is blue and $Z$ is green.  Notice the fractal structure in the spacetime diagram (where time goes up).
\label{fig:CQCA}}
}
\end{figure}

The Clifford constraint is quite strong, and as a result to understand the dynamics we need only look at a CA.  Let us see how this works for the example above.  We can represent Pauli products by strings of integers.  We label $\openone, X, Y, Z$ by $0, 1, 2, 3$ respectively.  Given an initial string of Paulis, we represent it by a string $\mu=(...,\mu_{-1},\mu_0,\mu_1,...)$, where $\mu_m\in\{0,1,2,3\}$ for each $m\in\mathbb{Z}$.  To find out the value of the string at position $m=0$ after one timestep we need only know $\mu_{-1}$, $\mu_{0}$ and $\mu_{1}$.  Suppose, e.g., we have $(...,\mu_{-1},\mu_{0},\mu_{1},...)=(...,2,1,2,...)$, which represents an operator that has the form $A_{\{-\infty,...,-2\}}\otimes Y_{-1}\otimes X_0 \otimes Y_{1} \otimes B_{\{2,...,\infty\}}$, where $A_{\{-\infty,...,-2\}}$ and $B_{\{2,...,\infty\}}$ are some products of Paulis on the rest of the qubits.  After one timestep we have
\begin{equation}
 u\left(A_{\{-\infty,...,-2\}}\otimes Y_{-1}\otimes X_0 \otimes Y_{1} \otimes B_{\{2,...,\infty\}}\right)=A_{\{-\infty,...,-1\}}^{\prime}\otimes Z_0 \otimes B_{\{1,...,\infty\}}^{\prime}
\end{equation}
by using equation (\ref{eq:Cliff_ex}), and $A_{\{-\infty,...,-1\}}^{\prime}$ and $B_{\{1,...,\infty\}}^{\prime}$ are strings of Paulis now localized on regions ${\{-\infty,...,-1\}}$ and ${\{1,...,\infty\}}$ respectively.  Thus, we get $\mu^{\prime}_0=3$ after the update.  By finding $\mu^{\prime}_0$ for all possible products of Paulis on the sites $-1,0,1$, we infer the corresponding CA rule.  In this case, the CA consists of a discrete line of cells where each cell has four possible states in ${0,1,2,3}$, and the update rule for a site depends only on its nearest neighbours.  Note that we disregard the overall phase:\ a Clifford operation could map, e.g., $X_n$ to $-X_n$, but we ignore this.  This CA description is enough to tell us how strings of Paulis spread out.  An example of the dynamics of the ensuing CA is shown in figure \ref{fig:CQCA}.  In \cite{SVW08,GUWZ10}, far more sophisticated methods are used for studying Clifford QCAs, but we can already understand a lot of the structure by deriving CAs for the evolution of Pauli strings as we have done here.

Furthermore, Clifford QCAs on a line can be classified into three classes \cite{GUWZ10,Gutschow10}.  There are those with gliders, which are observables that simply move along the line each timestep; there are periodic Clifford QCA, which are periodic in time; and there are fractal Clifford QCA, which are self similar on large scales of a spacetime diagram, an example of which can be seen in figure \ref{fig:CQCA}.  The QCA in equation (\ref{eq:Cliff_ex}) with $l=1$ has examples of gliders.  Consider the observable $X_0\otimes Z_{1}$.  After one timestep, this is mapped to $u(X_0\otimes Z_{1})=X_1\otimes Z_{2}$, so it is just translated to the right.  Similarly, $Z_0\otimes X_{1}$ gets translated to the left by the QCA.  The fractal structure in spacetime plots of fractal Clifford QCAs was studied in further detail in \cite{GNW10,Gutschow10}.  This was made possible by analysing the resulting fractal structures in the corresponding CAs.

Additionally, entanglement generation in Clifford QCAs was studied in \cite{Gutschow10,GUWZ10}.  In \cite{SVW08}, it was also shown that every unique translationally invariant stabilizer state in one dimension can be generated by the action of a Clifford QCA acting on a product state.  \one{And Clifford QCAs play a role in understanding the properties of stabilizer codes in more general systems \cite{Haah16}.}  More recently, the eigenstates of Clifford QCAs have been analysed in \cite{GZ18}, where it was seen that the eigenstate thermalization hypothesis does not apply to these models.  The eigenstate thermalization hypothesis is the hypothesis that, when we look at the reduced state of an energy eigenstate on a small subsystem, we approximately get a thermal state with a temperature that does not vary too quickly with the energy of the eigenstate.  (There are a few different versions  of the eigenstate thermalization hypothesis, and this is just one of them.)  On the one hand, it may not be so surprising that Clifford QCAs, which are discrete-time systems with a lot of structure, do not satisfy the eigenstate thermalization hypothesis.  But on the other hand, there is evidence that periodically driven Hamiltonian systems satisfy the hypothesis \cite{KIH14}, and QCAs are a useful approximate model for such systems, something we will discuss further in section \ref{sec:Ham}.

As we just saw, the constraint that the evolution is a Clifford operation is quite powerful, and the simplification arising from this constraint can be generalized \cite{SVW08}:\ one can go beyond qubits and Pauli operators, instead looking at $p$ dimensional qudits and discrete Weyl operators, provided $p$ is prime.  Fixing an orthonormal basis for such a qudit to be $\ket{q}$, with $q\in\{0,...,p-1\}$, then the Weyl (or generalized Pauli) operators, which generalize $X$ and $Z$, are defined by
\begin{equation}
\begin{split}
 \mathcal{X}\ket{q} &=\ket{q+1}\\
 \mathcal{Z}\ket{q} &=e^{2\pi i q/p}\ket{q},
 \end{split}
\end{equation}
where the addition in the first line is defined modulo $p$.  Again we have the constraint that products of Weyl operators, e.g.,  $\mathcal{X}_n\otimes \mathcal{Z}_{n+1}$ get mapped to products of Weyl operators.  Many of the results for qubit systems extend to these systems, though at the cost of being a bit more complicated \cite{SVW08}.

Remarkably, the QCAs in equation (\ref{eq:Cliff_ex}) for different values of $l$, together with on-site Clifford operations and the shift generate all possible translationally invariant qubit Clifford QCAs in one dimension \cite{SVW08}.
\one{And much of the general structure of qudit Clifford QCAs has been understood recently in \cite{Haah19}, where it was shown that the fourth power of any translationally invariant Clifford QCA is trivial, in the sense that it can be decomposed into a constant-depth circuit followed by shifts.}  Instead of the Clifford constraint, we could just restrict the QCA neighbourhood and the dimension of the systems at each site.  This was done in \cite{SW04}, where it was shown that nearest-neighbour \emph{qubit} QCAs are always decomposable in terms of shifts, on-site unitaries and controlled phase gates.  \one{Furthermore, in two dimensions \emph{every} QCA is known to be trivial \cite{FH19,Haah19}, a fact that was already known in one dimension as a consequence of index theory \cite{GNVW12}.}

\section{QCAs as quantum computers}
QCAs were originally envisaged as a model of quantum computation, useful for simulating physics or implementing more general quantum algorithms.  One of the earliest breakthroughs in this regard was the proof that QCAs can efficiently simulate quantum Turing machines with only constant slowdown \cite{Watrous95}.  The proof used the Watrous partitioned QCAs of the previous section.

If we consider instead the circuit model of quantum computation, it is clear from a result that we will see in section \ref{sec:uni-caus-local} that the circuit model can efficiently simulate any QCA.\footnote{Of course, another way to see this is by using a quantum circuit to simulate a quantum Turing machine \cite{MW18} that in turn simulates a QCA.  This would work in one dimension at least, since quantum Turing machines can simulate these Watrous partitioned QCAs with linear slowdown \cite{Watrous95}.}  And we know that the converse is true because we can use a QCA to simulate a quantum Turing machine, which can then simulate the circuit model.  Nevertheless, it will be interesting to look at an explicit example in the following section where we directly simulate a quantum circuit with a QCA.  Note that we only consider translationally invariant QCAs in this section.  (Relaxing this constraint would make the results trivial.  Besides, the main appeal of quantum computation via QCAs is that ideally one need only apply global operations without worrying about implementing many different single- or few-qubit operations.  This advantage may be lost in the translationally non-invariant case.)

\subsection{A QCA efficiently simulating quantum circuits}
Let us look at a prescription from \cite{Raussendorf05} for simulating quantum circuits using a translationally invariant QCA \two{on a finite lattice}.  The QCA system is a two-dimensional discrete torus (i.e., a two-dimensional lattice with periodic boundary conditions), and the program itself, as well as the input data for the algorithm, are encoded in the initial state of the QCA.  In this section it will be more convenient to work in the Schr\"{o}dinger picture.

We label the sites of the torus with coordinates $\vec{n} \in \{0,...,2s-1\}\times \{0,...,2r-1\}$, and each site has a single qubit.  We label columns on the torus by the second component of $\vec{n}$, which ranges from $0$ to $2r-1$.  The qubits in the odd-numbered columns will encode what gates we wish to implement.  So for the purposes of the algorithm, these qubits will always be in one of the the two computational basis states $\ket{0}$ and $\ket{1}$.  The qubits on the even-numbered columns will encode the data that our quantum circuit is to act upon.  In particular, at time $t=0$ column $0$ has the initial input data for the quantum algorithm.
\begin{figure}[!ht]
{\centering
\resizebox{9cm}{!}{\includegraphics{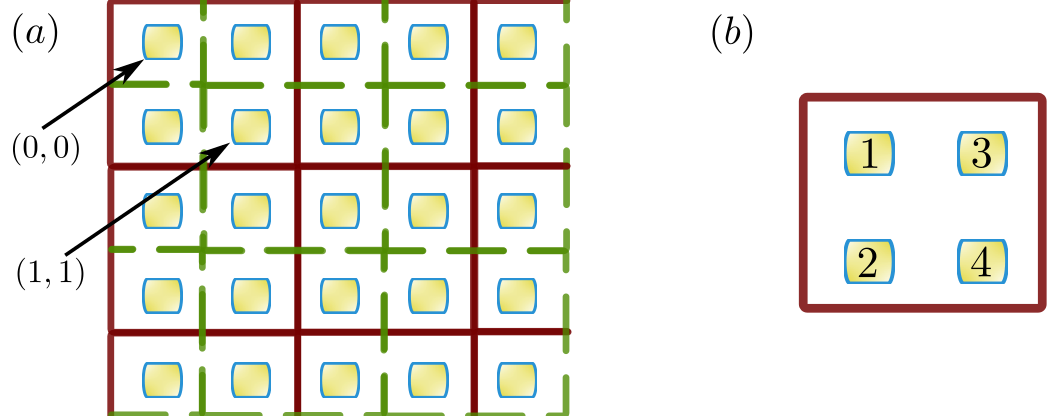}} \caption[Partitioning for Simulation]{The two different partitions used to define the QCA are $2\times2$ squares as shown in subfigure (a).  The first is given by the squares with solid red lines (with upper left site having both coordinates even, e.g., $(0,0)$), and the second corresponds to the green boxes with dotted lines (with upper left site having both coordinates odd).  Subfigure (b) shows the simpler labelling of qubits in one of the supercells used to define the unitary in equation (\ref{eq:supercelleq}). \label{fig:F}}
}
\end{figure}

We will use the supercell partitioning scheme of section \ref{sec:partitioning}, with two simple partitions of the torus, but we will consider the application of the unitaries to both partitions to correspond to two timesteps, meaning the evolution is different for even and odd timesteps.  The partitions are as shown in figure \ref{fig:F} (a).  The first consists of $2\times 2$ squares of cells with the cell in the top left corner having both coordinates even, e.g., $\vec{n}=(0,0)$.  The second consists of $2\times 2$ squares of cells but now with the cell in the top left corner having both coordinates odd, e.g., $\vec{n}=(1,1)$.  So the supercells in the first partition are labelled by vectors with both components even, and those in the second partition are labelled by vectors with both components odd.

Next we define the unitaries that act on these supercells, which all have the same form:
\begin{equation}
 V_{\vec{n}}=S_{\vec{n}}W_{\vec{n}},
 \end{equation}
where $S_{\vec{n}}$ just swaps the states of the two left qubits in the square with the two qubits on the right.  This part of the gate swaps the data qubits with the program qubits after applying $W_{\vec{n}}$.  To make the notation simple, label the four sites as in figure \ref{fig:F} (b).  For $W$ we take 
 \begin{equation}\label{eq:supercelleq}
 W=H_1C_{3}[U_{1}]C_{4}[R_{12}].
\end{equation}
Let us go through each of the components on the right hand side.  Two of these unitaries are controlled by the program qubits:\ $C_{3}[U_{1}]$ and $C_{4}[R_{12}]$ are unitaries applied to data qubits controlled by the program qubits $3$ and $4$.  So if, e.g., qubit $3$ is in the $\ket{1}$ state, the unitary $U_1$ is applied to qubit $1$, and if it is in the $\ket{0}$ state, the identity is applied.  $H_1$ is the Hadamard gate applied to qubit $1$, i.e., $H_1=(X_1+Z_1)/\sqrt{2}$.  The choice made in \cite{Raussendorf05} for the gates $U_1$ and $R_{12}$ is
\begin{equation}
 \begin{split} \label{eq:Raussgates}
  U_1 & =\exp\left(-i\frac{\pi}{8}Z_1\right)\\
  R_{12} & = C_1[Z_2],
 \end{split}
\end{equation}
so $R_{12}$ is simply a Pauli Z operator on qubit $2$ controlled on qubit $1$.  (These are both data qubits.)

The evolution at even and odd timesteps is then given by the products
\begin{equation}
\begin{split}
 V_{\mathrm{even}} & =\prod_{\vec{n}\ \mathrm{even}}V_{\vec{n}}\\
 V_{\mathrm{odd}} & =\prod_{\vec{n}\ \mathrm{odd}}V_{\vec{n}},
 \end{split}
\end{equation}
where the product in the first/second line is over all $\vec{n}$ with both entries even/odd.  Now consider the initial state
\begin{equation}
 \ket{\psi(0)}=\bigotimes_{i=0}^{r-1} \ket{D_i(0)}_{2i}\ket{P_{i+1}}_{2i+1}.
\end{equation}
Here $\ket{D_i(0)}_{2i}$ is the state of the data qubits in column $2i$.  And $\ket{P_{i+1}}_{2i+1}$ is the state of the qubits in column $2i+1$, which encode the program, i.e., the unitaries that should be implemented on the data qubits by the dynamics.  Note that the subscript label on the kets denotes the column where the qubits are, whereas the subscript label within the qubits specifies the state.  Let us use $[m]_k$ to denote $m$ modulo $k$.  After $t$ timesteps, the state is given by
\begin{equation}
\label{eq:raus}
 \ket{\psi(t)}=\bigotimes_{i=0}^{r-1} \ket{D_i(t)}_{[2i+t]_{2r}}\otimes\ket{P_{[i+t+1]_{r}}}_{[2i+t+1]_{2r}},
\end{equation}
where, in particular,
\begin{equation}
 \ket{D_0(t)}=G(P_{t})...G(P_1)\ket{D_{0}(0)},
\end{equation}
where $G(P_i)$ are unitaries composed of the gates applied at each timestep as the input data column moves across the torus encountering new program qubits.  The other vectors $\ket{D_i(t)}$ in equation (\ref{eq:raus}) with $i\neq 0$ will have unitaries $G(P_i)$ applied in some other order, but this is unimportant, as we are only interested in implementing the quantum algorithm on the input data that started at $t=0$ in the first column.  And to do this, we encode the necessary gates into the program qubits $P_i$.  For example, $G(P_1)$ is given by
\begin{equation}
 \prod_{\substack{n_1\ \mathrm{even}\\ n_2=0}}W_{\vec{n}}\ket{D_{0}(0)}_0\ket{P_{1}}_{1}=G(P_1)\ket{D_{0}(0)}_0\ket{P_{1}}_{1}.
\end{equation}
Note that the product above only involves unitaries on the first two columns.  Recall from earlier that these $W_{\vec{n}}$ unitaries were specifically chosen to implement desired gates on the data qubits controlled on the program qubits.

After $r$ timesteps (any longer and the data qubits encounter program qubits twice) the output of the program can be read from column $r$.  It was shown in \cite{Raussendorf05} that with this setup, the choice of gates in equation (\ref{eq:Raussgates}) is sufficient to implement a universal gate set on the data qubits and hence simulate any quantum circuit efficiently.

A similar idea was used in \cite{PC07}, where again the circuit to be simulated is, in a sense, drawn onto a two-dimensional QCA.

\subsection{Other ideas}
\label{sec:other}
A drawback of the method of the previous section, which maps the circuit onto a two-dimensional QCA, is simply that it is two dimensional.  This may be difficult to do in a laboratory, so it may be more desirable for practical reasons to work with a one-dimensional QCA.  That quantum circuits can be efficiently simulated by a one-dimensional QCA was shown in \cite{SFW06}, which gave a construction for a universal one-dimensional QCA with $12$ dimensional quantum systems at each cell.  A stepping stone to get this result was a proof that QCAs with external classical control can be simulated by those without.  In this context, a \hlt{classically controlled QCA} is a QCA where an external party can choose to evolve the system via one of some number of automorphisms at each timestep.  (This is also sufficient for quantum computation \cite{VC06}.)  By encoding which automorphism is to be applied at each timestep into the initial state, it is possible to get a truly autonomous QCA \cite{SFW06}, albeit with a bigger state space.

An interesting related concept to the quantum computational universality we are interested in here is so-called \hlt{intrinsic universality}.  For QCAs this amounts to finding a QCA that, after some regrouping of cells, can simulate any other QCA.  This has been studied in detail in \cite{AFW09,AG10,AG12,AG12again}, where $n$-dimensional intrinsically universal QCAs were found \cite{AG10,AG12}.

One might speculate on the possibility of using fermionic QCAs as quantum computers instead of qudit QCAs.  This would be a QCA analogue of using fermionic quantum circuits for quantum computation instead of qubit circuits \cite{BK02}.  (Also, the opposite direction, simulating fermionic systems with qubit quantum circuits, is highly relevant for quantum simulations of physics \cite{SW17}.)  For the quantum circuit model, it was shown in \cite{BK02} that quantum computers using qubits can efficiently simulate quantum computers composed of fermions and vice versa.  A similar result is true for QCAs:\ fermionic QCAs can simulate qudit QCAs efficiently and vice versa directly (without, e.g., simulating the circuit model first) \cite{Farrelly15}.  In this case, the slowdown is constant in both directions.

\subsection{Implementations}
\label{sec:implementations}
In contrast to the circuit model of quantum computing, which uses single and (typically) few qubit gates to build a global unitary circuit, quantum cellular automata involve global unitary operations, without necessarily requiring the ability to directly apply, e.g., single-qubit gates.  This is often put forward as an advantage because single-site addressability can be harder in practice \cite{Benjamin00}.  

One interesting possibility, first proposed in one of the earliest papers on QCAs \cite{Lloyd93}, is to arrange three types of atomic species periodically in an $\mathcal...{ABCABC}...$ pattern.  Each species can be addressed independently with electromagnetic pulses tuned to have the   resonant frequency of a species' ground to excited state transition.  (Note that for each atom, the resonant frequency can be affected by the states of its neighbours, which is a useful feature in the proposal as it allows, e.g., conditional operations to be implemented.)  With various different combinations of pulses, one can implement various different gates.  The input and readout of data occurs only on the end qubits of the line.  Later it was shown in \cite{Benjamin00} that two species in an $\mathcal...{ABAB}...$ pattern suffices for universal quantum computation.  (In fact, even a single species should be sufficient \cite{Raussendorf05a,VC06}.)  These kinds of models are what some authors mean by the name quantum cellular automata (e.g., in \cite{Jones11}), whereas, from our point of view, a better name would be classically controlled QCAs, as they typically involve different unitaries applied at each timestep.

This scheme was studied further in \cite{Benjamin01,BB04}, and in \cite{Twamley03} an implementation of this idea using endohedral fullerenes was analyzed theoretically.  Endohedral fullerenes are $C_{60}$ fullerenes (these are molecules made of $60$ carbon atoms that form a ball, sometimes called buckyballs) that have, e.g., an atom trapped inside the cage formed by the carbon atoms.  And the idea of using these molecules as parts of a quantum computer met with a degree of experimental success in \cite{BAB06}.  In that case, Nitrogen atoms were trapped in the fullerenes and the levels considered arose from a spin $3/2$ electron coupled to $^{14}N$ nuclear spin $I=1$.

Most recently, a proposal has been made to use ultracold atoms excited to Rydberg states to implement QCAs in practice \cite{WWL20}.  There it was also shown that such QCAs would be well suited to variational quantum optimization as well as quantum state engineering, which works by tuning the QCA to get steady states that have a high degree of entanglement.

While quantum computing via quantum cellular automata has its advantages (only global operations are necessary), the vast majority of work regarding quantum computing is focussed on other models, especially the circuit model.  Perhaps because of this, there is, e.g., no well-developed theory of error correction in quantum cellular automata, in contrast to the quantum circuit model.  (It may be straightforward to develop such a theory, and it would be interesting to see if it could be done without requiring few-qubit operations.)
Furthermore, the current trend in building quantum computers is oriented towards quantum computers with a focus on few-qubit addressability in order to use the quantum circuit model.

\section{Structure of QCAs}
\label{sec:structure}
There are many results that allow one to understand the structure of QCA dynamics.  In one dimension, QCAs are particularly well understood.  For example, the index theory (discussed in section \ref{sec:ind}) classifies when two QCAs can be smoothly deformed into one another while preserving locality.  In two dimensions, all qudit QCAs can be classified in terms of indices and the topology of the underlying space, as discussed in section \ref{sec:higherindex}.  And in any dimension, a powerful result is given in the following section, where we see that QCAs, together with some local ancillas, allow us to view any QCA as a local finite-depth circuit.  Finally, QCAs can be viewed as tensor-network operators, something we will see in section \ref{sec:MPU}.

\subsection{Unitarity plus causality implies localizability}
\label{sec:uni-caus-local}
This result was first proved for \two{qudit QCAs on infinite lattices with no assumption of translational invariance in \cite{ANW11}.}  It was later extended to QCAs with fermions and bosons in \cite{FS13}.  It is a little easier to understand the intuition if we prove this for finite systems as we always have a unitary implementing the QCA dynamics.  Nevertheless, the proof is almost as straightforward for infinite systems and automorphisms.  The trick is to consider the original QCA $A$ with dynamics $U_A$ and an identical copy $B$, evolving via the inverse dynamics $U_B^{-1}$.  So for a qudit at site $\vec{n}$ in the $A$ system, there is also a copy qudit from the $B$ system (of the same dimension) at the same site $\vec{n}$.  Next, we need to introduce the global swap operation:
\begin{equation}
\begin{split}
 S_{AB} =\prod_{\vec{n}}S_{\vec{n}},
 \end{split}
\end{equation}
where $S_{\vec{n}}$ swaps the state of the qudits of system $A$ and $B$ at site $\vec{n}$.  Using this swap, we can decompose the dynamics of the $A$ and $B$ system:
\begin{equation}
\begin{split}
 U_{A} U^{-1}_B & = S_{AB} U_{B} S_{AB} U^{-1}_B\\
 & = \left(\prod_{\vec{m}}S_{\vec{m}}\right) U_{B} \left(\prod_{\vec{n}}S_{\vec{n}}\right) U^{-1}_B\\
 & =  \left(\prod_{\vec{m}}S_{\vec{m}}\right) \left(\prod_{\vec{n}}U_B S_{\vec{n}}U^{\dagger}_B\right).
 \end{split}
\end{equation}
Crucially, $S_{\vec{n}}$ is a local unitary at site $\vec{n}$.  Then, $U_B S_{\vec{n}}U^{\dagger}_B$ must also be a local unitary localized on the inverse neighbourhood\footnote{The inverse neighbourhood is the neighbourhood of site $\vec{n}$ with respect to the \textit{inverse} unitary.} of site $\vec{n}$.  This follows because $U_B$ is a locality preserving unitary and hence so is $U_B^{-1}$.  (For a simple proof of this, see, e.g., \cite{ANW11}.)  Therefore, we see that the dynamics of the joint system can be implemented as a finite-depth circuit.  All the swaps $S_{\vec{m}}$ can be implemented in parallel, and any $U_B S_{\vec{n}}U^{\dagger}_B$ unitaries can be implemented in parallel as long as they act on non-overlapping regions.  The minimum depth circuit is then given by $\max_{\vec{n}} |\m{N}_B(\vec{n})| + 1$.  Here $\m{N}_B(\vec{n})$ denotes the neighbourhood of site $\vec{n}$ with respect to the unitary $U_{B}$.

The result was also later generalized to dynamical graphs in \cite{AM17}, where the QCA neighbourhood scheme could change over time.  This was done with a view towards discrete models of quantum gravity, where spatial geometry may not be fixed in time.

\subsection{Index theory in one dimension}
\label{sec:ind}
In the following sections, we discuss the \hlt{index for one dimensional QCAs}, introduced for qudits in \cite{GNVW12}.  This was inspired by an analogous index for quantum walks \cite{Kitaev06,GNVW12}, which was called the flow in \cite{Kitaev06}. 
\one{Essentially, the index is a number that quantifies how much quantum information moves along the \two{line}.  Interestingly, the QCAs considered need not be translationally invariant, yet one can still assign an index to them.  A key property of the index is that it can be computed by examining the dynamics locally (on a few sites), as we will see.  For qudits, the index is a non-zero rational number, whereas interestingly for fermions the index is a non-zero rational number times a power of $\sqrt{2}$.  In a very rough sense, the index is determined by how much of a local algebra of operators (\emph{any} local algebra corresponding to two or more contiguous sites) of the QCA moves to the left at each timestep.  The details are spelled out more precisely in the next few sections, though this requires a few definitions in order to write down a formula that is comprehensible.
}

\subsubsection{Qudit index theory}
\label{sec:Qudit_index}
We first need to introduce a few concepts to define the index.  In this section, we assume that the QCAs are nearest neighbour QCAs, which we can always achieve by regrouping sites.  Regrouping works by grouping together sites on the original lattice into a single site on a new lattice.  For example, we could regroup by defining $\m{B}_n=\m{A}_{2n}\otimes \m{A}_{2n+1}$ for all $n$ for a one-dimensional system.  Then $\m{B}_n$ are the cell algebras for the new regrouped system, each of which contain two cells of the original system.

Let us start by introducing \hlt{support algebras} \cite{Zanardi01,SW04}.  The intuition underlying support algebras is that they describe, in some sense, the extent to which one algebra has overlap with another algebra.  Note that, for most of our purposes, these algebras are finite dimensional and so are equivalent to matrix algebras.
\begin{Definition}
 Let $\m{A}$, $\m{B}_1$ and $\m{B}_2$ be C*-algebras, with $\m{A}\subseteq \m{B}_1\otimes \m{B}_2$.  We denote the support of $\m{A}$ in $\m{B}_1$ by $\g{S}(\m{A},\m{B}_1)$.  This is defined to be the smallest C*-subalgebra in $\m{B}_1$ needed to build elements of $\m{A}$.  More concretely, for any $a\in\m{A}$, we can decompose it as $a=\sum_{ij}b^1_i\otimes b^2_j$, where $b_i^1\in \m{B}_1$ and $b_j^2\in \m{B}_2$.  Then $\g{S}(\m{A},\m{B}_1)$ is generated by all $b_i^1$ arising in this way.
\end{Definition}

Two algebras $\m{A}$ and $\m{B}$ commute if each element $a\in \m{A}$ and each element $b\in \m{B}$ commute.  If this is the case, we write $[\m{A},\m{B}]=0$.  The following lemma (see, e.g., \cite{GNVW12} for a simple proof) is a useful result, which shows that, if two algebras commute, their support algebras also commute.
\begin{lemma}\label{lem:supp}
 Suppose we have the C*-algebras $\m{B}_1$, $\m{B}_2$ and $\m{B}_3$.  Let $\m{A}_{12}$ and $\m{A}_{23}$ be subalgebras of $\m{B}_1\otimes\m{B}_2\otimes\m{B}_3$ satisfying
 \begin{equation}
  \begin{split}
   \m{A}_{12} & \subseteq \m{B}_1\otimes \m{B}_2\\
   \m{A}_{23} & \subseteq \m{B}_2\otimes \m{B}_3.
  \end{split}
 \end{equation}
 Then $[\m{A}_{12},\m{A}_{23}]=0$ implies
\begin{equation}
 [\g{S}(\m{A}_{12},\m{B}_2),\g{S}(\m{A}_{23},\m{B}_2)]=0.
\end{equation}
\end{lemma}
\begin{figure}[!ht]
{\centering
\resizebox{9cm}{!}{\includegraphics{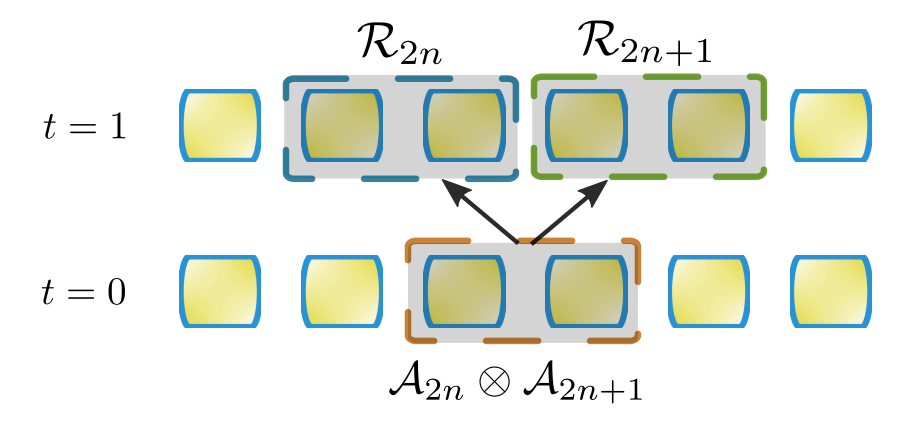}} \caption[Visualization of the even and odd algebras]{Visualization of even and odd algebras.  Note that, e.g., $\m{R}_{2n}$ rarely equals $\m{A}_{2n-1}\otimes\m{A}_{2n}$.\label{fig:EvenOdd}}
}
\end{figure}
Another tool we need are the so-called even and odd algebras, which are defined to be
\begin{equation}
\begin{split}
 \m{R}_{2n} & =\g{S}\left(u\left(\m{A}_{2n}\otimes\m{A}_{2n+1}\right),\left(\m{A}_{2n-1}\otimes\m{A}_{2n}\right)\right)\\
 \m{R}_{2n+1} & =\g{S}\left(u\left(\m{A}_{2n}\otimes\m{A}_{2n+1}\right),\left(\m{A}_{2n+1}\otimes\m{A}_{2n+2}\right)\right).
\end{split}
\end{equation}
These are best understood by looking at figure \ref{fig:EvenOdd}.

Recall that $\m{A}_n$ is the algebra at lattice site $n$.  So after evolving $\m{A}_{2n}\otimes\m{A}_{2n+1}$ by $u$, the image $u(\m{A}_{2n}\otimes\m{A}_{2n+1})$ is a subalgebra of $\m{R}_{2n}\otimes\m{R}_{2n+1}$.  These even and odd algebras roughly describe what information moves right and what information moves left along the \two{line}.  Next, we need the following important result about the even and odd algebras.
\begin{lemma}\label{lem:rr}
Each $\m{R}_m$ satisfies
\begin{equation}
 \m{R}_m\simeq \m{M}_{r_m},
\end{equation}
which is the algebra of $r_m\times r_m$ complex matrices.  Note that $r_m$ and hence the size of the algebra depends on the position $m$.  Furthermore,
 \begin{equation}\label{eq:rr}
 \m{R}_{2n}\otimes\m{R}_{2n+1}=u(\m{A}_{2n}\otimes\m{A}_{2n+1}).
\end{equation}
\end{lemma}
\one{This is proved in, e.g., \cite{GNVW12}.}
We see from equation (\ref{eq:rr}) that
\begin{equation}\label{eq:k1}
 \mathrm{dim}[\m{R}_{2n}]  \mathrm{dim}[\m{R}_{2n+1}]= \mathrm{dim}[\m{A}_{2n}] \mathrm{dim}[\m{A}_{2n+1}],
\end{equation}
where $\mathrm{dim}[\cdot]$ denotes the dimension of the algebras as vector spaces.  Also, we have $\m{R}_{2n+1}\otimes\m{R}_{2n+2}=\m{A}_{2n+1}\otimes\m{A}_{2n+2}$, which follows because the $\m{R}$ algebras generate the \two{full system's} algebra, and $\m{R}_{2n+1}$ and $\m{R}_{2n+2}$ are the only ones that have support on sites $2n+1$ and $2n+2$.  Then we get
\begin{equation}\label{eq:k2}
 \mathrm{dim}[\m{R}_{2n+1}]  \mathrm{dim}[\m{R}_{2n+2}]= \mathrm{dim}[\m{A}_{2n+1}] \mathrm{dim}[\m{A}_{2n+2}].
\end{equation}
We may then define the index of a QCA to be
\begin{equation}
 \mathrm{ind}[u]= \sqrt{\frac{\mathrm{dim}[\m{R}_{2n}]}{\mathrm{dim}[\m{A}_{2n}]}},
\end{equation}
which we can see from equations (\ref{eq:k1}) and (\ref{eq:k2}) is actually independent of $n$.
And because $\m{A}_n$ is isomorphic to $\m{M}_{d_n}$ and $\m{R}_m$ is isomorphic to $\m{M}_{r_n}$, we get the equivalent formula
\begin{equation}\label{eq:index}
 \mathrm{ind}[u]=\frac{r_{2n}}{d_{2n}}.
\end{equation}
(Note that some authors define the log of $\mathrm{ind}[u]$ to be the index.)

Let us consider some examples.  When each cell of the QCA has a quantum system of the same dimension, a possible dynamics is given by a shift $s$.  When we have $d$-dimensional qudits at each site, it follows from equation (\ref{eq:index}) that $\mathrm{ind}[s]=d$.  If instead we had a shift to the left, given by $s^{-1}$, then we get $\mathrm{ind}[s^{-1}]=1/d$.  Another interesting choice of dynamics is a local finite-depth circuit QCA.  These always have index $\mathrm{ind}[u]=1$.

Below we list some basic properties of the index.  These are proved in \cite{GNVW12}, where most properties follow straightforwardly from the definition of the index in equation (\ref{eq:index}).  However, the last two highly interesting properties of the index listed below are harder to prove and tell us something about its robustness with respect to perturbations of the dynamics.\\
\\~\\
\textbf{Properties of the index}
\begin{enumerate}
 \item The index $\mathrm{ind}[u]$ is locally computable and independent of how we regroup/block sites.
  \item The index is a group homomorphism from the group of locality-preserving automorphisms (QCAs) to the group of strictly positive rationals $\mathbb{Q}_+$.  The identity automorphism gets mapped to $1$, and  $\mathrm{ind}[u v]=\mathrm{ind}[u]\, \mathrm{ind}[v]$.
 \item The index is multiplicative under tensor products: $\mathrm{ind}[u\otimes v]=\mathrm{ind}[u]\, \mathrm{ind}[v]$.
 \item $\mathrm{ind}[u]=1$ if and only if $u$ is locally implementable with
 \begin{equation}
 \begin{split}
  u& = w\circ v\\
  v(\cdot) & = \left(\prod_{n\in\mathbb{Z}}V^{\dagger}_{2n}\right) \cdot \left(\prod_{m\in\mathbb{Z}}V_{2m}\right)\\
  w(\cdot) & = \left(\prod_{k\in\mathbb{Z}}W^{\dagger}_{2k+1}\right) \cdot \left(\prod_{l\in\mathbb{Z}}W_{2l+1}\right),
 \end{split}
 \end{equation}
 where $V_{2n}$ are unitaries acting on sites $2n$ and $2n+1$, and $W_{2m+1}$ are unitaries acting on sites $2m+1$ and $2m+2$.  So $u$ is a finite-depth circuit.  This need not be translationally invariant.
 \item Two QCAs $u_0$ and $u_1$ with the same C*-algebras (i.e., the same cell structure) have the same index if and only if one can be continuously deformed into the other.  This means that there exists a strongly continuous\footnote{Strong continuity means that, for any strictly local $A\in\m{A}$, $u_t(A)$ is continuous in the norm of the C*-algebra.} path $u_t$ with $t\in[0,1]$ connecting $u_0$ to $u_1$ such that $u_t$ are QCAs with the same cell structure and neighbourhoods as $u_0$ and $u_1$.
\end{enumerate}
Property $(2)$ and $(4)$ together tell us that applying additional local unitaries cannot change the index, e.g., $\mathrm{ind}[u]=\mathrm{ind}[v\circ u]$, where $v(\cdot)=V^{\dagger}(\cdot)V$ for some local $V$.  Property $(5)$ tells us that smooth deformations (in terms of the strong operator topology) also cannot change the index, so it is robust against these two types of perturbations.  This robustness to perturbations is important in the application of the index to classifying chiral Floquet phases (section \ref{sec:top_phases}).
 

\subsubsection{Another formula for the index}
There is a different formula for the index which is quite useful \cite{GNVW12}.  To write it down, we need another method for quantifying the overlap of two algebras.  The algebra of the whole QCA $\m{A}$ has a normalized trace, i.e., a state $\tau$, such that
\begin{equation}
\begin{split}
 \tau[\openone] & =1\\
 \tau[AB] & =\tau[BA],
 \end{split}
\end{equation}
where $A,B\in\m{A}$.  This state is essentially just the maximally mixed state but for an infinite \two{line}.  For any $A,B\in\mathcal{A}$, $\tau(A^{\dagger}B)$ defines an inner product.  Because $\m{A}$ is a complex vector space, we can think of any two subalgebras $\m{B}_1$, $\m{B}_2$ as vector subspaces of $\m{A}$.  This allows us to quantify the overlap of these two subspaces.  We pick orthonormal bases for $\m{B}_1$ and $\m{B}_2$, and denote them by $e_{\alpha}$ and $f_{\beta}$ respectively.  Then we define the overlap of the algebras to be
\begin{equation}
 \eta(\m{B}_1,\m{B}_2)=\sqrt{\sum_{\alpha \beta}\left(\tau[e^{\dagger}_{\alpha}f_{\beta}]\right)^2},
\end{equation}
which is independent of the choice of orthonormal bases.  We can also think of this as
\begin{equation}
 \eta(\m{B}_1,\m{B}_2)=\sqrt{\tr{P_1P_2}},
\end{equation}
where $P_1$ and $P_2$ are the projectors onto the subspaces $\m{B}_1$ and $\m{B}_2$.  The trace here is the trace of operators on $\m{A}$ viewed as an inner product space.\footnote{In fact, this inner product space is the GNS representation of $\m{A}$ generated by the state $\tau$.}  Note that $\eta(\m{B}_1,\m{B}_2)\geq 1$ if both subalgebras contain the identity.

To get our second formula for the index, define
\begin{equation}
 \begin{split}
  \m{A}_L & =\m{A}_{n-N}\otimes...\otimes \m{A}_{n}\\
  \m{A}_R & =\m{A}_{n+1}\otimes...\otimes \m{A}_{n+M},
 \end{split}
\end{equation}
where \two{$n$ is arbitrary, and} $N$ and $M$ can be any integer greater than zero.  (Actually, the case $N=M=\infty$ makes sense and is also allowed.)  Then we have
\begin{equation}
 \mathrm{ind}[u]=\frac{\eta\big(u(\m{A}_L),\m{A}_R\big)}{\eta\big(u(\m{A}_R),\m{A}_L\big)}.
\end{equation}
This formula for the index is useful for applications to topological phases and is also needed to prove some properties of the index in \cite{GNVW12}.

It is worth mentioning that there is yet another formula for the QCA index, which is in terms of the Jones index \cite{Vogts09}.  The Jones index \cite{JS97} essentially quantifies the multiplicity of one C*-subalgebra $\m{B}$ in a larger algebra $\m{C}$.  To be precise, $\m{B}$ and $\m{C}$ must be type $\mathrm{II}_1$ factors (von Neumann algebras with trivial center).  The theory of factors and subfactors is quite involved, but only a few properties are needed to write the QCA index in terms of Jones indices \cite{Vogts09}.

\subsubsection{Fermionic index theory}
\label{sec:fermion_index}
We might guess that, via the Jordan-Wigner transformation \cite{JW28,Nielsen05}, an \hlt{index for one-dimensional fermionic QCAs} should basically be the same as for qudit QCAs.  In fact, this is not quite the case.  The index theory does extend to fermions but it is somewhat richer \cite{FPP17}.  Here we consider the case where each lattice site has some fermion modes and possibly also qudits.  Most of the construction runs along the same lines as for qudits in section \ref{sec:Qudit_index}, with some complications arising because fermion operators on different sites do not necessarily commute.  \two{These results apply to finite or infinite systems, and the QCAs need not be translationally invariant.}

The support algebra construction from section \ref{sec:Qudit_index} also carries over, and so does lemma \ref{lem:supp}, with the only difference being commutators are replaced by graded commutators.  (See appendix \ref{app:fermions} for more on fermion QCAs, graded commutators and graded tensor products.)  It is also straightforward to prove the analogue of lemma \ref{lem:rr} \cite{FPP17}, namely
\begin{equation}\label{eq:fr}
 \m{R}_{2n}\otimes_g\m{R}_{2n+1}=u(\m{A}_{2n}\otimes_g\m{A}_{2n+1}),
\end{equation}
where $\otimes_g$ is the graded tensor product, and
\begin{equation}\label{eq:fr2}
 \m{R}_{2n+1}\otimes_g\m{R}_{2n+2}=\m{A}_{2n+1}\otimes_g\m{A}_{2n+2}.
\end{equation}
From equation (\ref{eq:fr}), we get
\begin{equation}
 \mathrm{dim}[\m{R}_{2n}]  \mathrm{dim}[\m{R}_{2n+1}]= \mathrm{dim}[\m{A}_{2n}] \mathrm{dim}[\m{A}_{2n+1}],
\end{equation}
where $\mathrm{dim}[\cdot]$ denotes the dimension.  Also, from equation (\ref{eq:fr2}), we get
\begin{equation}
 \mathrm{dim}[\m{R}_{2n+1}]  \mathrm{dim}[\m{R}_{2n+2}]= \mathrm{dim}[\m{A}_{2n+1}] \mathrm{dim}[\m{A}_{2n+2}].
\end{equation}
We then arrive at the formula
\begin{equation}\label{eq:fermind}
 \mathrm{ind}[u]= \sqrt{\frac{\mathrm{dim}[\m{R}_{2n}]}{\mathrm{dim}[\m{A}_{2n}]}},
\end{equation}
which is independent of $n$, and looks just like what we had in the qudit case.

So far, the extension to graded commutators has not caused us any problems.  However, finding $\mathrm{dim}[\m{R}_{n}]$ is a little harder.  In the qudit case, $\m{R}_{n}$ was a \textit{simple} finite dimensional algebra and hence was isomorphic to the algebra of $r_n\times r_n$ complex matrices $\m{M}_{r_n}$ (a consequence of the Artin-Wedderburn theorem).  In \two{the fermionic} case, however, we can only guarantee that $\m{R}_n$ are \textit{semisimple} (since any finite-dimensional C*-algebra is semisimple).  And one can show that each $\m{R}_{n}$ has trivial graded center \cite{FPP17}.  Luckily, this is enough to ensure that a generalization of the Artin-Wedderburn theorem holds \cite{Baez14,FPP17}.
\begin{lemma}
 A semisimple graded algebra with trivial graded center is isomorphic to either
 \begin{enumerate}
  \item $\mathbb{C}(p|q)$, the graded algebra of matrices acting on \two{the graded Hilbert space} $\mathbb{C}^{p|q}$\two{.  (These are discussed in more detail in appendix \ref{app:fermions}, but are a generalization of even and odd operators and subspaces for fermion systems, where $p$ and $q$ are the dimensions of the even and odd subspace respectively.)}  This algebra has (complex vector space) dimension $(p+q)^2$.
  \item $Cl_1(p|q)$, the graded algebra of matrices acting on $\mathbb{C}^{p|q}$ with matrix entries in $\mathbb{C}\oplus\mathbb{C}$, which we can think of as $a\openone_2+b\sigma_z$, where $a,b\in\mathbb{C}$.  This algebra has (complex vector space) dimension $2(p+q)^2$.
 \end{enumerate}
\end{lemma}
So we know that $\mathrm{dim}[\m{R}_{2n}]=m_{2n}(p_{2n}+q_{2n})^2$ where $p_{2n},q_{2n}$ are positive integers, and $m_{2n}\in\{1,2\}$.  Then the index takes the form
\begin{equation}
 \mathrm{ind}[u]=\frac{p_{2n}+q_{2n}}{d_{2n}}\sqrt{m_{2n}},
\end{equation}
since $\mathrm{dim}[\m{A}_{n}]=d_{n}^2$.  (Note that, in the special case where $\m{A}_n$ just corresponds to, e.g, a \emph{single} Majorana mode, $\mathrm{dim}[\m{A}_{n}]=2$, which is not equal to $d_{n}^2$ with $d_{n}\in \mathbb{N}$.)

So we see that, whereas the qudit QCA index took rational values, the fermion QCA index can also include a factor of root two.  For purely fermionic systems (i.e., with no qudits but just some number of fermion modes on each site), $\mathrm{ind}[\two{u}]=2^{n/2}$, where $n\in\mathbb{Z}$.  In contrast, for hybrid systems of qudits and fermions, any value given in equation (\ref{eq:fermind}) can be achieved.

A particularly illuminating example is the Majorana shift, introduced in \cite{FPP17}.  Take a system that has a single fermion mode at each site with annihilation operators $a_{\two{n}}$.  Then consider the corresponding Majorana operators $c_{2\two{n}}=a_{\two{n}}^{\dagger}+a_{\two{n}}$ and $c_{2\two{n}+1}=i(a_{\two{n}}^{\dagger}-a_{\two{n}})$.  Define $s_M$ to be the automorphism\footnote{On a finite dimensional system with periodic boundary conditions (site $N+1$ identified with site $1$), we would need to require that, e.g., $s_M$ satisfies equation (\ref{eq:Maj_shift}) for all $n\neq 2N$ and $s_M(c_{2N})=-c_{2N+1}$.  This is to ensure unitarity, since the matrix taking the vector of $c_n$ operators to new Majorana operators must be in $\mathrm{SO}(2N)$ and hence have determinant one.}
\begin{equation}\label{eq:Maj_shift}
 s_M(c_{n})=c_{n-1}.
\end{equation}
Clearly, $s_M$ applied twice is just the fermionic shift, which would act as $s(a_n)=a_{n-1}$.  We can verify from equation (\ref{eq:fermind}) that $s_M$ has index $\sqrt{2}$.  So we can view $s_M$ as a square root of the usual fermionic shift, which has index $2$.  This is in stark contrast to a shift for qubits.  There exists no QCA $u$, such that $u^2=s$, where $s$ is the shift by one site for qubits.  This itself is a consequence of the index theory for qudits in section \ref{sec:Qudit_index}.  The automorphism $s$ has index $2$ and so $u$ would have to have index $\sqrt{2}$, which is impossible for qudits because their QCA index is always a positive rational.

Other consequences of the fermionic index theory are not quite the same as in the qudit case.  For example, two QCAs with the same index are not necessarily equivalent up to local unitaries in this setting.  We can see this with the following example.  Consider a system with a fermion mode and a qubit at each site.  Now consider two different QCA dynamics:\ a shift of the fermions one site to the right but acting like the identity on the qubits $s_f$, and a shift of the qubits one site to the right acting like the identity on the fermions $s_q$.  These both have index $2$.  And we would hope that they would be related by a finite-depth circuit, but this is not true, as shown in \cite{FPP17}.

\subsection{Index theory in higher dimensions}
\label{sec:higherindex}
Very recently, classifying QCAs using the index theory has been extended to higher dimensional systems \cite{FH19}.  The classification is complete for two dimensional qudit QCAs on finite lattices and depends on the topology of the control space (described in section \ref{sec:dyna}).  The completeness of the classification is not expected to extend to higher dimensional lattices (due to the probable existence of non-trivial QCAs in higher dimensions\one{, discussed in more detail in section \ref{sec:top_phases_2}} \cite{HFH18}), and it is not clear how it applies to fermionic QCAs.  Nevertheless, let us now take a look at the ideas behind the classification in a little more detail.  \two{Again, these results make no assumption of translational invariance of the QCAs involved.}

First, we need to discuss the notion of equivalence used in \cite{FH19}.
\begin{Definition}
 Two QCAs $u$ and $v$ are \hlt{stably path equivalent} with range $l$ if we can locally add ancillas to each of them, such that there is a continuous path of QCAs with range no greater than $l$ from $u\otimes \openone$ to $v\otimes \openone$.
\end{Definition}
(In the following section, we will think of the ancillas as \emph{additional} sites, and we will require that there is a one-to-one correspondence between the sites of $u\otimes \openone$ and those of $v\otimes \openone$.)
This is a more general notion of equivalence than that used in section \ref{sec:ind}, as we are allowed to add ancillas to each site.
Note that this includes the case where, possibly by adding ancillary qudits to each site of the system, we have $u\otimes \openone=w\circ v\otimes \openone$, where $w$ is a finite-depth local unitary circuit.
\begin{figure}[!ht]
{\centering
\resizebox{7cm}{!}{\includegraphics{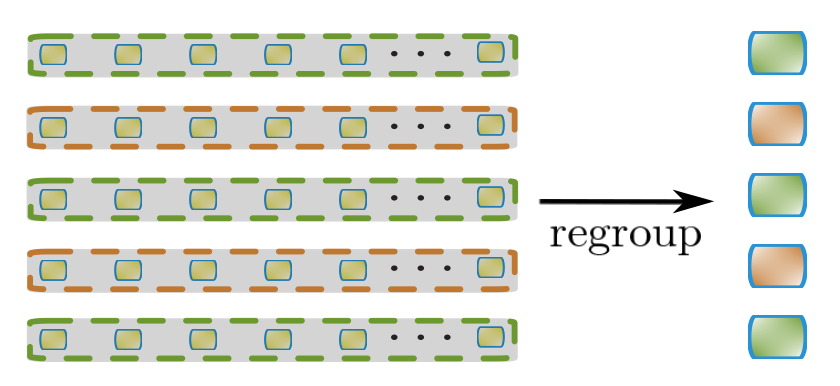}} \caption[Dimensional reduction]{Regrouping all sites along one lattice direction reduces the QCA to a QCA on a lattice with lower dimension.  The dimension of the Hilbert space corresponding to each cell in the regrouped QCA will be higher as a consequence of the regrouping.\label{fig:Ga}}
}
\end{figure}

The classification works by using the one-dimensional index together with dimensional reduction, which is an extreme form of regrouping:\ each site along a lattice direction is regrouped into a single site, so that the resulting regrouped QCA acts on a lattice system of lower dimension, as illustrated in figure \ref{fig:Ga}.  Then the classification depends on one-dimensional indices (corresponding to dimensional reductions along different directions) as well as the \hlt{topology of the underlying spatial manifold in the form of the first homology group and the torsion}.\footnote{A very rough description of the first homology group is that it classifies the topology of the manifold via one-dimensional closed loops that cannot be smoothly contracted to a point.  A similarly rough description of torsion is that it is a property of some non-orientable manifolds where a non-contractable loop in one direction, after being shifted all the way along a path in the manifold can become a loop in the opposite direction.}

Let us look at some useful examples.  Suppose our QCA lives on a discrete torus, with $d$-dimensional qudits at each site and coordinates $(n,m)$, where $n,m\in\{0,...,N-1\}$.  Now consider some possible dynamics for the QCA.  One very simple example is the following:\ $s_{K}$ acts like the identity on all qudits, except those with coordinates $(K,m)$ for all $m$ and fixed $K$.  On these $s_{K}$ shifts the qudit at site $(K,m)$ to site $(K,m+1)$.  Note the periodic boundary conditions with $N\equiv 0$.  We can see that $s_{K}$ is not equivalent to the identity because we can consider the dimensional reduction in the $x$-direction to a one-dimensional QCA, which is a partial shift on a circle.  This has index $d$ and so is not equivalent to the identity.  However, if we reduce this to a QCA along the other direction, we get an index one QCA.  More interesting behaviour is given by the QCA that shifts qudits from site $(n,m)$ to site $(n+1,m+1)$ for all $n$ and $m$.  Whichever direction we reduce, we get a QCA with index not equal to $1$.
\begin{figure}[!ht]
{\centering
\resizebox{14cm}{!}{\includegraphics{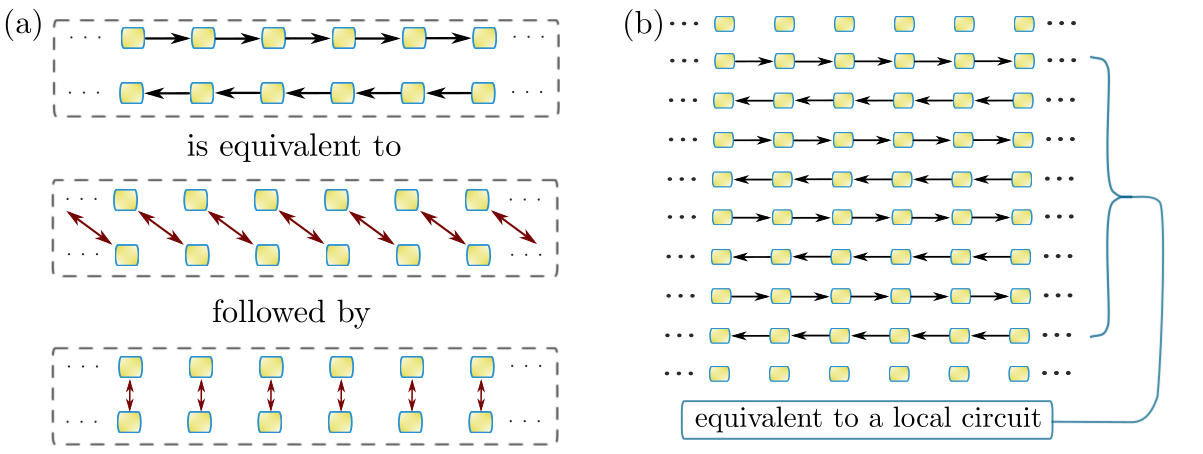}} \caption[Shifts and swaps]{Subfigure (a) shows how two shifts in opposite directions (black arrows) are equivalent to a depth-two local circuit of swaps (red arrows).  Subfigure (b) depicts dynamics involving many shifts in opposite directions.  This is also equivalent to a depth-two local circuit, as long as there is a shift to the left for every shift to the right.\label{fig:H}}
}
\end{figure}

Much like the loops arising in the first homology group, shifts around loops play a key role in this classification.
A useful argument shows us that any such ``shift loop'' on the torus can be contracted to any other (with the same index along the same direction) via a finite-depth local circuit.  Let us sketch how this works.  First, consider a QCA $s_{0}\circ s_{1}^{-1}$ which is just a shift of the qudits with coordinates $(0,m)$ for each $m$ and a shift in the opposite direction of all qudits with coordinates $(1,m)$ for each $m$.  Everywhere else it acts like the identity.  As seen in figure \ref{fig:H} (a), this can be written as a product of swaps:\ first apply the swaps between sites $(0,m)$ and site $(1,m+1)$ for each $m$.  And then apply the swaps between sites $(0,m)$ and site $(1,m)$ for each $m$.  So this QCA, which consists of two shifts in opposite directions that are beside each other, is equivalent to a finite-depth circuit.  Now define a generalization of $s_{0}\circ s_{1}^{-1}$ by
\begin{equation}
 v_{a,b}=\prod_{K=0}^{(b-a)/2}\left(s^{\ }_{a+2K\,}s^{-1}_{a+2K+1}\right).
\end{equation}
So if $a=b=0$, we get $s_{0}\circ s_{1}^{-1}$.  Note that $v_{a,b}$ is also clearly equivalent to a finite-depth circuit by a similar trick, as depicted in figure \ref{fig:H}.  Now notice that
\begin{equation}
 w:=v_{1,N/2-1}\circ v^{\ }_{0,N/2}=s^{\ }_0\circ s^{-1}_{N/2}.
\end{equation}
But $w$ is also a finite-depth circuit of swaps.  Then we have $s_{0}=w\circ s_{N/2}$, so $s_{0}$ and $s_{N/2}$ are (perhaps surprisingly) equivalent up to finite-depth circuits despite being shift loops far away from each other.

This idea also works in, e.g., the two-dimensional plane (possibly by adding local ancillas), though determining which loops are equivalent will depend on any holes in the plane.

Therefore, for the two-dimensional classification of QCAs, the first homology group of the covering manifold plays a big role, since it classifies loops in the manifold that cannot be contracted to a point.    And intuitively, the same thing happens for shift loops around any ``holes'' in the manifold.
The reason that torsion plays a role is also rather interesting.  Roughly, a manifold has torsion if we can move a loop along the manifold so that it changes its direction (so this can only happen for non orientable manifolds).  This means that we can use this topological property of the manifold to contract some seemingly non trivial pairs of shift loops to the identity.  This can happen for a QCA with, e.g., the Klein bottle as its control space.

\one{
\subsection{Group theory of QCAs}
\label{sec:GroupQCA}
We have already seen some hints at the group structure of QCAs in some contexts.  For example, the one-dimensional index is a group homomorphism from the group of QCAs to a subgroup of the reals (just the rationals in the case of qudit QCAs).  One point to bear in mind is the following.  For QCAs to form a group, the product of two QCAs must also be a QCA.  For infinite systems, this is simple:\ if the range of QCA $u$ is $l_u$ and the range of QCA $v$ is $l_v$, then the range of the QCA $u\circ v$ is at most $l_u+l_v$.  So for the product of any two QCAs on an infinite system, we get a QCA with a finite range.  This becomes a little more tricky for finite systems.  For a fixed finite system, we can eventually take enough products so that the resulting QCA has range comparable to the size of the system, which is clearly not what we mean by a QCA.  If we insist on working with finite systems, then this is another place where the idea of control spaces discussed in section \ref{sec:dyna} comes in handy.

In the case of finite systems, one option is to consider families of QCAs that are defined in terms of the control spaces \cite{FHH19}, as discussed in section \ref{sec:dyna}.  However, one of the interesting points made in \cite{FHH19} was that it is often natural to restrict to \hlt{coherent families} of QCAs.  Coherent families are families of QCAs $u_i$ with the property that $u_i$ is stably path equivalent with range $l_i=l2^{-1}$ to $u_{i+1}$ for every $i$.  This is best understood by looking at an example.  Consider the family with control space given by the circle with radius one.  The QCA $u_i$ is chosen to be a shift of all cells along the circle (in the same direction for each $i$), and the sites of QCA $u_i$ live at positions $n/2^i$ on the circle, where $n\in\{0,1,...,2^i-1\}$.  Then we can show that $u_i$ is stably equivalent to $u_{i+1}$ in the following way.  Consider adding an ancillary qubit to each of $u_i$'s lattice sites, so for each site $n/2^i$, there is another qubit that we think of as living at site $(2n+1)/2^{i+1}$.  Now this system has the same lattice system as $u_{i+1}$, but $u_i\otimes\openone$ is not the same as $u_{i+1}$.  However, notice that $u_i\otimes\openone$ shifts the qubit at site $n/2^i$ to $(n+1)/2^i$.  Then, if we subsequently apply the swaps between the neighbouring sites $(n+1)/2^i$ and $(2n+1)/2^{i+1}$ for all $n$, which we can call $w$, we get that $w\circ (u_i\otimes\openone) = u_{i+1}$, which means that $u_i$ and $u_{i+1}$ are indeed stably path equivalent, since $w$ is a finite-depth local circuit.

These coherent families have the property that different members are related (as they are stably path equivalent).  There was no such requirement for families of QCAs in general, which means that one could come up with rather unnatural families.  E.g., consider modifying the shift example above so that it is the same except every $u_i$ with even $i$ is just the identity automorphism, as opposed to a shift.  These QCAs do not have the same index for every $i$, which seems quite unnatural.  The restriction to coherent families is one option to ensure that the automorphisms in the family are related in a meaningful way.  However, it is not clear whether all seemingly natural families of QCAs are also coherent families.  In the case when the control space is a torus, however, it can be shown that any translationally invariant Clifford QCA in any dimension and any translationally invariant QCA in three dimensions defines a coherent family \cite{FHH19}.  (Strictly speaking, the families associated to these translationally invariant QCAs may only contain an infinite \emph{subsequence} that is coherent.)

Analogously to coherent families of QCAs, one can define coherent families of states \cite{FHH19}, which are entanglement renormalization group fixed points.

One of the most interesting results regarding the group theory of QCAs is that the group of QCAs modulo finite-depth quantum circuits is abelian.  This was shown for QCAs in arbitrary dimensions in \cite{HFH18} with the caveat that we may have to append ancillas at each lattice site.  Remarkably, in \cite{FHH19} the result was strengthened by showing that for most reasonable control spaces the ancillas are no longer necessary.  This is a consequence of a general method called ancilla removal:\ given a QCA that can be implemented by a finite-depth local circuit that acts like the identity on ancillary sites, then there is another finite-depth circuit implementing the same automorphism where \emph{each element} of the circuit acts like the identity on the ancillary sites.  Note that the depth and range of the new circuit may be larger.  The idea behind this is to break a single timestep up into a few sub-timesteps and to break the circuit up into blocks acting at different sub-timesteps.  The trick then is to use some of the original sites not in use at a given sub-timestep as the ancillas at that sub-timestep.
}

\subsection{Margolus partitioning}
\label{sec:Margolus}
A generalized \hlt{Margolus partitioning} method for translationally invariant QCAs was first introduced in \cite{SW04}.  It gives us a different method of constructing a QCA from local unitaries.  A crucial point about this decomposition is that the local unitaries are not generally unitaries \textit{on} subsystems but rather local unitary maps \textit{from one subsystem to another}.  This is a significant departure from, e.g., the local finite-depth circuit QCAs we have already seen (sections \ref{sec:ex} and \ref{sec:partitioning}).  \two{The ideas apply to QCAs on finite or infinite lattices.}

\begin{figure}[!ht]
{\centering
\resizebox{7cm}{!}{\includegraphics{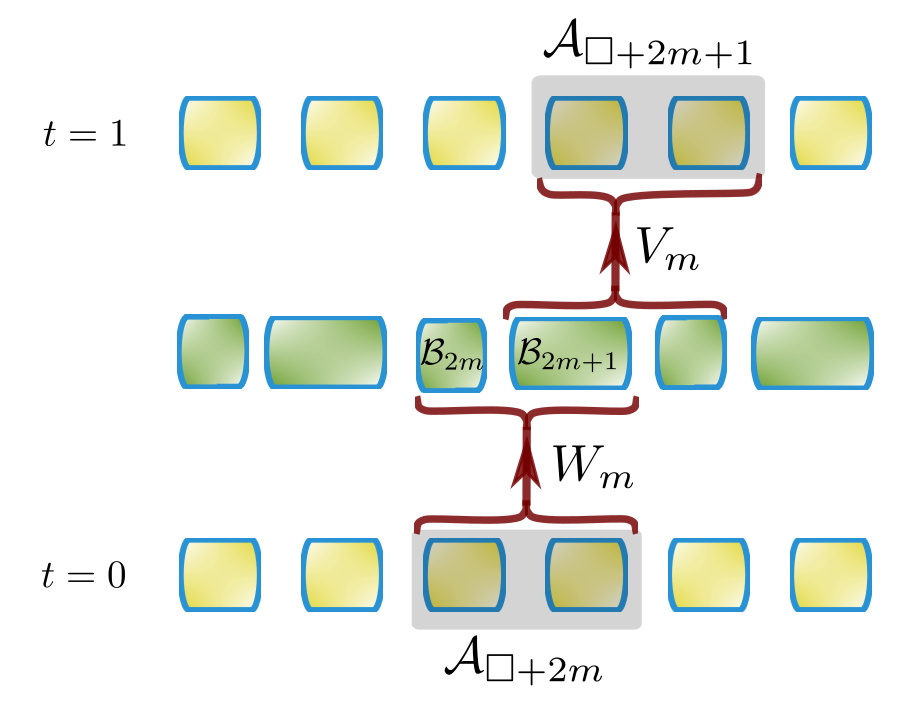}} \caption[QCAs by Margolus partitioning]{Constructing QCAs via Margolus partitioning involves an intermediate step with new algebras $\mathcal{B}_{n}$.  The unitaries in the decomposition are not unitaries on subsystems but rather unitary maps $W_m$ from subsystems (yellow boxes) to other ancillary systems (green boxes) followed by unitary maps from the ancillas back to the QCA subsystems $V_m$.\label{fig:J}}
}
\end{figure}

Let us introduce the construction for one-dimensional QCAs with $d$-dimensional qudits at each site.  (The generalization to higher dimensions is straightforward.)  First, we introduce a generalized Margolus partitioning (given in \cite{SW04} and generalized from \cite{TM90}).  We regroup so that the QCA is nearest-neighbour, and then we define the supercell $\square$ to contain two sites:
\begin{equation}
 \square =\{0,1\}.
\end{equation}
In higher dimensions, we would have a larger cube as our supercell.  Then we define $\m{A}_{\square}=\m{A}_0\otimes\m{A}_1$, and more generally we define $\m{A}_{\square+2m}=\m{A}_{2m}\otimes\m{A}_{2m+1}$, where $m\in\mathbb{Z}$.
We define the algebras $\mathcal{B}_{2m}\subseteq \m{A}_{\square+2m -1}$ and $\mathcal{B}_{2m+1}\subseteq \m{A}_{\square+2m +1}$, with $\mathcal{B}_n\cong\mathcal{M}_{b(n)}$, the $b(n)\times b(n)$ complex matrices.  Consider the algebras $\mathcal{B}_{2m+1}$ and $s^{-2}(\mathcal{B}_{2m})$, where $s$ is the shift.  Notice that these are both subalgebras of $\m{A}_{\square+2m +1}$.  The central assumption necessary for the construction of QCAs via Margolus partitioning is that these algebras commute and they also generate $\m{A}_{\square+2m + 1}$.  We may then define an automorphism using this partitioning via two steps:\ first, for each $m\in\mathbb{Z}$, we have a map
\begin{equation}
 w_m: \mathcal{A}_{\square+2m}\rightarrow \m{B}_{2m}\otimes \m{B}_{2m+1},
\end{equation}
where $w_m(\cdot)=W^{\dagger}_m \cdot W_m$ and $W_m$ is a unitary.  This is followed by	
\begin{equation}
 v_m: \m{B}_{2m+1}\otimes \m{B}_{2m+2}  \rightarrow \mathcal{A}_{\square+2m+1},
\end{equation}
with $v_m(\cdot)=V^{\dagger}_m \cdot V_m$ and $V_m$ unitary.
Then the QCA automorphism is given by
\begin{equation}\label{eq:Margolus}
 u(\cdot)=\prod_m V_m^{\dagger} \prod_n W_n^{\dagger}\left( \cdot \right)\prod_l W_l \prod_k V_k.
\end{equation}
See figure \ref{fig:J} for an illustration of how this works.  It helps to consider how this looks for a shift QCA.  In that case, $\mathcal{B}_{2m}=\mathcal{A}_{\square+2m -1}$ and $\mathcal{B}_{2m+1}$ is just $\mathbb{C}\openone$.  So the shift can be written in terms of local unitary maps in this way.  But as we saw in section \ref{sec:ind}, there are QCAs (such as a shift) that cannot be written as a local unitary circuit.  The point is that those circuits involve unitaries \emph{on} systems, as opposed to the more general unitary maps we have considered here.  So QCAs written using this partitioning scheme may not be implementable by a constant-depth quantum circuit.

In fact, any one-dimensional QCA with $d$-dimensional qudits at each site can be written in this Margolus-partitioned form (which is proven in \cite{SW04} for translationally invariant QCAs in one dimension).  The proof follows by defining $\m{B}_n:=\m{R}_n$, where $\m{R}_n$ denotes the support algebras from section \ref{sec:ind}.  The proof in \cite{SW04} is actually stated as applying to higher dimensions too, but this is not true, as a counterexample was found in two dimensions in \cite{ANW08}.  Nevertheless, one can still \emph{construct} QCAs in higher dimensions via a Margolus partitioning scheme, and this found application in, e.g., \cite{Raussendorf05,CPS17}.

\subsection{Tensor-network unitaries}
\label{sec:MPU}
Our final structure result for QCAs involves tensor networks.  For example, in one dimension it turns out that translationally invariant qudit QCAs are actually equivalent to a type of tensor network known as \hlt{matrix product unitaries} \cite{CPS17,SSB17}.  (For a good review of tensor network methods, see \cite{BC17}.)  This equivalence is not true if we relax the requirement of translational invariance:\ QCAs are still MPUs, but an MPU need not be a QCA because it may no longer be a locality-preserving operation.  \two{(We will give an example showing why this is true below.)}  The motivation for this decomposition of QCAs in terms of tensor networks is that it is useful for characterizing symmetries of QCAs \cite{CPS17,GSS18}, and also topological phases of Floquet models \cite{PFM16}.  \two{Note that these results only apply to finite lattices.}

\begin{figure}[!ht]
{\centering
\resizebox{13cm}{!}{\includegraphics{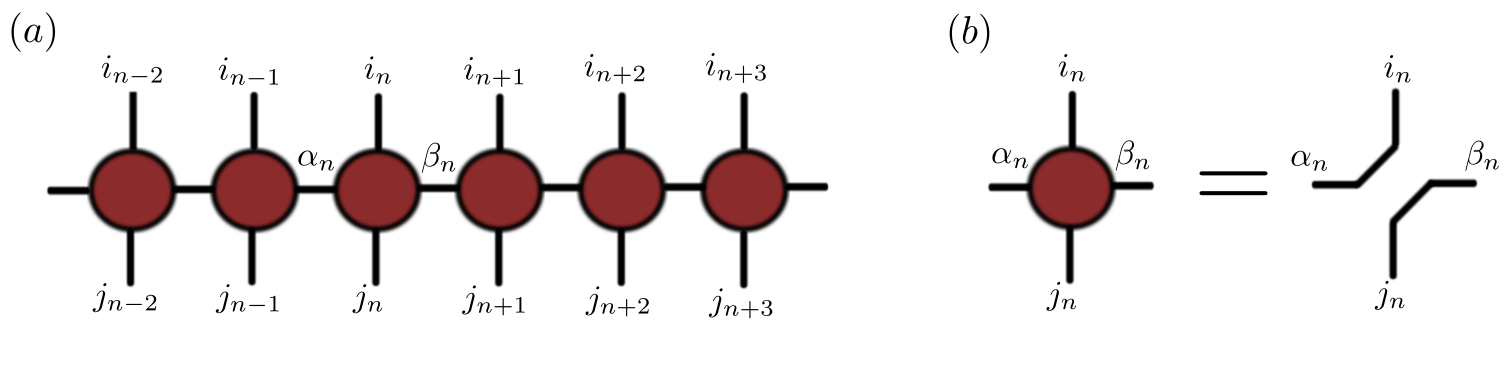}} \caption[Matrix product unitaries]{Subfigure (a) shows a general matrix product unitary.  The $n$th tensor, for example, has an input leg with index $j_n$, output leg with index $i_n$ and bond indices with indices $\alpha_n$ and $\beta_n$.  Subfigure (b) shows the choice of tensor needed for a matrix product unitary to implement the shift QCA.  The straight black lines represent delta functions like $\delta_{\alpha_ni_n}$. \label{fig:K}}
}
\end{figure}

First, let us introduce translationally invariant matrix product unitaries.  (Note that we work in the Schr{\"o}dinger picture in this section.)  We have a one-dimensional system with periodic boundary conditions and $d$ dimensional qudits at each site.  A translationally invariant matrix product unitary is constructed out of tensors $w^{ij}_{\alpha\beta}$.  Here $i,j\in\{0,...,d-1\}$ are the physical indices and $\alpha,\beta\in\{0,\two{...,}D-1\}$ are known as bond indices, where $D$ is a positive integer, known as the bond dimension.  It helps to think of indices $i$ and $j$ as labelling different matrices and $\alpha$ and $\beta$ as indices specifying the matrix elements.  The associated matrix product unitary is then
\begin{equation}
 U=\sum_{\substack{i_1,...,i_N\\j_1,...,j_N}}\mathrm{tr}\left[w^{i_1j_1}...w^{i_Nj_N}\right]\ket{i_1,...,i_N}\bra{j_1,...,j_N},
\end{equation}
where the $w^{ij}$ matrices are being multiplied inside the trace (i.e., the bond indices are contracted), and the trace is also over the bond indices.  The ket $\ket{i_1,...,i_N}$ is a basis for the $N$-qudit Hilbert space.  As usual for tensor networks, this is best understood in terms of a picture, as in figure \ref{fig:K} (a).  We can consider a simple example, a shift.  In this case, the simple tensor
\begin{equation}
 w^{ij}_{\alpha\beta}=\delta_{i\alpha}\delta_{j\beta}
\end{equation}
does the job, as displayed in figure \ref{fig:K} (b).  Notice that the bond dimension is equal to the qudit dimension in this case.

\begin{figure}[!ht]
{\centering
\resizebox{12cm}{!}{\includegraphics{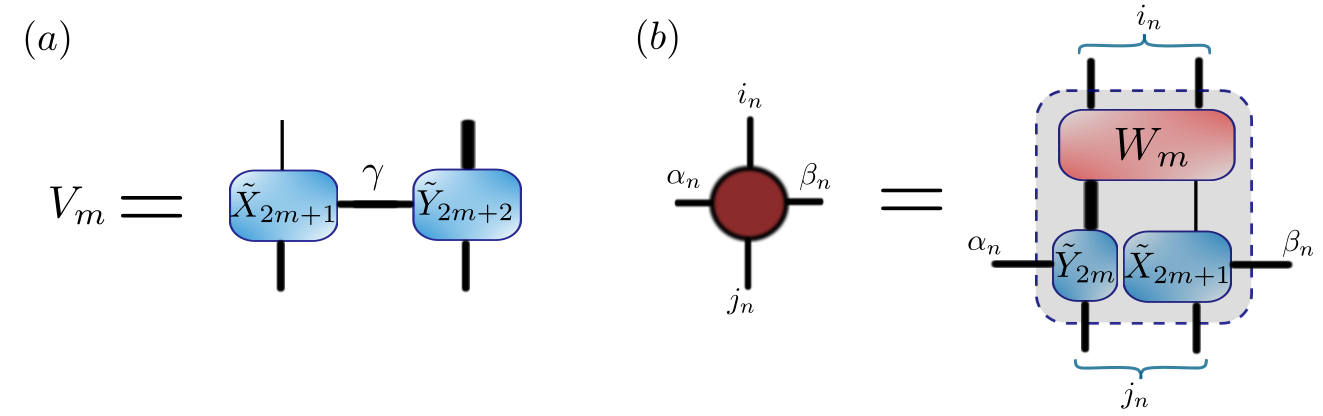}} \caption[Tensor for MPU decomposition of a QCA]{In subfigure (a) we see the decomposition of the unitary map $V_n$ into a sum over $\gamma$ of tensor products of maps $\tilde{X}_{2m+1}^{\gamma}\otimes \tilde{Y}_{2m+2}^{\gamma}$.
The thicker and thinner input and output lines highlight that the dimensions of the local vector spaces may be different.  Subfigure (b) shows how the tensor building the matrix product unitary representation of the QCA unitary is put together.  Note that the input and output indices act on two neighbouring qudits of the QCA. \label{fig:L}}
}
\end{figure}

Next, let us convince ourselves that a one-dimensional QCA of $d$-dimensional qudits is always a matrix product unitary with finite bond dimension.  To do this, we will regroup so that the QCA is nearest neighbour.
As we saw in section \ref{sec:Margolus}, we can write the QCA in terms of local unitary maps.
The trick now is to rewrite each $V_m$ in the local map decomposition of the QCA from equation (\ref{eq:Margolus}) in an operator basis.  We know that $v_m(\cdot)=V_m^{\dagger}\cdot V_m$ maps $\m{B}_{2m+1}\otimes \m{B}_{2m+2}$ to $\m{A}_{2m+1}\otimes \m{A}_{2m+2}$ (in the Heisenberg picture).  Let us drop the position index for a moment, so we have that $v(\cdot)=V^{\dagger}\cdot V$ takes $\m{B}\otimes\m{B}^{\prime}$ to $\m{A}\otimes\m{A}^{\prime}$, where, e.g., $\m{B}=\m{B}_{2m+1}$.  In terms of Hilbert spaces, $V$ is a unitary from $\m{H}_{\m{A}}\otimes\m{H}_{\m{A}^{\prime}}$ to $\m{H}_{\m{B}}\otimes\m{H}_{\m{B}^{\prime}}$.  Note that, e.g., $\mathrm{dim}(\m{H}_{\m{A}})$ need not equal $\mathrm{dim}(\m{H}_{\m{B}})$, but we must have that $\mathrm{dim}(\m{H}_{\m{A}})\times\mathrm{dim}(\m{H}_{\m{A}^{\prime}})=\mathrm{dim}(\m{H}_{\m{B}})\times\mathrm{dim}(\m{H}_{\m{B}^{\prime}})$.  Let $X^r$ denote a basis of linear maps from $\m{H}_{\m{A}}$ to $\m{H}_{\m{B}}$ and let $Y^r$ denote a basis of linear maps from $\m{H}_{\m{A}^{\prime}}$ to $\m{H}_{\m{B}^{\prime}}$.  Then we can write $V$ in the form
\begin{equation}
 \begin{split}
  V & =\sum_{r=1}^{d_{\m{A}\m{B}}}\sum_{s=1}^{d_{\m{A}^{\prime}\m{B}^{\prime}}}c^{rs}X^{r}\otimes Y^{s}
 \end{split}
\end{equation}
where $c^{rs}\in \mathbb{C}$, and $d_{\m{X}\m{Y}}$ is the dimension of the space of linear maps from $\m{H}_{\m{X}}$ to $\m{H}_{\m{Y}}$.  This can be condensed by using singular value decomposition to get
\begin{equation}
 \begin{split}
  V & =\sum_{\gamma=1}^{D}\tilde{c}^{\gamma}\tilde{X}^{\gamma}\otimes \tilde{Y}^{\gamma},
 \end{split}
\end{equation}
where $\tilde{c}^{\gamma}\in \mathbb{C}$ and $D\leq\max\{d_{\m{A}\m{B}},d_{\m{A}^{\prime}\m{B}^{\prime}}\}$.  Absorbing the number $\tilde{c}^{\gamma}$ into the operator $\tilde{X}^{\gamma}$ and plugging back in the position dependence, we get
\begin{equation}
 \begin{split}
  V_{m} & =\sum_{\gamma=1}^{D(m)}\tilde{X}_{2m+1}^{\gamma}\otimes \tilde{Y}_{2m+2}^{\gamma}.
 \end{split}
\end{equation}
Now we define the tensors
\begin{equation}
 w^{ij}_{\alpha\beta}[m] = \left[W_m\, \tilde{Y}_{2m}^{\alpha}\otimes\tilde{X}_{2m+1}^{\beta}\right]_{ij},
\end{equation}
\two{where $W_m$ is the other unitary map from the decomposition of the QCA in equation (\ref{eq:Margolus}).}
So now we have a decomposition of the QCA in terms of local tensors (see figure \ref{fig:L}).
Thus, any QCA in one dimension can be viewed as matrix product unitary.  This matrix product unitary approach allows a different derivation of the QCA index \cite{CPS17}.
It can be shown (in the translationally invariant case) that the converse of this result is also true, namely that any matrix product unitary with finite bond dimension is equivalent to a QCA \cite{CPS17}.  \two{To see why translational invariance is necessary, consider the simple matrix product unitary that describes a generalized CNOT on the whole lattice controlled on one lattice site.  Note that this works for any lattice dimension and the MPU (or tensor-network unitary more generally) is defined by just two tensors.  The tensors, which now depend on position $n$, are given by 
\begin{equation}
  w_{\alpha\beta}^{ij}[n]=\begin{cases}
                                 \delta_{\alpha\beta ij}\mathrm{\ if\ }n=0\\
                                 \delta_{\alpha\beta}(\sigma_x^{ij})^{\alpha}\mathrm{\ otherwise,}
                                \end{cases}
\end{equation}
where all indices take values in $\{0,1\}$.  Here $\delta_{\alpha\beta ij}$ is one if all indices are equal and is zero otherwise.  Also note that $(\sigma_x^{ij})^{\alpha}$ is the Pauli X matrix with matrix indices $i,j$ to the power of $\alpha$.  If we consider the operator $\sigma_x^{0}$, then $U^{\dagger}\sigma_x^{0}U=\prod_{n}\sigma_x^{n}$, so $U$ is clearly not locality preserving.}

These MPUs can be further classified in the presence of on-site unitary symmetries \cite{GSS18}.  This works similarly to the classification of matrix product \emph{states} in the presence of symmetries \cite{BC17}.  The main result of \cite{GSS18} is that one-dimensional matrix product unitaries are completely classified by (i) the QCA index and (ii) the cohomology class of the representation of the symmetry group acting on the bond indices\two{, as conjectured in \cite{Hastings13}}.  The classification considers two matrix product unitaries $U_0$ and $U_1$ to be equivalent if, possibly by adding ancillas transforming under \emph{arbitrary} representations of the symmetry group, there is a continuous path of matrix product unitaries $U_t$ joining $U_0$ and $U_1$.  If we restrict our attention to equivalence under only appending local ancillas that all transform under the \emph{same} representation, then the classification can be further refined, though it remains incomplete.  Non-unitary symmetries, relevant for, e.g., the tenfold way in condensed matter physics, are discussed in \cite{CPS17}, where (i) complex conjugation, (ii) transposition and (iii) time-reversal symmetric MPUs are classified to some extent.  In this case, the classification of matrix product unitaries that are symmetric under complex conjugation is complete.

\three{Further developments have been made recently relating tensor networks and QCAs.  First, the connection between one-dimensional fermionic QCAs and fermionic matrix product states is now understood \cite{PTSC20}.  This is trickier than in the qudit case:\ a very natural definition for fermionic matrix product states was shown \emph{not} to capture all fermionic QCAs \cite{PTSC20}.  In fact, it is the Majorana shift that we saw in equation (\ref{eq:Maj_shift}) in section \ref{sec:fermion_index} that provides a counterexample.  To circumvent this problem, the definition of fermionic matrix product states is extended to include ancillary degrees of freedom at each site.  Then a subset of these tensor networks does include all fermionic QCAs.  This allows a re-derivation of the fermionic QCA index we saw in section \ref{sec:fermion_index}.

Another recent development was that qudit QCAs in \emph{any} spatial dimension can be viewed as tensor-network unitaries with bond dimension independent of the system size \cite{PC20}.  The only requirement is that the tensors satisfy a property called \hlt{simplicity}, which ensures that the action of the tensor-network must be locality-preserving.  (For most tensor networks, this property seems difficult to verify, as it is not a local criterion but rather requires contracting the entire tensor network.)
A consequence of this equivalence is that the entanglement created by QCAs always obeys an area law.  Furthermore, this formalism also extends naturally to the non-unitary case, where the dynamics is a completely positive trace-preserving map.  In that case, however, tensor networks with bond dimension independent of the system size are \emph{not} always enough to describe the dynamics.}

\section{QCAs in physics}
Aside from being a model for quantum computation, QCAs arise in physics, for example, as effective models for Floquet dynamics or as proposals for quantum simulators.  One of the reasons this is natural is because QCAs have an inherent maximum speed of propagation of information.  Depending on the setting, this speed can be identified either with the speed of light if we are approximating a relativistic system with a QCA; or we can identify it with the speed of sound or Lieb-Robinson velocity if we are approximating non-relativistic systems.  In the following sections, we will look at both of these applications, but first let us look in more detail at the differences between QCA dynamics and continuous-time Hamiltonian evolution.

\subsection{QCAs vs Hamiltonian dynamics}
\label{sec:Ham}
We can simulate the dynamics of continuous-time systems on lattices with local Hamiltonians by using the Suzuki-Trotter decomposition \cite{Trotter59,Suzuki90,Nielsen00}.  By ordering the unitary gates in the right way \cite{Farrelly15}, we get a QCA.  A different approach uses Lieb-Robinson bounds \cite{LR72} to ensure that we can approximate continuous-time dynamics with high accuracy using a QCA \cite{Osborne06,HP17}.  For any such simulation methods to work, it is essential that the continuous-time model has some \emph{approximate} locality-preserving property, e.g., it obeys Lieb-Robinson bounds.  On the other hand, continuous-time evolution cannot be strictly locality preserving.  Let us convince ourselves of this fact.

Consider qudits on a lattice, evolving via the \two{local} time-independent Hamiltonian $H$.  \two{Here, local means that the Hamiltonian is a sum of local terms, i.e., $H = \sum_{X\in \mathcal{X}}h_X$, where $\mathcal{X}$ are all $d$-dimensional hypercubes on the lattice of size at most $l^d$, where $l$ is a positive integer and $d$ is the lattice dimension, and $h_X$ is an operator acting non-trivially only on qudits in $X$.}  And consider an operator localized at site $\vec{0}$ at $t=0$, so $A\in\mathcal{A}_{\vec{0}}$.  Now suppose the dynamics is locality preserving in a strict sense.  This means that there would be some time $T$ such that, for all $t<T$, the evolved operator $e^{iHt}Ae^{-iHt}$ is strictly contained within some finite region $R$ containing the point $\vec{0}$.  So, for any $B$ localized outside of $R$ and any $t<T$, we must have $[B,e^{iHt}Ae^{-iHt}]=0$.  Taylor expanding for small $t$, we get
\begin{equation}
 [B,e^{iHt}Ae^{-iHt}]  = [B,A] + i[B,[H,A]]t +O(t^2)=0,
\end{equation}
where we can use $[B,A]=0$ to get
\begin{equation}
 [B,[H,A]] +O(t)=0.
\end{equation}
We can take $t$ to be as small as we wish, so the second term on the second line can be made arbitrarily small.  Therefore, the first term $[B,[H,A]]$ must be zero.  We can do the same for the higher order terms, and we see that all the nested commutators must vanish, e.g., $[B,[H,[H,A]]]=0$.  But this means that $[B,e^{iHt}Ae^{-iHt}]=0$ for \emph{all} $t$, which tells us that the operator $e^{iHt}Ae^{-iHt}$ never actually has support outside of the region $R$ for any time $t$.  And this is true for any $A\in\m{A}_{\vec{0}}$.  Hence, if information is ever going to spread from $\vec{0}$ to outside $R$, it does so immediately, though with a very small amplitude in general.

This simple argument shows us why we need to consider discrete time for lattice systems if we want our dynamics to be strictly locality preserving.

Another interesting question regarding the connection between Hamiltonians and QCAs is whether there is a natural way to associate a Hamiltonian to a QCA.\footnote{The analogous question of finding a Hamiltonian for continuous-time dynamics obeying a Lieb-Robinson bound has a positive answer \cite{WW20}.}  For finite lattices, we always have a unitary $U$ implementing the dynamics, so we can define $H$ by $U=e^{-iH}$.  A drawback of this prescription is that it is not unique because we can add multiples of $2\pi$ to the eigenvalues of $H$ and still get the same $U$.  Furthermore, this Hamiltonian will generally not be strictly local\footnote{Another option for finding a Hamiltonian corresponding to a QCA is to consider $i(U-U^{\dagger})$, which is self adjoint though generally not a sum of local terms.} (i.e., a sum of local operators on the lattice), as can be seen by writing $H=i\ln(U)$, defined as a power series:
\begin{equation}\label{eq:lnU}
 H=i\ln(U)=-i\sum_{n=1}^{\infty}\frac{(\openone - U)^n}{n},
\end{equation}
which holds as long as $\|U-\openone \|<1$ \cite{Hall03}.  This contains interactions between sites arbitrarily far away from each other, though the strength of the interaction between sites decays with distance.  Interestingly, one can construct QCAs such that \textit{all} Hamiltonians $H$ satisfying $U=e^{-iH}$ are fully non-local, meaning that the interaction terms in $H$ between different regions do not decay with distance \cite{ZFF20}.  For this to be possible, the unitary must violate $\|U-\openone \|<1$, so that the right hand side of equation (\ref{eq:lnU}) does not converge.

A related question for QCAs is whether there is a meaningful notion of a ground state or energy ordering.  Because the eigenvalues of a unitary are complex phases, there is no natural way to order the eigenvalues.  This means that there is no clear way to decide which state has the lowest energy.  For simulating dynamics via QCAs, this may not be much of a problem, but this is important to take into consideration if we wish to view QCAs as something more fundamental.  This is because understanding low energy and ground state properties of physical systems plays a huge role in, e.g., condensed matter and high energy physics.  But for Hamiltonians there is a natural ground state and a meaningful notion of an energy gap.  At the moment, there appears to be nothing similar for QCAs.

There is one final setting where QCAs arise naturally:\ as approximations to or toy models for periodically driven quantum systems, also known as \hlt{Floquet quantum systems} \cite{DHI98}.  Suppose we have a lattice system and a time-dependent Hamiltonian $H(t)$ that is periodic with period $T$, i.e., for all $t$,
\begin{equation}
 H(t+T)=H(t).
\end{equation}
The Schr{\"o}dinger equation is
\begin{equation}
 i\frac{\partial}{\partial t}\ket{\psi(t)}=H(t)\ket{\psi(t)}.
\end{equation}
This has solutions \cite{DHI98}
\begin{equation}
\begin{split}
 & \ket{\Psi_{\alpha}(t)} =e^{-i\epsilon_{\alpha}t}\ket{\Phi_{\alpha}(t)}\\
 \mathrm{where}\ & \ket{\Phi_{\alpha}(t+T)} =\ket{\Phi_{\alpha}(t)}.
\end{split}
\end{equation}
Here, $\epsilon_{\alpha}$ are found by solving $(H(t)-i\partial/\partial t)\ket{\Phi_{\alpha}(t)}=\epsilon_{\alpha}\ket{\Phi_{\alpha}(t)}$.  These are called quasi-energies, and they are real and unique up to adding integer multiples of $2\pi/T$.  If we restrict our attention to times $t=nT$, where $n\in \mathbb{Z}$, then we have a discrete-time system evolving via $U^n$, where $U$ is a unitary operator, given by
\begin{equation}
 U=\mathcal{T}\exp\left(-i\int_0^T\!\!\mathrm{d}t\, H(t) \right),
\end{equation}
where $\mathcal{T}$ is the time-ordering operator.  Typically, $U$ will not be locality preserving, but even in the time-dependent case, Lieb-Robinson bounds are enough to guarantee that we can approximate the dynamics well with a QCA \cite{Osborne06,HP17}.  This allows us to use QCAs as toy models for Floquet systems and to use QCA techniques to understand some of their properties, as we will see in the following section.

\subsection{Dynamical topological phases in Floquet systems}
\label{sec:top_phases}
In general, \hlt{topological phases of matter} are phases that do not fit well into the classification of phases in terms of symmetry breaking \cite{Wen17}.  Instead, they are classified by topological properties of the system.  A key example is the integer quantum Hall effect, where the topological classification arises from Chern numbers in the TKNN formula for the Hall conductivity \cite{Tong16}.  The quantized Hall conductivity can be observed when such a system (which consists of non-interacting particles) is at temperatures low compared to the gap, and so the system is close to its ground state.  Then one observes the existence of chiral edge modes:\ in the two-dimensional bulk of the material the system is actually an insulator, but along the boundary these chiral edge modes allow the material to conduct.  A requirement for this phenomenon to occur is that there is some disorder in the system, which leads to localization in the material.

The integer quantum Hall effect involves the low temperature physics of \emph{static} systems.  But we may also ask whether similar phenomena can occur in \emph{dynamical} systems.  To answer this question, dynamical topological phases have been introduced and classified in Floquet systems \cite{KBR10,RLB13,TLR15,16TBR} for non-interacting particles.  These are called anomalous Floquet-Anderson insulators, and one again sees the presence of propagating edge modes on a two-dimensional material.  It became clear later in \cite{PFM16,FPP17} that this phenomenon can persist even in the presence of interactions.  Here we will look at the models given in \cite{PFM16}, which display this behaviour, admitting information propagation only along the boundary.

\begin{figure}[!ht]
{\centering
\resizebox{13cm}{!}{\includegraphics{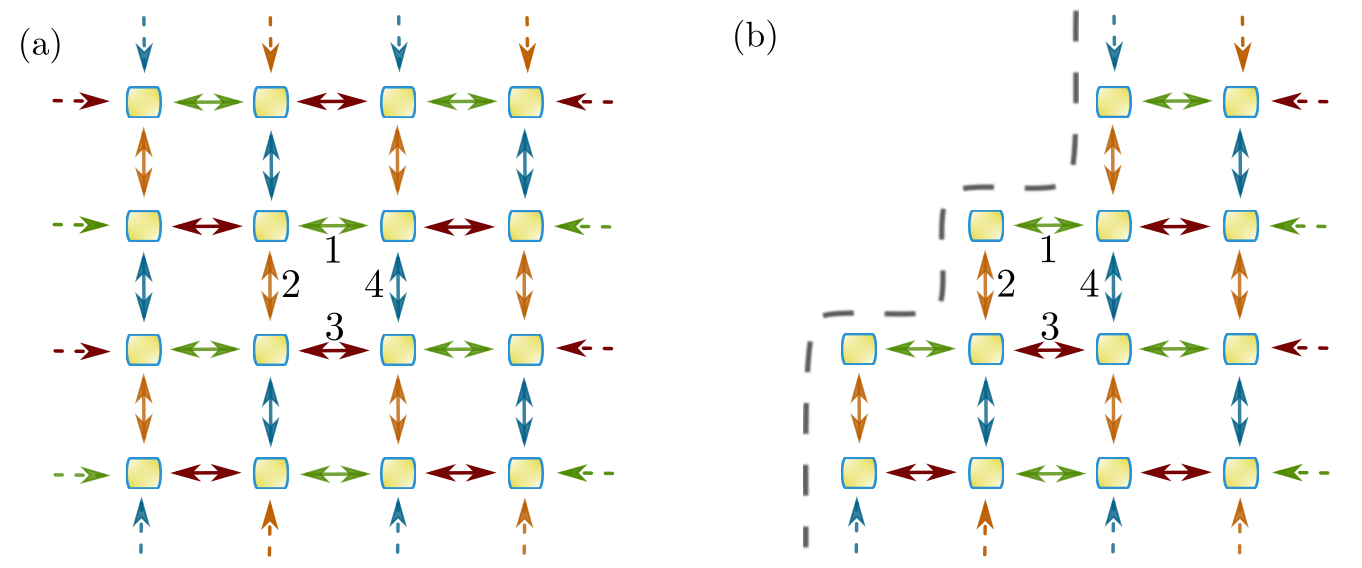}} \caption[Example of a non trivial Floquet phase]{Subfigure (a) shows the Floquet dynamics on the infinite plane.  The green swaps are applied first, orange are second, purple third and blue swaps are applied last.  By following the path of a qudit, we see that the dynamics after all four swaps is the identity.  In subfigure (b) however, a boundary has been introduced, and we can see that some of the qudits at the boundary have nontrivial dynamics. \label{fig:M}}
}
\end{figure}

The paradigmatic example is quite intuitive and works as follows.  Take a two-dimensional lattice $\Gamma$, where each site has $d$ dimensional qudits.  Suppose the system evolves by a simple QCA, which is a depth-four quantum circuit, where each unitary in the circuit is simply a swap, as depicted in figure \ref{fig:M} (a).  Let $\vec{e}_1$ and $\vec{e}_2$ be lattice basis vectors and denote the swap between sites $\vec{n}$ and $\vec{n}+\vec{e}_i$ as $S_{\vec{n},\vec{n}+\vec{e}_i}$.  Define the sublattice
\begin{equation}
 \begin{split}
  \Gamma_{\mathrm{even}} & =\{(n,m)\in\Gamma|n+m\ \mathrm{is\ even}\}.
 \end{split}
\end{equation}
We consider evolution via
\begin{equation}\label{eq:swapFloq}
 U=U_{[+\vec{e}_2]}U_{[-\vec{e}_1]}U_{[-\vec{e}_2]}U_{[+\vec{e}_1]},
\end{equation}
where
\begin{equation}
 \begin{split}
 \label{eq:swapfloq}
  U_{[\pm \vec{e}_i]} =\prod_{\vec{n}\in \Gamma_{\mathrm{even}}}S_{\vec{n},\vec{n}\pm\vec{e}_i}.
 \end{split}
\end{equation}
This is easiest to understand by looking at figure \ref{fig:M} (a).  If $\Gamma$ is the infinite plane or a torus, then it is clear that $U$ is actually just the identity:\ a qudit gets moved around a plaquette back to where it started.  On the other hand, if we introduce a boundary, then qudits in the bulk still stay put after one timestep, but qudits at the edge propagate.  See figure \ref{fig:M} (b).  Notice that in this case we have a one-dimensional QCA at the boundary:\ a shift of $d$-dimensional qudits each timestep.  By layering two of these two-dimensional systems on top of each other, where one has $d_1$-dimensional qudits at each site undergoing the evolution given above, and the other layer has $d_2$-dimensional qudits at each site undergoing the inverse evolution, we get a trivial bulk theory but a one-dimensional QCA on the boundary with index $d_1/d_2$.  The central idea of \cite{PFM16} is that this index classifies the chiral propagation on the boundary and hence the topological phase.  The examples here with qudits are then representatives for each different phase, as we can get a QCA with any boundary index with the right choice of $d_1$ and $d_2$.

Of course, any phase classification is physically meaningless without some sort of stability against perturbations.  We already know that the index is stable with respect to local perturbations (see section \ref{sec:ind}), so this classification should be stable with respect to local defects at the boundary, but what about bulk perturbations?  This is where assuming localization comes in.  Just as was the case for the integer quantum Hall effect, we need to demand that the systems under consideration have some form of localization to ensure that we get well-defined phases.  In this case, because we are dealing with interacting systems, we make the requirement that the system is many-body localized.  Let us take a quick detour to introduce \hlt{many-body localization in Floquet systems}.

We take as our definition for a many-body localized Floquet unitary $U$ that (for an infinite lattice system) there exists a localization length $\xi$, such that
\begin{equation}
 U=\prod_{\vec{n}}V_{\vec{n}},
\end{equation}
where $[V_{\vec{n}},V_{\vec{m}}]=0$ and $V_{\vec{n}}$ is supported on a region $N(\vec{n})$ centred on $\vec{n}$ of radius $\xi$.\footnote{Note that, if we relax the commutation criterion, then any unitary arising from finite-time Hamiltonian evolution can be approximated in this form as a finite-depth local unitary circuit.}  (Actually, we need only require that $V_{\vec{n}}$ is approximately supported on $N(\vec{n})$, allowing some exponential decay outside.)  As a very simple example, consider qubits with
\begin{equation}
 \label{eq:loc}
 W=\prod_{\vec{n}}\exp\left(-i\pi J_{\vec{n}}Z_{\vec{n}}\right),
\end{equation}
where $J_{\vec{n}}$ are real numbers.  This is just a product of on-site unitaries and so is clearly many-body localized in this sense.

We will use this many-body localization property to define the phases in a way that is stable against bulk perturbations.  To do this, we look at two unitary operators that act on the infinite system $\Gamma_{\mathrm{inf}}$ and the system with boundary $\Gamma_{\mathrm{fin}}$.  First, for dynamics on the unbounded system $\Gamma_{\mathrm{inf}}$, we have a decomposition of the dynamics into local commuting unitaries because we are assuming many-body localization.  Then, for the system with boundary, we may \emph{define} a unitary operator from a product of these local unitaries restricted to $\Gamma_{\mathrm{fin}}$ to be
\begin{equation}
 U_{\mathrm{lc}}=\prod_{\substack{\vec{n}\\ N(\vec{n})\subset \Gamma_{\mathrm{fin}}}}U_{\vec{n}},
\end{equation}
where the product only includes terms that are localized on the system $\Gamma$.  Importantly, this is not the actual dynamics of the system.  Instead, that is given by considering which physical Hamiltonian terms act on the system
\begin{equation}
 U_{\mathrm{dyn}}=\mathcal{T}\exp\left(-i\int_0^T\!\!\mathrm{d}t\, H_{\mathrm{fin}}(t) \right),
\end{equation}
where $H_{\mathrm{fin}}(t)$ is the Hamiltonian consisting of terms contained in $\Gamma_{\mathrm{fin}}$.  The exact details of $H_{\mathrm{fin}}(t)$ do not matter, but $U_{\mathrm{dyn}}$ itself does.  For example, in the simple four-step swap dynamics we considered in equation (\ref{eq:swapFloq}), $U_{\mathrm{dyn}}$ just involves swaps on the lattice between subsets of vertices.  In that case, where there is a boundary, some swaps do not occur because there is nothing to swap with, as in figure \ref{fig:M} (b).  

Now we define $Y=(U_{\mathrm{lc}})^{\dagger}U_{\mathrm{dyn}}$.  Because of the many-body localization property, this can only be a non trivial unitary \emph{close to the boundary}, as the local unitaries in $(U_{\mathrm{lc}})^{\dagger}$ cancel the bulk dynamics in $U_{\mathrm{dyn}}$.  We classify the phases according to the QCA index of $Y$ as a one-dimensional boundary QCA.  Let us see how this works for the swap example.

Suppose we augment the swap evolution defined in equation (\ref{eq:swapfloq}) with local commuting unitaries $W=\prod_{\vec{n}} W_{\vec{n}}$, so that now the dynamics are given by
\begin{equation}
U=WU_{[+\vec{e}_2]}U_{[-\vec{e}_1]}U_{[-\vec{e}_2]}U_{[+\vec{e}_1]}.
\end{equation}
On an infinite lattice or discrete torus, we know that $U_{[+\vec{e}_2]}U_{[-\vec{e}_1]}U_{[-\vec{e}_2]}U_{[+\vec{e}_1]}=\openone$.  Therefore,
\begin{equation}
 U_{\mathrm{lc}}=\prod_{\vec{n} \in\Gamma_{\mathrm{fin}}} W_{\vec{n}}.
\end{equation}
As we saw earlier, on the system with boundary $U_{[+\vec{e}_2]}U_{[-\vec{e}_1]}U_{[-\vec{e}_2]}U_{[+\vec{e}_1]}$ is just a shift along the boundary, which we will denote by $S_{B}$, so that $U_{\mathrm{dyn}}=WS_B$.  But then $Y=(U_{\mathrm{lc}})^{\dagger}U_{\mathrm{dyn}}$ is simply the shift of qudits along the boundary $S_B$.  This has index $d$.

There is an important point regarding this phase classification that we have glossed over.  In reality, a QCA will only be an approximation to the actual dynamics.  However, by choosing the neighbourhood size of the QCA to be at least as big as the Lieb-Robinson cone after one period of the dynamics $T$, this will be a very good approximation \cite{Osborne06}.  One might speculate that the classification via the index theory may not work simply because the index may not stop changing as we consider better and better approximations to the true dynamics (which will have larger and larger neighbourhoods), and it is conceivable that the index could converge to an irrational number.  It may be that this concern is unwarranted, especially if one considers, e.g., the case of qubits, where the index must always be an integer power of two.  Then the allowed index values are not dense in $\mathbb{R}$, so the index cannot converge to anything other than a power of two (or zero).  Nevertheless, it would be a worthwhile pursuit to extend the index theory to almost locality-preserving unitaries to back up this intuition.

A proposal for actually measuring the index characterizing information flow along the boundary of a two-dimensional Floquet system has been made \cite{DDP18}.

\one{
\subsection{Understanding phases of matter}
\label{sec:top_phases_2}
As mentioned before, a central problem in condensed matter physics is understanding quantum phases of matter.  Essentially, two systems described by local or quasi-local Hamiltonians $H_0$ and $H_1$ are in the same phase if there is a continuous path of Hamiltonians $H_s$ with $s\in [0,1]$ joining them.  However, there are further requirements that the path must preserve.  Typically, $H_0$ and $H_1$ will be gapped Hamiltonians, so the path $H_s$ must also be gapped.  In some further cases, $H_0$ and $H_1$ will have some symmetry, such as time-reversal or chiral symmetry, which must also be preserved along the path.  These latter cases are \hlt{symmetry-protected phases}.

Classifying such phases of matter has seen a lot of success in one dimension, with all Hamiltonians being classified.  In two dimensions, there is no full classification, but there are examples of Hamiltonians displaying non-trivial topological order, such as Levin-Wen models \cite{LW05}.  In higher dimensions, there are more exotic examples still, suggesting the possible phases are even more rich.  One such example in three dimensions, where QCAs play a role, is a specific case of the Walker-Wang model \cite{WW12,KBS13}.  This model is only topologically interesting at boundaries, and with boundary conditions chosen to include all possible boundary terms still supported in the system, it is believed that topological order appears.  The 3-fermion Walker-Wang model consists of a three-dimensional lattice with two qubits on the links (analogous to lattice gauge theories, where physical degrees of freedom live on the links between sites, rather than at the sites themselves).  Much like the toric code, an example of a lattice gauge theory, the Hamiltonian is a sum of commuting local Pauli operators.  Furthermore, with periodic boundary conditions, the ground state is unique \cite{HFH18}.

One concept that plays a role here is triviality of QCAs, introduced in section \ref{sec:partitioning}.  Recall that a QCA $u$ is called trivial if it is a product of local finite-depth circuits and shifts possibly by adding local ancillas.  In other words, if $u$ is trivial, then $u\otimes\openone$ can be decomposed in terms of shifts and a local finite-depth circuit.

In \cite{HFH18} a QCA $u_{WW}$ is explicitly constructed as a \hlt{disentangler} (defined below) of the ground state of this Walker-Wang model.  The consequences of this are interesting.  If this QCA is trivial, then there exists a \emph{two-dimensional} commuting projector Hamiltonian that realizes the topological order of the Walker-Wang model.  This works by considering the Walker-Wang model on a lattice that is infinite in the $x$ and $y$ directions but semi-infinite in the $z$ direction (with $z$ coordinate in $(-\infty,0]$).  If the QCA is trivial, then we could use it to disentangle the ground state in the two-dimensional boundary plane at $z=0$ from the bulk.  Then we could construct a commuting projector model for this two-dimensional plane, which has topological order.  

However, realizing the topological order of the Walker-Wang model with a two-dimensional commuting projector Hamiltonian is believed to be impossible \cite{HFH18}.  This in turn suggests that $u_{WW}$ cannot be trivial.  Further evidence is given in \cite{HFH18} for this:\ $u_{WW}$ is a Clifford QCA, but is nontrivial as a \emph{Clifford} QCA, meaning it cannot be written as a product of shifts and local-finite depth Clifford circuits.  Also if $u_{WW}$ is trivial, then there must exist a fermionic QCA in two dimensions that is nontrivial.  So there is a guarantee that some nontrivial QCA exists.

Let us look at some of the methods used to arrive at these results.  A useful concept introduced in \cite{HFH18} is that of \hlt{locally flippable separators}.  These are a generalization of the Clifford elements $\mathcal{X}_{\vec{n}}$ and $\mathcal{Z}_{\vec{n}}$ for qudits, introduced in section \ref{sec:Cliff}.  To understand this, first we need to define \hlt{separators}, which generalize the qudit $\mathcal{Z}_{\vec{n}}$ operators.
\begin{Definition}
 A separator for a lattice system consists of some label set $\mathcal{L}$ and unitary operators $\mathbf{Z}_{a}$ with $a\in\mathcal{L}$ that satisfy the following properties.
 \begin{enumerate}
  \item All $\mathbf{Z}_{a}$ have even parity (only necessary for fermionic degrees of freedom).
  \item $[\mathbf{Z}_{a},\mathbf{Z}_{b}]=0$ for all $a,b\in\mathcal{L}$.
  \item There is a $l\in\mathbb{Z}$, which is small compared to the dimensions of the lattice, such that every $\mathbf{Z}_{a}$ is contained in a hypercube with length of side $l$.
  \item For each $a$, there is an integer $D_a$ with $\mathbf{Z}_{a}^{D_a}=\openone$ and $D_a\geq 2$.
  \item For every possible choice of map $a\rightarrow \omega(a)$, where $\omega(a)$ is a $D_a$th root of one, there is a single state satisfying $\mathbf{Z}_{a}\ket{\psi}=\omega(a)\ket{\psi}$.
 \end{enumerate}
\end{Definition}
The states $\ket{\psi}$ in the definition are analogous to qudit stabilizer states.  In fact, a separator is called \hlt{trivial} if the labels can be chosen such that lattice sites $\vec{n}$ and labels $a\in\mathcal{L}$ are in one-to-one correspondence with $\mathcal{Z}_{\vec{n}}=\mathbf{Z}_{a}$ (for fermion modes, this is replaced by $2a^{\dagger}_{\vec{n}}a^{\ }_{\vec{n}}-1=\mathbf{Z}_{a}$).

Then a \hlt{local flipper} is analogous to the $\mathcal{X}_{\vec{n}}$ operators for qudits.
\begin{Definition}
 A local flipper $\mathbf{X}_{a}$ for a separator $\mathbf{Z}_{a}$ is a set of unitary operators with the same label set $\mathcal{L}$ that satisfies the properties below.
 \begin{enumerate}
  \item $\mathbf{X}_{a}$ has odd parity if $\mathbf{Z}_{a}$ is a fermionic operator.
  \item $[\mathbf{X}_{a},\mathbf{Z}_{b}]=0$ for all $a,b\in\mathcal{L}$ with $a\neq b$.  Otherwise, $\mathbf{Z}_{b}\mathbf{X}_{a}=e^{2\pi i/D_a}\mathbf{X}_{a}\mathbf{Z}_{b}$
  \item There is a $l\in\mathbb{Z}$, which is small compared to the dimensions of the lattice, such that every $\mathbf{X}_{a}$ is contained in a hypercube with length of side $l$.
 \end{enumerate}
\end{Definition}
Note that there is no assumption about the commutation or anticommutation relations between $\mathbf{X}_{a}$ and $\mathbf{X}_{b}$ for $a\neq b$.  A locally flippable separator is a separator that has a local flipper.

A central result in \cite{HFH18} is the following theorem for which we only state the qudit case.  The case with fermionic degrees of freedom involves some additional complexity because it is necessary to add ancillary degrees of freedom.
\begin{theorem}
 Any locally flippable separator can be mapped to a trivial separator by a QCA, which is said to \hlt{disentangle} the locally flippable separator.
\end{theorem}
The role this plays for the Walker-Wang model is that the ground state can be viewed as a locally flippable separator, and hence this theorem guarantees that the ground state can be disentangled via a QCA, namely $u_{WW}$.

These disentangling methods play a role also in \cite{Haah19}, where they are used to construct nontrivial Clifford QCA in three dimensions with qudits with \emph{any} prime dimension.  The examples again arise from topological models, in this case via coupled toric codes that have toplogical order on a surface.

QCAs play another role in classifying topological phases.  In \cite{SNB18}, the symmetries of subsystems of a two-dimensional system are described by QCAs.  In this case, this allows a deeper understanding of the connection between symmetry-protected topological order and measurement-based quantum computation.

In \cite{FHH20}, an exactly solvable $4+1$ dimensional model with a $\mathbb{Z}_2$ symmetry is introduced which is outside the classification by cohomology.  The Walker-Wang model again plays a role here, and we also see the effective realization of the symmetry on a subsystem (in this case a $3+1$ dimensional boundary) by a QCA.  Interestingly, this QCA is inequivalent to a finite-depth quantum circuit and is in fact nontrivial.
}

\subsection{QCAs and particles}
\label{sec:QCAs_and_particles}
In this and the following sections we will construct QCAs that have a clear interpretation in terms of particles.  To do this, it is natural to first look at the single-particle analogues of QCAs, which are called \hlt{discrete-time quantum walks}.  These consist of a single quantum particle evolving on a lattice unitarily, with the restriction that the particle can hop at most by a bounded distance each timestep.\footnote{In some contexts, e.g., quantum search algorithms, discrete-time quantum walks are defined without the restriction on the hopping range.}  This is the single-particle version of locality preservation.  These discrete-time quantum walks have many nice properties:\ with the right choice of unitary, we recover relativistic dynamics in the continuum limit, e.g., \cite{Bial94}; the symmetry-protected topological phases of these models can be classified \cite{KRBD10,CGG16,CGS16,CGS18}; and they also have more direct applications in quantum computing \cite{Venegas-Andraca2012}.

With this in mind, it is interesting to consider the possibility of constructing QCAs that are simply many-particle versions of these quantum walks.  As a consequence, in the non-interacting setting much of the analysis that applies for quantum walks (e.g., the classification of topological phases) should also apply to these QCAs.  On the other hand, we can also add interactions and arrive at entirely new physics.  Before considering how to upgrade quantum walks to QCAs, let us look at a particularly important example of a quantum walk. 

Consider a quantum particle on a discrete line with coordinates $n\in\mathbb{Z}$ and an extra two-dimensional degree of freedom with basis states $\ket{l}$ and $\ket{r}$.  Then an orthonormal basis of the whole Hilbert space is simply given by $\ket{n}\ket{l}$ and $\ket{n}\ket{r}$ where $n\in\mathbb{Z}$.  We define the unitary $\sigma_x=\ket{l}\bra{r}+ \ket{r}\bra{l}$, and then we can consider the following unitary implementing the evolution:
\begin{equation}
 U=e^{-im\sigma_x a}\bigg(\ket{r}\bra{r}S+ \ket{l}\bra{l}S^{\dagger}\bigg),
\end{equation}
where $m\in \mathbb{R}$, $a$ is the lattice spacing, and $S$ is a shift for a single particle, i.e., $S\ket{n}=\ket{n+1}$ for all $n$.  \two{Notice that the dynamics is translationally invariant.}  In the continuum limit, $m$ will be the mass.  The term on the right in brackets is a conditional shift:\ if the particle is in the $r$ state, it moves one step to the right, and if it is in the $l$ state, it moves one step to the left.  If we switch to momentum space, we have
\begin{equation}
 \ket{p}=\sqrt{a}\sum_{n\in\mathbb{Z}}e^{ipna}\ket{n},
\end{equation}
where $p\in(-\pi/a,\pi/a]$ is the lattice momentum and $\langle p | q \rangle = 2\pi\delta(p-q)$.  In this basis, the shift is just a phase $S\ket{p}=e^{-ipa}\ket{p}$.  Then, defining $\sigma_z=\ket{r}\bra{r}-\ket{l}\bra{l}$, we can rewrite the conditional shift as
\begin{equation}
 \ket{r}\bra{r}S+ \ket{l}\bra{l}S^{\dagger}=e^{-iP\sigma_za},
\end{equation}
where $P=\int \f{\mathrm{d}p}{2\pi} p\ket{p}\bra{p}$ is the lattice momentum operator.  This allows us to write the evolution operator as
\begin{equation}\label{eq:DQW}
 U=e^{-im\sigma_x a}e^{-iP\sigma_za}.
\end{equation}
Acting on smooth states (those with bounded momenta $|p|\leq \Lambda\ll \pi/a$), we can look at many timesteps $N=t/a$, to get
\begin{equation}
\begin{split}
 U^N & =\left(e^{-im\sigma_x a}e^{-iP\sigma_za}\right)^N\\
 & = \openone + -i(P\sigma_z +m\sigma_x)t + O(\Lambda^2 a^2)\\
 & = e^{-i(P\sigma_z +m\sigma_x)t}  + O(\Lambda^2 a^2).
 \end{split}
\end{equation}
As long as we choose our cutoff $\Lambda$ to grow sufficiently slowly as $a\rightarrow 0$, we get a sensible continuum limit, with the last line above corresponding to evolution via the \hlt{one-dimensional Dirac equation with Hamiltonian} $H=P\sigma_z +m\sigma_x$.  A lot of interesting work has been done using this simple yet powerful quantum walk \cite{Strauch06,BES07,dMD12}.  This includes studying, e.g., how mass mixes chiralities and Zitterbewegung \cite{Kurz08}.

We can also construct quantum walks in higher dimensional spaces that give rise to relativistic particles in the continuum limit \cite{Bial94}, which is in some sense generic \cite{FS14}.  For example, to discretize the Weyl equation, we take a cubic lattice with coordinates $\vec{n}\in\mathbb{Z}^3$ and let the particle have a two dimensional extra degree of freedom with orthonormal basis states $\ket{\!\uparrow_z}$ and $\ket{\!\downarrow_z}$, which correspond to spin in the continuum limit.\footnote{This is in contrast to the one-dimensional case, where $\ket{l}$ and $\ket{r}$ correspond to chirality in the continuum limit.}  The evolution operator is
\begin{equation}
\begin{split}
 U & =T_xT_yT_z\\
 & = e^{-iP_x\sigma_xa}e^{-iP_y\sigma_ya}e^{-iP_z\sigma_za},
 \end{split}
\end{equation}
where $T_b=\ket{\!\uparrow_{b}}\bra{\uparrow_{b}\!}S_b+ \ket{\!\downarrow_{b}}\bra{\downarrow_{b}\!}S_b^{\dagger}$ is a conditional shift along the $b$ axis depending on the spin in the $b$ direction.\footnote{$\ket{\uparrow_b}$ and $\ket{\downarrow_b}$ are the $+1$ and ${-}1$ eigenvectors of the $b$th Pauli matrix respectively.}  We can take the continuum limit of this quantum walk just as we did for the one-dimensional case, which is done rigorously (an explicit proof of convergence for regular initial conditions is given) for this quantum walk and more in \cite{AFN13}.  In the continuum limit, the dynamics is described by the \hlt{Weyl equation with Hamiltonian} $H=\vec{P}.\vec{\sigma}$ \cite{Bial94}.  It is straightforward to extend this to, e.g., the massive Dirac equation in three dimensional space.

There are at least two natural ways to upgrade single-particle quantum walk dynamics to QCAs.  One way goes via quantum lattice gases, and the other is by second quantization.  It is worth highlighting that we can discretize many physical models with quantum walks, e.g., Dirac or Weyl particles in curved spacetime \cite{dMB13,AFF16,AF16,MMK19} or in external gauge fields \cite{CRW13,BHS16,CGW19} or both \cite{dMB14}, and so these models can also be upgraded to QCAs.  As an aside, it is also possible to construct the Feynman propagator in some cases for quantum walks \cite{DMP15}.

\subsubsection{Lattice gases}
\label{sec:lattice}
The study of \hlt{quantum lattice gases} dates back to the early days of QCAs \cite{SB93,Meyer96,Meyer97a,BT98,YB02}, and a nice discussion can be found in \cite{16MS}.  The dynamics, which is generally taken to be translationally invariant, consists of two steps:\ a step where particles on the lattice propagate without interaction, and an interaction (or scattering) step between particles on the same site.  So the idea is that each lattice site has up to $d$ particles.  For each particle, there is a $k$-dimensional qudit, with algebras $\m{B}^j_{\vec{n}}\cong\m{M}_k$, which describes the states of the $j$th particles at that site.  As $\m{B}^j_{\vec{n}}$ corresponds to a $k$-dimensional quantum system, each particle has $k-1$ possible internal states, with the remaining state corresponding to the empty state $\ket{q_{\vec{n},j}}$, which represents the absence of the particle.  The algebra for each site is then $\mathcal{A}_{\vec{n}}=\otimes_j\m{B}^j_{\vec{n}}$.

To define the dynamics, we choose $d$ lattice vectors $\vec{e}_i$, and the propagation step simply shifts the $j$th particle by $\vec{e}_j$, meaning
\begin{equation}
 \sigma:\mathcal{B}_{\vec{n}}^j\rightarrow \mathcal{B}_{\vec{n}+\vec{e}_j}^j.
\end{equation}
The interaction step is a product of on-site unitaries, which are the same at each site:
\begin{equation}
 v_{\vec{n}}:\mathcal{A}_{\vec{n}}\rightarrow \mathcal{A}_{\vec{n}}.
\end{equation}
The dynamics comprises the propagation step followed by the scattering step, $u=v\sigma$.
Quantum lattice gas automata are usually defined with a quiescent state.  For finite systems this has the form
\begin{equation}
 \ket{q}=\bigotimes_{\vec{n}\in\Gamma}\bigotimes_{j=1}^k\ket{q_{\vec{n},j}},
\end{equation}
where $\Gamma$ denotes the spatial lattice (this state also makes sense for infinite systems, where it is defined by its expectation values).  This carries the interpretation of the completely empty state, i.e., the state with no particles at all present.

Let us consider an example in one dimension that corresponds to a lattice gas version of the single-particle evolution in equation (\ref{eq:DQW}).  \two{(Because that single-particle evolution is translationally invariant, the dynamics makes sense on a finite system with periodic boundary conditions, as long as the length of the system is large compared to the maximum distance the particle can hop over a single timestep, but the lattice gas approach also works for infinite systems.)}  Suppose that each lattice site has two qubits labelled by $l$ and $r$.  So the on-site algebra is $\mathcal{A}_{n}= \mathcal{B}^l_{n}\otimes\mathcal{B}^r_{n}$.  Let the quiescent state be the zero state of all qubits
\begin{equation}
 \ket{q}=\bigotimes_{n\in\Gamma}\bigotimes_{j=1}^k\ket{0},
\end{equation}
where $\Gamma=\{0,...,N-1\}$.  And if any qubit is in the $\ket{1}$ state, then there is a particle present.  For example if the $l$ qubit at site $n$ is in the one state, there is an $l$ type particle at site $n$.  We choose the propagation step to simply shift the $l$ qubits left and the $r$ qubits right, so
\begin{equation}
\sigma :
\begin{cases}
 & \mathcal{B}_{n}^l\rightarrow \mathcal{B}_{n+1}^l\\
 &\mathcal{B}_{n}^r\rightarrow \mathcal{B}_{n-1}^r.
 \end{cases}
\end{equation}
The interaction term at site $n$ is then implemented by the unitary (dropping the $n$ subscripts)
\begin{equation}
 V = \ket{00}\bra{00}+\cos(ma)\bigg[\ket{01}\bra{01}+\ket{10}\bra{10}\bigg]-i\sin(ma)\bigg[\ket{01}\bra{10}+\ket{01}\bra{10}\bigg]+e^{i\phi}\ket{11}\bra{11},
\end{equation}
where $\ket{ab}$ denotes a state with $a\in\{0,1\}$ $l$ particles present and $b\in\{0,1\}$ $r$ particles present.  Here we see that $V$ preserves the empty state, as $\ket{00}$ is taken to $\ket{00}$, while the left and right handed particles are mixed to some degree by the mass term.  We have some freedom left (after demanding that the single-particle sector evolves as the original quantum walk), which involves choosing the phase that is applied when both particle types are present $e^{i\phi}$.  We can use this to model an on-site interaction between the particles.

A general recipe for upgrading quantum walks to QCAs in a way similar to quantum lattice gases also appeared in \cite{SW04,Vogts09,CPM18}.  It is also interesting to look at this from a different perspective:\ we may ask when are QCAs equivalent to quantum lattice gases \cite{SL13}, in other words, when do they allow a particle interpretation in the way we have described here?  It has been shown that one can construct QCAs that cannot be written as quantum lattice gases despite propagating information \cite{16MS}.  Even taking products of quantum lattice gas automorphisms does not generate all QCAs \cite{16MS}.  Finally, it is worth mentioning that a general proposal for discretizing fermions interacting with gauge fields via quantum lattice gases was given in \cite{Yepez16,Yepez16b}.

\subsubsection{Fermions}
\label{sec:2ndQQCA}
Next we consider the second option, which is \hlt{second quantization}.\footnote{Incidentally, second quantization is not a particularly good name.  We are just constructing a many-particle quantum system that acts like many copies of a single-particle one; we are not making the theory more quantum!}  We will only look at fermionic second quantization here, but it is also possible for bosons.  In fact, the lattice gases of the previous section could be considered as hardcore bosons, which are bosons with an on-site repulsive interaction that is so strong that only one boson of that type may be present at each site.  Let us go through how fermionic second quantization works.  To every state $\ket{\alpha}$ of an orthonormal basis of our original single-particle system, we associate a fermion creation operator $\psi^{\dagger}_{\alpha}$, such that
\begin{equation}
\begin{split}
 \{\psi_{\alpha},\psi^{\dagger}_{\beta}\} & =\delta_{\alpha \beta}\\
 \{\psi_{\alpha},\psi_{\beta}\} & =0.
 \end{split}
\end{equation}
\two{I}n the many-particle Hilbert space there is an empty state $\ket{0}$, which is annihilated by all the annihilation operators, i.e., $\psi_{\alpha}\ket{0}=0$ for all $\alpha$.  Next we may define a linear map $\mathcal{Q}$ from single-particle states to creation operators, which we can also define for adjoints.  Given a single-particle state $\ket{v}=\sum_{\alpha}c_{\alpha}\ket{\alpha}$, $\mathcal{Q}$ is defined by
\begin{equation}
\begin{split}
 \mathcal{Q}\big(\ket{v}\big) & =\sum_{\alpha}c_{\alpha}\psi^{\dagger}_{\alpha}\\
 \mathcal{Q}\big(\bra{v}\big) & =\sum_{\alpha}c^{*}_{\alpha}\psi_{\alpha}.
 \end{split}
\end{equation}
This also extends to operators on the single-particle space in a natural way.  Consider the single-particle Hamiltonian $h=\sum_E E\ket{E}\bra{E}$, then we have
\begin{equation}
\begin{split}
 \mathcal{Q}\big(h\big) =\sum_{E}E\psi^{\dagger}_{E}\psi_E,
 \end{split}
\end{equation}
where $\psi_E=\m{Q}(\bra{E})$.  Now, we can upgrade our quantum walk operator in equation (\ref{eq:DQW}) to a QCA evolution operator.  First, denote a complete set of annihilation operators corresponding to the single-particle states $\ket{n}\ket{l}$ and $\ket{n}\ket{r}$ as $\psi_{nl}$ and $\psi_{nr}$.  For compactness, we can write this as a two-component operator $\psi_n= (\psi_{nr},\psi_{nl})^T$.  Analogously to the single-particle case, we can switch to momentum space, and we have the two-component momentum space annihilation operator
\begin{equation}
 \psi_{p}= \sqrt{a}\sum_n e^{-ipna}\psi_n.
\end{equation}
Then the \hlt{Dirac QCA} in one dimension has evolution operator given by
\begin{equation}\label{eq:DiracQCA}
 W= \exp\left(-ima\int\!\frac{\mathrm{d}p}{2\pi}\, \psi^{\dagger}_p\sigma_x\psi_p \right)\exp\left(-ia\int\!\frac{\mathrm{d}p}{2\pi}\,p\, \psi^{\dagger}_p\sigma_z\psi_p \right).
\end{equation}
It is straightforward to check that this gives the correct dynamics, e.g., $W^{\tau}\psi^{\dagger}_n\ket{0}$ evolves exactly as a one-particle state should over \two{$\tau$} timesteps.  The second term above is a conditional fermionic shift:\ it shifts $\psi_{nl}$ to the left and $\psi_{nr}$ to the right.  This is clearly locality preserving, as is the first term above:\ by switching to position space we see that
\begin{equation}
 \exp\left(-ima\int\!\frac{\mathrm{d}p}{2\pi}\, \psi^{\dagger}_p\sigma_x\psi_p \right)=\prod_n\exp\left(-ima\,\psi^{\dagger}_n\sigma_x\psi_n \right),
\end{equation}
which is a product of on-site unitaries.  Therefore, $W$ itself is also locality preserving.

Of course, the main reason for using second quantization is to look at interacting systems, which works via adding another unitary step to the dynamics, e.g.,
\begin{equation}\label{eq:int_ferm}
 V= \exp\left(-i\lambda\sum_n \psi^{\dagger}_{nl}\psi_{nl} \psi^{\dagger}_{nr}\psi_{nr} \right),
\end{equation}
which gives a phase if there are two particles at the same site, analogously to an energy penalty, e.g., from Coulomb repulsion of electrons.  Then the overall dynamics is given by $U=VW$, and this QCA is referred to as a \hlt{Thirring QCA} \cite{BDP18}.  This is essentially a second-quantized version of the interacting quantum walks considered in \cite{AAM12,BDM18}.  \two{For other studies of interacting two-particle quantum walks, see \cite{LVH12,KLM15}.}

\subsection{Quantum field theory}
A good reason to \hlt{discretize quantum field theories with QCAs} is that they are already algorithms for simulating the dynamics of QFTs on a quantum computer, which retain some of the nice properties of QFTs like strict causality, i.e., locality preservation.  However, any QCA discretization breaks Lorentz symmetry, something that also occurs for lattice quantum field theory.  Nevertheless, some attempts have been made to introduce a discrete version of Lorentz symmetry for QCAs \cite{AFF14}.  One approach is to consider a change of observer as a change of representation of the dynamics (without explicit reference to spacetime), which can be done in such a way as to reproduce the familiar Lorentz group in the low momentum regime for Dirac and Weyl QCAs \cite{BDP16}.  The emergence of a modification of special relativity, doubly special relativity (where there is a frame-invariant length as well as speed) was also studied for QCAs in one dimension \cite{BBD15}.

It is interesting that strongly interacting QFTs are often \emph{defined} to be the continuum limit of lattice quantum field theories (i.e., lattice models with local Hamiltonians) \cite{Creutz83}.  So QCAs may offer an alternative definition of some QFTs too.
An early work on the connection between QCAs and field theory is given for example in \cite{McGuigan03}, though this differs at least superficially from our notion of QCAs in the following way:\ to update the system to timestep $\tau+1$, one needs to know the state at times $\tau$ and $\tau-1$.
 
Recently, several quantum algorithms have been proposed that efficiently simulate some quantum field theories ($\phi^4$ theory \cite{JLP12} and the Gross-Niveau model \cite{JLP14}).  These algorithms did not use QCAs to simulate the dynamics but rather Suzuki-Trotter decompositions of the evolution operator ($e^{-iHt}$, where $H$ is the Hamiltonian of a suitable lattice QFT).
However, an important point to highlight here is that the difficult part of these proposals was not simulating the dynamics but rather the initial state preparation \cite{JLP12,JLP14}.  Nevertheless, it is interesting to consider QCAs as natural discrete models, which provide alternative options for quantum simulations.  One advantage of a QCA approach could be their simplicity:\ for example, to simulate the dynamics of Weyl fermions via a QCA, we need only implement swaps on the lattice \cite{BDT12,FS13}.  Interactions are then provided, as in equation (\ref{eq:int_ferm}), by on-site unitaries.

A nice example of a QCA was introduced in \cite{DdV87} that discretizes a fermionic QFT called the \hlt{massive Thirring model} (though the work actually pre-dates QCAs and so does not use the name QCA).  After taking a (not so rigorous) continuum limit of the QCA, the dynamics is given by the Hamiltonian
\begin{equation}\label{eq:Thirring}
 H=\int\!\frac{\textrm{d}p}{2\pi}\,\psi^{\dagger}_{p}(p\sigma_z+ m\sigma_x)\psi_{p}+2g\!\int\! \textrm{d}x\,\psi_l^{\dagger}(x)\psi_l^{\ }(x)\psi_r^{\dagger}(x)\psi_r^{\ }(x),
\end{equation}
where $g$ is a constant.  The rightmost term is the interaction term, so setting $g=0$ would simply give us free Dirac fermions in one dimension.
A QCA that discretizes this was already given in the previous section:\ the QCA is a lattice system with fermionic creation and annihilation operators with extra index $l$ and $r$, and the evolution operator is given by
\begin{equation}\label{eq:ThirringQCA}
 U= \exp\left(-i\lambda\sum_n \psi^{\dagger}_{nl}\psi_{nl} \psi^{\dagger}_{nr}\psi_{nr} \right)\exp\left(-ima\int\!\frac{\mathrm{d}p}{2\pi}\, \psi^{\dagger}_p\sigma_x\psi_p \right)\exp\left(-ia\int\!\frac{\mathrm{d}p}{2\pi}\,p\, \psi^{\dagger}_p\sigma_z\psi_p \right).
\end{equation}
Recall that $\psi_{pl}$ and $\psi_{pr}$ are the Fourier transforms of $\psi_{nl}$ and $\psi_{nr}$ respectively, and $\psi_p=(\psi_{pr},\psi_{pl})^{T}$.  Here, we can naively see that the continuum limit dynamics should correspond to the Hamiltonian in equation (\ref{eq:Thirring}).  How one shows this is given in detail in \cite{DdV87}, though the evolution operator for the QCA is slightly different in that case, but the basic idea is the same.  See also \cite{BDP18}.

A crucial point to bear in mind is that the discrete analogue of the physical ground state is \emph{not} the state $\ket{0}$ annihilated by all the annihilation operators $\psi_{n\alpha}$.  This is because the vacuum in QFT is a highly entangled state \cite{SW85}, and the QCA must reproduce this entanglement in the continuum limit.  Even in non-interacting QFTs, there is vacuum entanglement, so we cannot simulate the QFT ground state via the naive QCA vacuum state $\ket{0}$, which has no entanglement.
For the one-dimensional massless Dirac QCA \two{(equation (\ref{eq:DiracQCA}) with $m=0$)}, convergence of the QCA to the massless Dirac field was shown in \cite{Farrelly15}, a consequence of which was that there is a choice of \hlt{QCA vacuum} that converges to the quantum field vacuum.

There has been a lot of work on discretizing and simulating non-interacting fermionic fields by QCAs \cite{D'Ariano12a,D'Ariano12b,BDT12,FS13,BDP15,BDP15a,DP16,MC16,HSN20}, which consider Dirac and Weyl QCAs in higher dimensional spaces.  Furthermore, the one-dimensional Dirac QCA model was actually recently re-derived in a QFT context \cite{18DW}.  Of course, the more interesting cases are QCAs that discretize \emph{interacting} quantum field theories, as in the example given in equation (\ref{eq:ThirringQCA}).  Another  example of an interacting fermion QCA that discretizes an interacting quantum field theory, in this case quantum electrodynamics in one spatial dimension, was given in \cite{ABF19}.  Furthermore, for interacting bosonic quantum field theories, such as $\phi^4$ theory and non-abelian gauge theories, the discrete-time path integral from lattice QFT is actually \emph{exactly equivalent} to a QCA \cite{FS20}.

Another interesting discrete model of a quantum field theory is the so-called \hlt{QCA theory of light} \cite{16BDP}.  This is not dissimilar to a discretized version of the neutrino theory of light \cite{deBroglie34} and constructs a QCA approximating quantum electromagnetism (without matter) from two Weyl QCAs in three dimensional space.
At a very superficial level, the main idea is to use that the vector of Pauli matrices $\vec{\sigma}$ transforms like
\begin{equation}
 \exp\left(-\frac{i}{2}\vec{v}.\vec{\sigma}\right)\sigma_i\exp\left(\frac{i}{2}\vec{v}.\vec{\sigma}\right)=\sum_{j}\left[\exp\left(-i\vec{v}.\vec{J}\right)\right]_{ij}\sigma_j,
\end{equation}
where $J_i$ form a representation of the generators of the \two{Lie} algebra of $SO(3)$ given by $[J_i]_{jk}=-i\sum_{j,k}\varepsilon_{ijk}$,
where $i,j,k$ all take values in $\{1,2,3\}$.  Then, roughly speaking, the dynamics causes a bilinear fermion field $M_i({\vec{p}})=\psi_{\vec{p}}\,\sigma_i\Psi_{\vec{p}}$ to evolve as
\begin{equation}
M_i(p,t)=\sum_j[e^{-i2\vec{p}.\vec{J}t}]_{ij}\psi_{\vec{p}}\,\sigma_j\Psi_{\vec{p}}=\sum_j[e^{-i2\vec{p}.\vec{J}t}]_{ij}M_j(p,t),
\end{equation}
where $\psi_{\vec{p}}$ and $\Psi_{\vec{p}}$ are two different two-component fermion fields in momentum space.  This is equivalent to
\begin{equation}
 -i\partial_t \vec{M}({\vec{p}},t) = \vec{p}\times \vec{M}({\vec{p}},t).
\end{equation}
We then define the electric and magnetic field operators in terms of $M_i$
\begin{equation}
 \begin{split}
  E_i({\vec{p}},t) & =M_i^{\dagger}({\vec{p}},t)+M_i({\vec{p}},t)\\
  B_i({\vec{p}},t) & =i(M_i^{\dagger}({\vec{p}},t)-M_i({\vec{p}},t)).
 \end{split}
\end{equation}
If we also introduce the constraint that $\vec{p}.\vec{M}({\vec{p}},t)=0$, then we recover Maxwell's equations for $E_i$ and $B_i$.  Of course, there are many details that have been swept under the rug here, which are dealt with in \cite{16BDP}.  For example, one also has to ensure that the correct commutation relations for $E_i$ and $B_i$ are recovered.  Another point about the model is that it is not clear how to extend it to an interacting model via coupling to matter because the electric $\vec{E}(\vec{x})$ and magnetic fields $\vec{B}(\vec{x})$ are the fundamental quantities in the QCA, as opposed to the electromagnetic four-vector potential $A_{\mu}(\vec{x})$.  Due to phenomena like the Aharanov-Bohm effect, one would expect that any direct coupling of the $\vec{E}(\vec{x})$ and $\vec{B}(\vec{x})$ fields to matter would have to be non-local.

\section{Outlook and open problems}
We have looked at many interesting aspects of quantum cellular automata, including their use in quantum computation, QCA structure theorems, and their role in physics, e.g., as models of Floquet topological phases; in understanding phases of matter; or as quantum field theories.  \three{It is remarkable how versatile the QCA model has been, and how many varied applications it has had.  It is interesting then to speculate on the future of QCAs.  For example, currently there are several different efforts to discretize QFTs in terms of QCAs, and it is not inconceivable that the standard model of particle physics could have a QCA discretization.  An argument supporting this possibility is that many QFTs already have a QCA representation, from Dirac fermions to non-abelian gauge theories.  However, one roadblock to be overcome is fermion doubling, which to this day causes issues in classical and quantum simulations of lattice QFT.

Another fascinating topic is the connection between topological phases of matter and QCAs (e.g., as disentanglers).  It is fascinating to speculate on what can be gained by combining this approach with the tensor-network picture of QCAs, as the latter has proved to be a very useful representation for classifying phases of matter.  Progress in this regard has been fast:\ in the first version of this review, one of the open problems asked if one can generalize the correspondence between matrix product unitaries and QCAs to higher dimensional systems.  This was just solved in \cite{PC20}.  Because tensor networks help to understand the role of symmetries in phase classifications, it seems that there is a lot of potential to study phases further using QCAs.

With all this in mind, there are currently many fascinating open problems related to QCAs.  Below we list some concrete problems} as well as some possible ideas to tackle them.
\begin{enumerate}
\item In terms of quantum computation, one interesting question that arose before was whether there is a meaningful notion of QCA error correction?  If many different single or two-qubit gates are needed, as opposed to a small set of global operations, then it is not clear why QCA quantum computation would be sensible instead of simply focussing on the circuit model or the measurement-based model.  In other words, is there a comprehensive theory of quantum computation using QCAs including all necessary aspects?
 \item Fully classifying QCAs in dimensions greater than two remains open.  The classification in terms of the index and topology of the control space (section \ref{sec:higherindex}) was quite elegant and complete in two dimensions, so it is natural to wonder what extra invariants there are beyond two dimensions.  And how does the classification change depending on how we define equivalence of two QCAs (e.g., whether we allow appending ancillas at each site or not)?
 \item What can be said about essentially local QCAs, i.e., those with only approximately locality preserving dynamics, as opposed to the strict locality preservation we considered here?  If we allow tails with a suitable notion of decay, what results for strictly local QCAs survive?  This has relevance for, e.g., the use of QCAs as toy models for understanding Floquet topological phases in section \ref{sec:top_phases}.
 \item What can be said about irreversible QCAs, which had only received little attention in, e.g., \cite{RW96,BW03}?  At the moment, there are only constructive definitions.  For example, it makes sense to give examples of irreversible QCAs by taking a finite-depth circuit of local completely positive trace-preserving maps.  But is there any reasonable axiomatization that seems physically motivated and which also allows useful structure theorems?  Even for probabilistic classical cellular automata, this is not at all straightforward \cite{AFN11}.  \three{However, progress has been made very recently in \cite{PC20}, where the connection to tensor networks allowed some investigation of a very general definition of irreversible QCAs.  It seems likely that the methods considered there will allow further insight into irreversible QCAs, and maybe even some form of classification.}
 \item What can QCAs tell us about the structure of quasi-local algebras on infinite systems?  Boundary algebras (defined similarly to the support algebras of section \ref{sec:Qudit_index} \cite{Haah19}) can be used as a tool to define the one-dimensional index, but in higher dimensions, where the algebraic structure becomes very interesting, boundary algebras of QCAs give rise to nontrivial decompositions of quasi-local algebras \cite{Haah19}.  There may be deep connections between QCAs and the structure of infinite-lattice qudit or fermion algebras that remain to be found.
 \item When does it make sense to talk about ground states for QCAs?  A related question is whether there is any systematic way (other than exact diagonalization for finite systems) to find invariant states for QCAs, perhaps by getting some local constraints on the state.  One might then in some cases try to define a ground state for a QCA by finding an invariant pure state that has low entanglement.  For some ideas in this direction, see \cite{RW96}.
 \item What more can be said about QCAs with symmetries?  For example, when is there a symmetry-protected topological classification of QCAs in one dimension (with symmetries as in the tenfold way, which are a combination of time-reversal, chiral and particle-hole symmetry \cite{Ludwig15})?  Progress in this direction has already been made for one-dimensional QCAs via the matrix product unitary approach as discussed in section \ref{sec:MPU}, as these approaches allow some classification of some unitary and antiunitary symmetries for one-dimensional QCAs.
 \item How can we take continuum limits of interesting QCAs?  So far all models with continuum limits are integrable (solvable in some sense).  It is not clear how to go beyond this.  In lattice QFT, continuum limits are typically signalled by a second order phase transition.  Is there any analogue of this for QCAs?  \three{In some cases, the path integral formulation of lattice QFT is actually equivalent to a QCA \cite{FS20}, which might provide some insight.}  Another idea would be to use the framework for general continuum limits from \cite{Osborne19}.  \three{Also interesting in this regard are the coherent families of \cite{FHH19}, or the more abstract generalization of QCAs \cite{GSC20}.}
\end{enumerate}

\section*{Acknowledgements}
The author is grateful to Pablo Arrighi, C\'edric B\'eny, Alessandro Bisio, Tobias Geib, Jeongwan Haah, Norm Margolus, Tobias Osborne, Asif Shakeel and Reinhard Werner for useful suggestions and feedback.  The author was supported by the Australian Research Council Centres of Excellence for Engineered Quantum Systems (EQUS, CE170100009).

\appendix
\section{Infinite systems and quasi-local algebras}
\label{app}
In this section, we will look at the basics of C*-algebras, particularly those describing infinite spin lattices.  Some good references are \cite{Naaijkens13,Simon93,BR97}.  One of the key features of C*-algebras is the possible existence of inequivalent representations, which is the reason we deal with the abstract algebra itself rather than any particular representation in terms of operators on a Hilbert space.  This is related to the existence of superselection sectors, see \cite{Haag92}.

Let us start with the definition of a C*-algebra.
\begin{Definition}
A C*-algebra $\mathcal{A}$ is a complex algebra with a norm $\|\cdot\|$ in which it is complete as a vector space (making it a Banach space) and with an anti-linear map, $A\rightarrow A^*$, which is an involution, with the following properties:
\begin{enumerate}
\item $(AB)^*=B^*A^*$
\item $\|AB\|\leq \|A\|\|B\|$
\item $\|A^*\| = \|A\|$
\item $\|A^*A\|=\|A\|^2.$
\end{enumerate}
\end{Definition}
We will also assume that our C*-algebras have an identity.  It is useful to keep in mind that any closed set of bounded operators on a Hilbert space is a C*-algebra, where the * operation is just the adjoint.  And a special case of this is $\mathcal{M}_n(\mathbb{C})$, the set of $n\times n$ complex matrices.  Since it is finite dimensional, this example misses out on the subtleties associated to infinite dimensions.

We will focus on the C*-algebra description of quantum lattice systems.  We start with the lattice $\mathbb{Z}^d$ which has a finite dimensional quantum system associated to each point.  Then, to each lattice site there is an associated finite dimensional C*-algebra that is equivalent to $\mathcal{M}_n(\mathbb{C})$ for some $n$.  Next we associate a finite dimensional C*-algebra $\mathcal{A}_{\Lambda}$ to any finite region $\Lambda$ of the lattice, such that  $\mathcal{A}_{\Lambda}$ is isomorphic to a tensor product of the algebras associated to each site in $\Lambda$.  The C*-algebra of a subset $\Lambda^{\prime}$ of $\Lambda$ is identified with the elements in $\mathcal{A}_{\Lambda}$ that act like the identity on $\Lambda\backslash\Lambda^{\prime}$.  So far this is similar to the usual quantum theory with a finite number of subsystems, and it would be exactly equivalent if the lattice were finite.

For any finite $\Lambda$, the norm on elements of $\mathcal{A}_{\Lambda}$ is equivalent to the operator norm on the corresponding operators on the finite dimensional system associated to $\Lambda$.

When we consider the whole lattice $\mathbb{Z}^d$, the C*-algebra is a vector space that includes all elements that are non-trivial only on a finite subset of $\mathbb{Z}^d$.  And to get the full C*-algebra, we complete this vector space, by including Cauchy sequences of elements that converge in the norm.  So the total C*-algebra contains local elements that are non-trivial on a finite subset of $\mathbb{Z}^d$ and quasi local elements, which can be approximated arbitrarily well by sums of local elements.

Let us define isomorphisms of C*-algebras (*-isomorphisms).  First, a morphism (or *-morphism) $\pi$ between two C*-algebras $\mathcal{A}$ and $\mathcal{B}$ is a linear map preserving the algebraic structure:
\begin{enumerate}
 \item $\pi(A)\pi(B)=\pi(AB)\ \ \forall A\in\mathcal{A},\ \forall B\in\mathcal{B},$
 \item $\pi(A^*)=\pi(A)^*\ \ \forall A\in\mathcal{A}$.
\end{enumerate}
An isomorphism is an invertible morphism.  An example of an isomorphism on our quantum lattice system in the case of qubits is the conjugation of each site's algebra by $X_n$, i.e., $A\rightarrow X_n A X_n$ for any $A\in \m{A}_n$.  It follows that the elements of the algebra associated to larger finite regions are conjugated by tensor products of $X$s.  Intuitively, this looks like applying the map $A\rightarrow \mathbb{X} A \mathbb{X}$ for any $A\in\m{A}$, where $\mathbb{X}=\otimes_{m\in\mathbb{Z}}X_m$.  Notice that $\mathbb{X}$ is not in the C*-algebra itself.

As mentioned above, a closed set of bounded operators on a Hilbert space is an example of a C*-algebra.  Thanks to the following theorem by Gel'fand and Naimark, it is the only example.
\begin{theorem}
 Any C*-algebra is isomorphic to a C*-algebra of operators on a Hilbert space.
\end{theorem}
But the catch is that this Hilbert space may not be separable.

A positive element of a C*-algebra, written as $A\geq 0$, is self adjoint with a real positive spectrum.  Saying an element $A$ is positive is equivalent to saying that there exists a $B\in\mathcal{A}$ such that $A=B^*B$. This allows us to define states on a C*-algebra.
\begin{Definition}
 A state on a C*-algebra $\omega$ is a linear functional that is positive, meaning $\omega(A)\geq 0$ if $A\geq 0$ and normalised, meaning $\omega(\openone)=1$.
\end{Definition}

For the C*-algebra $\mathcal{M}_n(\mathbb{C})$ corresponding to an $n$-dimensional qudit, we can identify states with elements of $\mathcal{M}_n(\mathbb{C})$, since finite dimensional vector spaces are isomorphic to their dual spaces.  We recover the familiar density matrix formalism, with $\rho(A)=\textrm{tr}[\rho A]$, where $\rho$ is a positive semi-definite matrix in $\mathcal{M}_n(\mathbb{C})$, and $\textrm{tr}[\cdot]$ is the trace.

For our quasi-local algebra, we do not need to think in terms of abstract functionals.  It is equivalent to associate a density operator to each finite subregion of the lattice.  This works provided the family of states satisfies a natural consistency condition: given $\rho_{\Lambda^{\prime}}$ and $\rho_{\Lambda}$ are density operators associated to the finite regions $\Lambda^{\prime}$ and $\Lambda$, with $\Lambda^{\prime}\subseteq\Lambda$, then
\begin{equation}
 \rho_{\Lambda^{\prime}}= \textrm{tr}_{\Lambda\setminus\Lambda^{\prime}}[\rho_{\Lambda}].
\end{equation}


A representation of a C*-algebra $\mathcal{A}$ is a morphism $\pi$ from $\mathcal{A}$ to a set of bounded operators on a Hilbert space.  A faithful representation is one where $\pi$ is an isomorphism.  If we have a representation $\pi$ of a C*-algebra acting on the Hilbert space $\mathcal{H}$, then, for any $\phi\in\mathcal{H}$, the closure of the subspace spanned by $\pi(A)\phi$ with $A\in\mathcal{A}$, $\mathcal{H}_{\phi}$, is left invariant by the action of all $\pi(A)$.  So we may as well focus on representations with $\mathcal{H}_{\phi}=\mathcal{H}$.  These are called cyclic representations and $\phi$ is called a cyclic vector.  Note that if we normalise $\phi$ then it defines a state on $\mathcal{A}$ via
\begin{equation}
 \omega(A)=(\phi,\pi(A)\phi).
\end{equation}
The following theorem provides the GNS construction, which allows us to construct a representation using any state on $\mathcal{A}$.
\begin{theorem}
\label{th:1}
 Given a state $\omega$ on $\mathcal{A}$, there exists a cyclic representation of $\mathcal{A}$ with a cyclic vector $\phi$, such that
\begin{equation}
 \omega(A)=(\phi,\pi(A)\phi).
\end{equation}
Furthermore, this representation is unique up to unitary equivalence, meaning that, if there are two cyclic representations of $\mathcal{A}$ on $\mathcal{H}_1$ and $\mathcal{H}_2$ such that
\begin{equation}
  \omega(A)=(\phi_1,\pi_1(A)\phi_1)=(\phi_2,\pi_2(A)\phi_2),
\end{equation}
then there is a unique unitary map $U$ from $\mathcal{H}_1$ to $\mathcal{H}_2$ such that $\phi_2=U\phi_1$ and $\pi_2(A)=U\pi_1(A)U^{-1}$ for all $A\in\mathcal{A}$.
\end{theorem}
The proof entails using the state to define an inner product on the C*-algebra via\footnote{If $\omega(A)=0$ for some $A\geq 0$, then we have to work with the quotient of $\mathcal{A}$ by all $B\in\mathcal{A}$ with $\omega(B^*B)=0$.}
\begin{equation}
 \forall\ A,B\in\mathcal{A}\ \ (A,B)\equiv\omega(A^*B).
\end{equation}
This inner product gives us a Hilbert space with vectors $A\in\mathcal{A}$.  In particular, $\openone\in\mathcal{A}$ is thought of as $\phi$, since
\begin{equation}
 \omega(\openone A\openone)=\omega(A)\equiv(\phi,\pi_{\omega}(A)\phi).
\end{equation}
Then the representation of the elements in $\mathcal{A}$ as operators on this Hilbert space is simply
\begin{equation}
 \pi(B)\phi_A\equiv BA,
\end{equation}
where $A$ is thought of as a vector here.  Although mathematically we think of $A\in\mathcal{A}$ as vectors in a Hilbert space, it is also useful to think of these vectors as arising from the action of $A$ on the state $\omega\equiv\phi_{\omega}$.

A good example of a representation involves the spin lattice C*-algebra.  Take $\omega$ to be the state with all spins pointing up along the $z$-direction, meaning
\begin{equation}
 \omega_{\Lambda}\equiv\bigotimes_{i\in\Lambda}\begin{pmatrix}
 1 & 0 \\
 0 & 0
\end{pmatrix}
\end{equation}
for all finite regions $\Lambda$.  A basis for $\mathcal{H}_{\omega}$ is given by vectors we can think of as
\begin{equation}\label{eq:spin}
 \bigotimes_{i\in\Lambda}\ket{\downarrow}_{i}\bigotimes_{i\in\mathbb{Z}^d\backslash\Lambda}\ket{\uparrow}_{i},
\end{equation}
for all finite regions $\Lambda$.  So the representation is spanned by all states with only a finite number of spins pointing down.

Let us now look at automorphisms, which are just isomorphisms from the C*-algebra to itself.  The following interesting result is a consequence of theorem \ref{th:1}.
\begin{corollary}
 Given an automorphism $\alpha$ of a C*-algebra $\mathcal{A}$, and a state $\omega$ with cyclic representation $(\pi_{\omega},\mathcal{H}_{\omega},\phi_{\omega})$, then $\alpha$ can be implemented by a unitary on $\mathcal{H}_{\omega}$ such that
\begin{equation}
 \pi(\alpha(A))=U\pi(A)U^{-1}
\end{equation}
and
\begin{equation}
 U\phi_{\omega}=\phi_{\omega}
\end{equation}
if and only if
\begin{equation}
 \omega(A)=\omega(\alpha(A)),
\end{equation}
and $U$ is uniquely determined.
\end{corollary}

An example of an automorphism that is \textit{not} implementable in a representation as a unitary is the application of Pauli $X$s everywhere in the spin lattice representation given above in equation (\ref{eq:spin}).  This follows since flipping all the spins from up to down is not an operation that is allowed in the Hilbert space.

\section{QCAs with fermions}
\label{app:fermions}
For fermionic QCAs, instead of having qudits at each site we have some number of fermionic modes.  (We could also consider hybrid QCAs with both qudits and some fermion modes at each site.)  The simplest way to describe fermion systems is via their creation and annihilation operators (creation operators are just adjoints of annihilation operators).  A nice discussion of finite fermionic algebras is given in \cite{Nielsen05}.  We denote the annihilation operators by $a_{\alpha}$, where $\alpha$ is a label in some set.  For example, if we have a single fermion mode on each lattice site, we would take $\alpha$ to be a vector in $\mathbb{Z}^d$.  For simplicity, let us choose the index to run over $\mathbb{Z}$.  The creation and annihilation operators obey the anticommutation relations:
\begin{equation}
 \begin{split}
  \{a_{\alpha},a_{\beta}\} & =0\\
  \{a_{\alpha},a^{\dagger}_{\beta}\} & =\delta_{\alpha\beta}.
 \end{split}
\end{equation}
Furthermore, we can construct our Hilbert space by starting with a state $\ket{0}$ that satisfies
\begin{equation}
 a_{\alpha}\ket{0}=0
\end{equation}
for all $\alpha$.  This state has the interpretation as being empty.  The empty state also makes sense for infinite lattices of fermions, where it is defined to be a positive normalized functional $\omega_{0}$ satisfying
\begin{equation}
 \omega_{0}[a^{\dagger}_{\alpha}a_{\alpha}]=0
\end{equation}
for all $\alpha$.  The operator $a^{\dagger}_{\alpha}a_{\alpha}$ counts the number of fermions in the mode labelled by $\alpha$ and has eigenvalues $0$ and $1$.  The state $a^{\dagger}_{\alpha}\ket{0}$ has the interpretation as having a single fermion present.  Furthermore, it follows from the anticommutation relations that $a^{\dagger 2}_{\alpha}=0$, which means that we can have at most one fermion in each mode.

We can also define Majorana fermion operators, which are sometimes more convenient to work with.  These are defined to be
\begin{equation}
 \begin{split}
  c_{2\alpha} & =a_{\alpha}^{\dagger}+a_{\alpha}\\
  c_{2\alpha+1} & =i(a_{\alpha}^{\dagger}-a_{\alpha}).
 \end{split}
\end{equation}
These are self-adjoint and satisfy the anticommutation relations
\begin{equation}
  \{c_{n},c_{m}\}  =2\delta_{nm}.
\end{equation}

Most of the details of the quasi-local algebra approach to defining infinite systems of fermions are similar to the qudit case.  We assign fermionic algebras to regions of the lattice.  The main difference with the qudit case is that creation and annihilation operators from different regions or modes anticommute.  
See \cite{BR97} for a detailed discussion.


Another way to think about fermion systems is via graded algebras.  In a graded algebra we have some basis of operators with the following property:\ each element is either even or odd.  This also extends easily to hybrid systems of qudits and fermions.  For example, the fermion algebra we have already seen is a graded algebra, with, e.g., $\openone$ and $a^{\dagger}_{\alpha}a_{\alpha}$ being even elements whereas $a_{\alpha}$ are odd.  A linear combination of an even and an odd operator is neither even nor odd.

Let us take a step back and start with the graded Hilbert space $\m{H}$ (really we mean $\mathbb{Z}_2$ graded).  We decompose $\m{H}=\m{H}^e\oplus\m{H}^o$, where $\m{H}^o$ is the odd subspace and $\m{H}^e$ is the even subspace.  These will correspond to the subspaces with odd and even fermion number respectively.  One can also use the notation $\mathbb{C}^{p|q}$, where $p$ and $q$ are the dimensions of the even and odd subspaces respectively.  We define even operators to be those that preserve evenness or oddness, e.g., they map even states to even states.  Odd operators are then those that map the even subspace to the odd subspace and vice versa.

This leads naturally to the graded commutator.  We can write this as $[\cdot, \cdot\}$, and it is defined to be bilinear and satisfy
\begin{equation}
\begin{split}
[A,B\} & =\{A,B\}\ \ \mathrm{if}\ A\ \mathrm{and}\ B\ \mathrm{are\ odd}\\
 [A,B\} & =[A,B]\ \ \ \mathrm{if\ at\ least\ one\ of\ } A\ \mathrm{and\ } B\ \mathrm{are\ even.}
 \end{split}
\end{equation}
So, for example, if $A$ and $B$ are odd and $C$ is even, we have
\begin{equation}
[A,B+C\} =\{A,B\}+[A,C].
\end{equation}

We can introduce a graded tensor product too.  The tensor product of two graded Hilbert spaces is also a graded Hilbert space.  A tensor product of vectors that are both even or both odd is even.  A tensor product of vectors that have opposite grading is odd.  So we denote the graded tensor product of $\m{H}_1$ and $\m{H}_2$ by $\m{H}_1\otimes_g\m{H}_2$.  Then we have
\begin{equation}
\begin{split}
\left(\m{H}_1\otimes_g\m{H}_2\right)^e & =\left(\m{H}^e_1\otimes\m{H}_2^e\right)\oplus \left(\m{H}^o_1\otimes\m{H}_2^o\right)\\
\left(\m{H}_1\otimes_g\m{H}_2\right)^o & =\left(\m{H}^e_1\otimes\m{H}_2^o\right)\oplus \left(\m{H}^o_1\otimes\m{H}_2^e\right).
\end{split}
\end{equation}
The graded tensor product of graded algebras is similar:\ a graded tensor product of even operators is an even operator, for example.
After taking the graded tensor product of two graded algebras $\m{A}$ and $\m{B}$, even elements of, e.g., $\m{A}$ and all elements of $\m{B}$ commute.  Furthermore, we say two algebras $\m{A}$ and $\m{B}$ graded commute if $[A,B\}=0$ for all $A\in\m{A}$ and all $B\in\m{B}$.

\newpage
\addcontentsline{toc}{section}{References}

\bibliographystyle{unsrtnat}

\end{document}